\documentclass[
 aip,
 amsmath,amssymb,
 reprint
]{revtex4-1}

\usepackage{graphicx}
\usepackage{dcolumn}
\usepackage{cancel}
\usepackage[utf8]{inputenc}
\usepackage[T1]{fontenc}
\usepackage{bm}
\usepackage{etoolbox}
\usepackage{color}
\usepackage{soul}
\usepackage{empheq}
\usepackage{multirow}
\usepackage{comment}
\usepackage{url}
\usepackage{hyperref}

\newcommand{\vect}[1]{\boldsymbol{#1}}
\newcommand{\diff}{\mathrm{d}}
\newcommand{\im}{\mathrm{i}}
\newcommand{\e}{\mathrm{e}}
\newcommand{\ld}{\lambda_\mathrm{D}}
\newcommand{\taud}{\tau_\mathrm{D}}

\newcommand{\vecez}{\hat{\vect{e}}_z}
\newcommand{\veckext}{\vect{k}_{E}}
\newcommand{\kext}{k_{E}}
\newcommand{\bfk}{\vect{k}}
\newcommand{\bfK}{\vect{K}}
\newcommand{\Kext}{K_{E}}
\newcommand{\bfr}{\vect{r}}
\newcommand{\vecEext}{\vect{E}^{\rm ext}}
\newcommand{\mcE}{\mathcal{E}}
\newcommand{\bmmcE}{\tilde{\bm{\mathcal{E}}}}
\newcommand{\tdrhoN}{\tilde{\rho}_N}
\newcommand{\tdrhoZ}{\tilde{\rho}_Z}
\newcommand{\Npairs}{N_{\rm pairs}}
\newcommand{\Nions}{N_{\rm ions}}

\makeatletter
\def\@email#1#2{%
 \endgroup
 \patchcmd{\titleblock@produce}
  {\frontmatter@RRAPformat}
  {\frontmatter@RRAPformat{\produce@RRAP{*#1\href{mailto:#2}{#2}}}\frontmatter@RRAPformat}
  {}{}
}%
\makeatother
\begin{document}

\preprint{AIP/123-QED}

\title{Coupled concentration-charge dynamics in asymmetric 1:1 electrolytes, local transient response and fluctuations}

\author{Th\^e Hoang Ngoc Minh}
\affiliation{
Department of Civil and Environmental Engineering, Princeton University, Princeton, New Jersey 08544, USA
}
\affiliation{ 
Physicochimie des \'Electrolytes et Nanosyst\`emes Interfaciaux, Sorbonne Universit\'e, CNRS, Paris, 75005, France
}

\author{Sleeba Varghese}
\affiliation{ 
Physicochimie des \'Electrolytes et Nanosyst\`emes Interfaciaux, Sorbonne Universit\'e, CNRS, Paris, 75005, France
}

\author{Benjamin Rotenberg}
\affiliation{ 
Physicochimie des \'Electrolytes et Nanosyst\`emes Interfaciaux, Sorbonne Universit\'e, CNRS, Paris, 75005, France
}
\affiliation{Réseau sur le Stockage Electrochimique de l’Energie (RS2E), FR CNRS 3459, 80039 Amiens Cedex, France}

\date{20 January 2026}

\begin{abstract}
We investigate the coupled dynamics of concentration and charge in asymmetric 1:1 electrolytes, focusing on the interplay between diffusion asymmetry and external electric fields. Using Brownian dynamics simulations and linearized stochastic density functional theory (SDFT), we analyze the transient response of charge and number currents to inhomogeneous electric fields, as well as the steady-state spatio-temporal fluctuations under uniform fields. Our results reveal that asymmetry in ionic diffusion coefficients introduces a non-trivial coupling between charge and number transport, which modifies the two relaxation modes already present in symmetric electrolytes -- a fast one associated with charge relaxation and a slow one linked to ambipolar diffusion. The dynamics are further modulated by the applied field, which enhances diffusion, alters screening lengths, and induces oscillatory behavior in the relaxation modes. The SDFT framework provides closed-form expressions for the intermediate scattering matrix, capturing the dynamics of density fluctuations and cross-correlations between number and charge. These predictions are validated by simulations, demonstrating excellent agreement across a wide range of wave vectors, both at equilibrium and under a finite electric field. Our findings highlight the critical role of diffusion asymmetry and external fields in tuning the transport properties of electrolytes, with implications for nanofluidic devices, energy harvesting, and iontronic circuits. This work bridges theoretical insights with practical applications, offering a robust framework for understanding and controlling electrolyte dynamics in asymmetric systems.
\end{abstract}

\maketitle

\section{Introduction}
\label{sec:intro}

The dynamics of ions in electrolyte solutions are fundamental to a broad spectrum of natural and technological processes, including ion exchange in biology, pollutant retention, capacitive energy storage, osmotic energy harvesting, desalination, and electrochemical sensing~\cite{Hille2001-es, liu_molecular-level_2022, Appelo2005-qm, simon_perspectives_2020, siria_new_2017, laszlo_decoding_2014}. At the nanoscale, where confinement and interfacial effects dominate, ion transport is governed by complex couplings between geometry, electrostatics, thermal fluctuations, hydrodynamics, and surface interactions \cite{kavokine_fluids_at_the_nanoscale_2021}. These couplings give rise to a wide variety of phenomena, such as charge hyperuniformity \cite{blum_perfect_82, hoang_hyper_23}, Casimir forces \cite{lee_2018_casimir}, enhanced electrokinetic effects \cite{siria_giant_2013, Balos2020-gf, mondal_2021_anomalous_dielectric}, or low-frequency electrical noise in nanopores \cite{smeets_2008_noise_in_ssn, secchi_scaling_2016, knowles_2020_noise, gravelle_adsorption_2019, marbach_intrinsic_2021, robin_disentangling_1_f_2023}. 

A robust framework to describe these dynamics is provided by the intermediate scattering functions and their Fourier-Laplace transforms, the dynamical structure factors \cite{hansen_mcdonald_theory_of_simple_liquids_book}. These functions encode \textit{spatio-temporal correlations of ionic density fluctuations}, linking equilibrium properties to transport coefficients via fluctuation-dissipation relations. For example, the charge-charge intermediate scattering function is related to frequency-dependent electrical conductivity, dielectric relaxation, or quadrupolar NMR relaxation \cite{yamaguchi_theoretical_frequency_dependent_conductivity_2007, hoang_electrical_noise_2023, hoang_frequency_field_2023}, while mass-charge correlations are related to electro-acoustic effects~\cite{yamaguchi_theoretical_2007, sedlmeier_2014_charge_mass}. Although seldom directly measured in experiments, such functions have played a fundamental role in theory since the seminal work of Debye, H\"uckel and Onsager~\cite{DebyeHuckel1923,onsager_report_1927,onsager_theories_1933,onsager_relaxatio_1957}. They underpin key developments in implicit-solvent theories, including the mean spherical approximation (MSA) for Ornstein--Zernike closures and mode-coupling theory (MCT)~\cite{bernard_conductance_1992, bernard_self_diffusion_1992, bernard_binding_msa_1996, dufreche_2005_transport,dufreche_2002_ionic_self, dufreche_2008_electrostatic_relaxation, chandra_frequency_1993, chandra_frequency_2000, banerjee_ions_2019, jardat_self-diffusion_2012, yamaguchi_dynamic_2011}. 

More recently, dynamical density functional theory (DDFT)~\cite{te_vrugt_classical_2020} has enabled refined predictions for diffusion, fluctuation-induced forces, conductivity, dielectric response, and fluctuation spectra~\cite{bazant_diffuse-charge_2004, donev_fluctuating_hydro_2019, peraud_fluctuating_2017, mahdisoltani_transient_2021, illien_2024_stochastic, demery_conductivity_2016, bonneau_2023_temporal, bonneau_2024_frequency, bonneau_2025_stationary, hoang_hyper_23, poncet_2017_universal, berthoumieux_2024_nonlinear_conductivity, bernard_2023_analytical, zorkot_current_2016, zorkot_current_sPNP_2018}. While at the mean-field level, DDFT is closely related to Poisson-Nernst-Planck (PNP) theory, it can formally incorporate effects that become critical in concentrated systems, such as thermal fluctuations of the solvent (via Dean-Kawasaki coarse-graining or stochastic PNP~\cite{kawasaki_stochastic_1994,dean_langevin_1996, illien2025dean}), hydrodynamic interactions (\textit{e.g.} through Stokesian flows~\cite{peraud_fluctuating_2017}), and finite-size corrections (\textit{e.g.} via modified Coulomb potentials~\cite{avni_conductivity_2022}). In parallel, Brownian or molecular dynamics simulations can also predict the wave vector and time- or frequency–dependent fluctuations, that are generally compared to theory or experiments only in the macroscopic limit ($k\to0,\omega\to0$)~\cite{giaqinta_1978_generalized_hydro, caillol_theoretical_1986, caillol_dielectric_1987, felderhof_fluctuation_1980, pollock_frequency-dependent_1981, chandra_frequency_2000, yamaguchi_brownian_2011, cheng_computing_2020}.

In asymmetric electrolytes where cations and anions have different mobilities, the coupled relaxation of equilibrium concentration and charge fluctuations results in various modes, including the relaxation of the charge restoring local electroneutrality and the diffusion of the ion pair restoring concentration homogeneity with the so-called Nernst-Hartley diffusion coefficient~\cite{Robinson1959}. Studies within PNP theory highlighted the role of asymmetry in ion valencies or mobilities on the frequency-dependent impedance  of electrolytic double-layer capacitors~\cite{lelidis_effect_2005, Barbero2008, Antonova2020}, characterizing their linear response to small oscillatory voltages, or on their time-dependent response even beyond the linear regime~\cite{palaia_charging_2025, palaia_poisson-nernst-planck_2025}. SDFT has also been used to investigate ionic fluctuations in driven symmetric electrolytes, leading to Casimir-like long-range forces~\cite{mahdisoltani_long_2021, mahdisoltani_transient_2021,du_2024_correlation, du_repulsive_2025}, or to highlight the role of ionic correlations on their temporal or frequency-dependent response~\cite{demery_conductivity_2016, bonneau_2023_temporal, bonneau_2024_frequency, bonneau_2025_stationary}.

Here, we study the effect of the asymmetry in ion diffusivity on the 
coupled concentration and charge spatio-temporal fluctuations in driven electrolytes. Our study focuses on two complementary observables: (1) the transient response of charge and number currents to spatially inhomogeneous electric fields, and (2) the steady-state spatio-temporal fluctuations under uniform electric fields. By contrasting symmetric and asymmetric cases, we demonstrate how unequal ionic mobilities reshape relaxation modes, fluctuation-response behavior, and transport properties. Our SDFT framework, validated by simulations, provides closed-form expressions for the intermediate scattering functions capturing the dynamics across a wide range of wave vectors, both at equilibrium and under a finite electric field.

The paper is organized as follows: In Section~\ref{sec:methods}, we introduce the Brownian model of electrolyte solutions, the properties of interest, and the simulation and theoretical methods. Section~\ref{sec:results} presents our results, including the transient current response (Section~\ref{ssec:res_currents}) and the dynamical structure matrix at equilibrium and under non-equilibrium steady-state (NESS) (Section~\ref{ssec:res_Fkt}). Finally, Section~\ref{sec:conclusion} summarizes our findings and offers some perspectives on the implications and relevant extensions of this work.

\section{System and methods}
\label{sec:methods}

\begin{figure}[ht!]
    \begin{center}
        \includegraphics[width=0.49\textwidth]{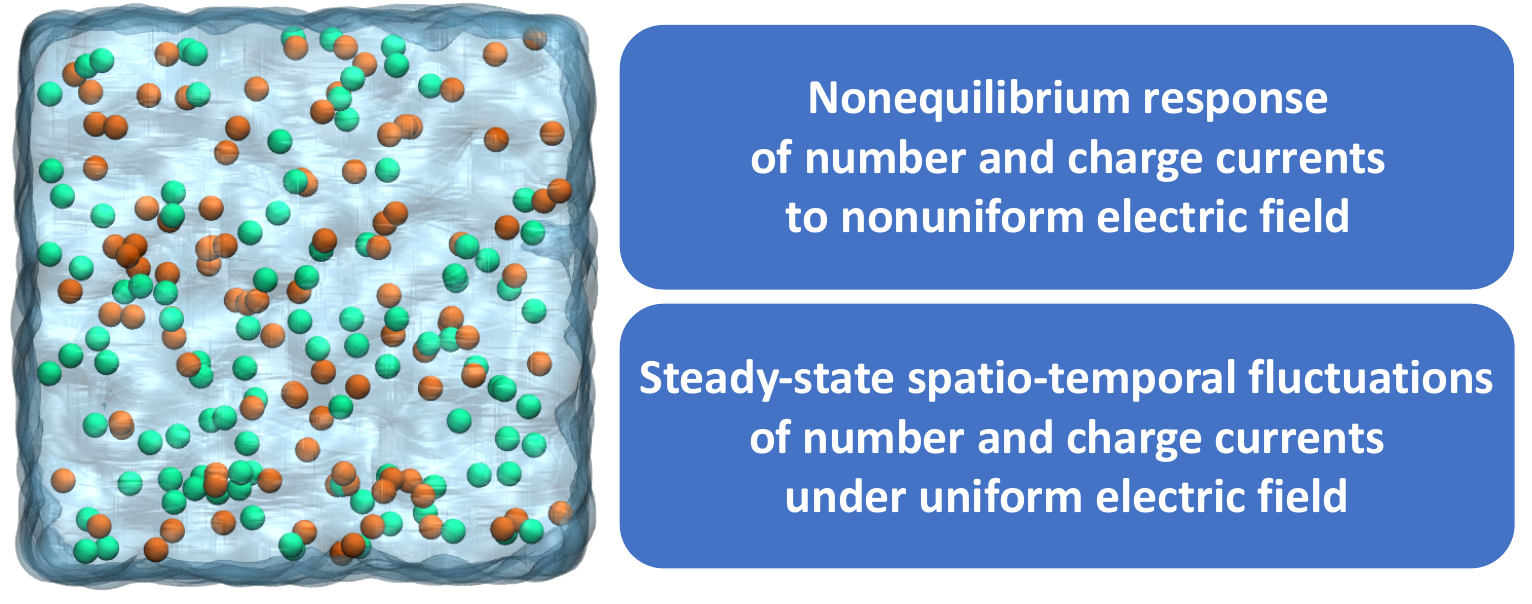} 
    \end{center}
    \caption{
    Overview of the present study of number and charge current fluctuations in electrolytes.}
    \label{fig:overview}
\end{figure}

We investigate the dynamics of number and charge current fluctuations in electrolytes, described by cations and anions in an implicit solvent with identical (symmetric case) or different (asymmetric case) diffusion coefficients, by considering (1) the non-equilibrium response to a nonuniform electric field characterized by a single wave vector $\kext$ and (2) the steady-state spatio-temporal fluctuations under a uniform electric field, for various wave vector $k$, as illustrated in Fig.~\ref{fig:overview}.

We first introduce the description of electrolyte solutions in Section~\ref{ssec:BD_model} and the properties of interest in Section~\ref{ssec:obs}. Simulation details are provided in Section~\ref{ssec:BD_sim}, while Stochastic Density Functional Theory is presented in Section~\ref{ssec:SDFT}.

\subsection{Brownian model of electrolyte solutions}
\label{ssec:BD_model}

We consider an aqueous solution of a 1:1 binary electrolyte at a concentration of $ C_s = 0.1~\mathrm{mol/L} $ and temperature $T=300$~K. The ions, with charges $ q_i = Z_i e $ where the valencies are $ Z_+ = +1 $ for cations and $ Z_- = -1 $ for anions and $ e $ is the elementary charge, are modeled as Brownian particles immersed in an implicit solvent. The diffusion coefficients $D_\pm$ of the ions are related to their mobility via the fluctuation–dissipation relation $ \mu_{\pm} = \beta D_{\pm} $, where $ \beta = 1 / k_{\rm B} T $, with $ k_{\rm B} $ the Boltzmann constant. The ion positions $ \vect{r}_i $ evolve according to the overdamped Langevin equation (Brownian dynamics)
\begin{equation}
    \dot{\vect{r}}_i = \mu_i \vect{F}_i + \sqrt{2 D_i} \, \vect{\eta}_i 
    \label{eq:BD}
\end{equation}
where the term $\vect{\eta}_i$ represents uncorrelated isotropic white noise vectors modeling the solvent's equilibrium fluctuations, with cartesian components $\alpha,\alpha'\in\{x,y,z\}$ satisfying 
\begin{equation}
    \left\langle \eta_{i,\alpha}(t) \right\rangle = 0, \quad 
    \left\langle \eta_{i,\alpha}(t) \, \eta_{j,\alpha'}(t') \right\rangle =  \, \delta_{ij} \, \delta_{\alpha\alpha'} \delta(t - t'),
    \label{eq:noise}
\end{equation}
where $\delta_{ij}$ and $\delta_{\alpha\alpha'}$ denote the Kronecker symbol to select identical particles and cartesian components, $\delta(t)$ is the Dirac distribution and brakets indicate an average over the realizations of the random noise. The deterministic conservative forces $\vect{F}_i$ acting on the ions in Eq.~\ref{eq:BD} include a drive from a time-- and space-- dependent external electric field $q_i \vect{E}^\mathrm{ext}(\vect{r}, t)$ and inter-ionic interactions. The inter-ionic interactions are derived from pairwise additive potentials comprising Coulomb interactions screened by the implicit solvent, and short-range repulsions:
\begin{equation}
    v_{ij}(r) = \frac{q_i q_j}{4 \pi \varepsilon_0 \varepsilon_r r} + v_{ij}^{\mathrm{WCA}}(r),
    \label{eq:PairPotential}
\end{equation}
where $\varepsilon_r=78.5$ is the solvent's relative permittivity, and $\varepsilon_0$ is the vacuum permittivity. The short-range interactions are modeled using the Weeks--Chandler--Andersen (WCA) potential:
\begin{equation}
    v_{ij}^{\mathrm{WCA}}(r) =
    \begin{cases}
        v_{ij}^\mathrm{LJ}(r) - v_{ij}^\mathrm{LJ}(r_\mathrm{min}), & r \leq r_\mathrm{min}, \\
        0, & r > r_\mathrm{min},
    \end{cases}
\end{equation}
which is a truncated and shifted Lennard--Jones (LJ) potential:
\begin{equation}
    v_{ij}^\mathrm{LJ}(r) = 4\epsilon_{ij} \left[ \left(\frac{\sigma_{ij}}{r}\right)^{12} - \left(\frac{\sigma_{ij}}{r}\right)^6 \right],
\end{equation}
where $r_\mathrm{min} = 2^{1/6} \sigma_{ij}$ is the position of the LJ potential's minimum.  We set the LJ energy and distance parameters for the model binary electrolyte as $\epsilon_{++} = \epsilon_{--} = \epsilon_{+-} = 0.1\ \textrm{kcal}\ \textrm{mol}^{-1}$, and $\sigma_{++} = \sigma_{--} = \sigma_{+-} = 3.0$\ \AA. While we consider here identical diameters for the cations and ions, which is a good approximation for electrolytes such as aqueous KCl, we note that this is the exception rather than the rule. In the following, we will however consider the effect of an asymmetry in the dynamical properties of the ions, \textit{i.e.} $D_+\neq D_-$. We therefore introduce the average diffusion coefficient
\begin{equation}
    D = \frac{D_+ + D_-}{2}
    \label{eq:def:D_ave}
\end{equation}
and the asymmetry factor
\begin{equation}
    \gamma = \frac{D_+ - D_-}{D_+ + D_-} \, .
    \label{eq:def:gamma}
\end{equation}
Specifically, we will consider two systems with the same average $ D = 1.50 \times 10^{-9}~\text{m}^2/\text{s} $ : the first one, with $D_+=D_-=D$, corresponds to the symmetric case $ \gamma = 0 $, while the second one, with $D_+=2.25 \times 10^{-9}~\text{m}^2/\text{s}$ and $D_-=0.75 \times 10^{-9}~\text{m}^2/\text{s}$ corresponds to $ \gamma = 0.5 $. This allows us to cover the experimental range of values for this asymmetric factor for typical alkaline chloride salts, including $ \gamma_{\mathrm{K^+Cl^-}} \approx -0.02 $, $ \gamma_{\mathrm{Na^+Cl^-}} \approx -0.21 $, $ \gamma_{\mathrm{Li^+Cl^-}} \approx -0.33 $

As indicated in Sec.~\ref{sec:intro}, a key quantity to characterize the system is the electrostatic screening length quantifying the screening of electrostatic interactions by mobile ions. A simple expression, valid for sufficiently dilute electrolytes, was obtained  by Debye and H\"uckel under the approximation of weak electrostatic couplings, treated at the mean-field level, resulting in the Debye screening length, which for a 1:1 electrolyte reads
\begin{equation}
    \ld = \kappa^{-1} = \sqrt{\frac{\varepsilon_0 \varepsilon_r \, k_B T}{2 \, C_s e^2}} \, .
    \label{eq:lambdad}
\end{equation}
For the present concentration $C_s = 0.1$~mol/L, $\ld\approx0.96$~nm. This length scale arising from the thermal fluctuations of the ionic cloud, can be combined with the thermal electrostatic potential $k_BT/e\approx 25$~mV at $T=300$~K to define a thermal electric field $E^{\rm th}=k_BT/e\ld$. The corresponding value is $E^{\rm th}\approx 2.68$~mV/\AA~ will be used in the following to discuss the transition between regimes of small/large applied fields.

\subsection{Properties of interest}
\label{ssec:obs}

In order to discuss the effect of the asymmetry in the diffusion coefficients on the dynamical properties of the electrolyte, we consider two complementary situations. In the first, we apply an inhomogeneous electric field and monitor the transient response of the electric and number currents. In the second, we monitor the fluctuations of the charge and number densities at equilibrium and in the presence of a steady-state homogeneous electric field. Those two situations are described in Sections~\ref{ssec:obs:transient} and \ref{ssec:obs:ness}.

\subsubsection{Transient response of the charge and number currents to an inhomogeneous external field}
\label{ssec:obs:transient}

In Sec.~\ref{ssec:res_currents}, we will examine the response induced by inhomogeneous electric fields. Starting from an equilibrium electrolyte (no applied field), we apply for $t\ge0$ a spatially-dependent electric field characterized by a wave vector $\veckext$ and magnitude $E$: 
\begin{equation}
    \vect{E}^\mathrm{ext}_{\veckext}(\vect{r}, t) = \operatorname{Im}\left[ H(t) \, E   \, \mathrm{e}^{\mathrm{i} \veckext \cdot \vect{r}} \,  \vecez \, \right] ,
    \label{eq:Ert_complex}
\end{equation}
with $H(t)$ the Heaviside step function and $\operatorname{Im}$ denoting the imaginary part. Even though this is simply a sinusoidal wave, the complex notation will be used below when discussing the linear response in terms of Fourier modes. Furthermore, we restrict the study to longitudinal electric field where the field is aligned with the direction of the inhomogeneity, \textit{i.e.} $\veckext=\kext \vecez$, such that $\veckext \cdot \vect{r}=\kext z$.

Within Brownian dynamics, currents are defined using the hydrodynamic velocities of the ions, $\vect{v}_i^\mathrm{BD} = \beta D_i \, \vect{F}_i$ (involving the conservative forces only). Specifically, the number and electric currents (normalized by the elementary charge) are conveniently expressed as the (6-dimensional) vector
\begin{equation}
    \vect{J}(t) =
    \begin{bmatrix}
        \vect{J}_N(t)\equiv \sum_i \, \beta D_i \, \vect{F}_i(t)  \\
        \vect{J}_Z(t)\equiv \sum_i Z_i \, \beta D_i \, \vect{F}_i(t)
    \end{bmatrix}
    \label{eq:current_def}
\end{equation}
where the sums run over ions $i$ with valencies $Z_i$. In order to refine the analysis to local currents, we define the corresponding current densities as
\begin{equation}
    \vect{j}(\vect{r},t) =
    \begin{bmatrix}
        \vect{j}_N(\vect{r},t) \equiv \sum_i \, \beta D_i \, \vect{F}_i(t) \, \delta^3(\vect{r}-\vect{r}_i(t)) \\
        \vect{j}_Z(\vect{r},t) \equiv \sum_i Z_i \, \beta D_i \, \vect{F}_i(t) \, \delta^3(\vect{r}-\vect{r}_i(t))
    \end{bmatrix}.
    \label{eq:current_dens}
\end{equation}
where $\delta^3(\vect{r})$ is the three-dimensional Dirac delta distribution. Lastly, we define the corresponding Fourier component of the current densities at the wave vector $\veckext$ of the applied field,
\begin{equation}
    \vect{J}_{\veckext}(t) = \tilde{\vect{j}}(\veckext,t) =
    \begin{bmatrix}
    \sum_i \, \beta D_i \, \vect{F}_i(t) \, \mathrm{e}^{-\mathrm{i} 
   \veckext \cdot \vect{r}_i(t)} \\
    \sum_i Z_i \, \beta D_i \, \vect{F}_i(t) \, \mathrm{e}^{-\mathrm{i} 
   \veckext \cdot \vect{r}_i(t)}
    \end{bmatrix},
    \label{eq:current_k}
\end{equation}
where we have used the convention
\begin{equation}
    \left\lbrace
    \begin{aligned}
        & \tilde{f}(\vect{k}) = \int_{\mathbb{R}^3} f(\vect{r}) \, \mathrm{e}^{-\mathrm{i} \vect{k} \cdot \vect{r}} \, \mathrm{d}^3\vect{r}, \\
        & f(\vect{r}) = \frac{1}{(2\pi)^3}\int_{\mathbb{R}^3} \tilde{f}(\vect{k}) \, \mathrm{e}^{+\mathrm{i} \vect{k} \cdot \vect{r}} \, \mathrm{d}^3 \vect{k}.
    \end{aligned}
    \right.
    \label{eq:space_FT}
\end{equation}
For a periodic system (such as in simulations) with volume $V$, the accessible wave vectors are discrete, the inverse Fourier transform is defined as:
\begin{equation*}
    f(\vect{r}) = \frac{1}{V}\sum_{\vect{k}} \tilde{f}(\vect{k}) \, \mathrm{e}^{+\mathrm{i} \vect{k} \cdot \vect{r}}
\end{equation*}
and the Dirac delta reduces to $(2\pi)^{3}\delta^3\left(\vect{k}-\vect{k}'\right) \to V \delta_{\vect{k},\vect{k}^{'}}$.

\noindent
While in the general case, the currents can exhibit inhomogeneities at different wave vectors than that of the applied electric field, here we only consider the currents 
single mode $\vect{k}=\veckext$ (Eq.~\ref{eq:current_k}) identical to the one of the perturbation (Eq.~\ref{eq:Ert_complex}), due to the properties of the linear response regime in bulk systems~\cite{hansen_mcdonald_theory_of_simple_liquids_book}. In the following, we will report the component of the number and electric currents in the direction of the applied field, denoted as $J_N^{\kext}(t)$ and $J_Z^{\kext}(t)$, respectively.

\subsubsection{Charge and number density fluctuations at equilibrium and under a steady homogeneous external field}
\label{ssec:obs:ness}

In Sec~\ref{ssec:res_Fkt}, we will consider the dynamics of the number and charge fluctuations at all wave vectors at equilibrium ($\vect{E}^\mathrm{ext}=\vect{0}$) and under a steady homogeneous external field ($\veckext=\vect{0}$):
\begin{equation}
    \vect{E}^\mathrm{ext}_{\vect{0}}(\vect{r}, t)=  E \, \vecez \, .
    \label{eq:E_bias}
\end{equation}
These fluctuations are encoded in the intermediate scattering matrix $\hat{\vect{F}}(\vect{k},t)$ defined as
\begin{equation}
    \hat{\vect{F}}(\vect{k},t) = \frac{1}{\Nions} \left\langle \tilde{\vect{\rho}}(\vect{k},t) \, \tilde{\vect{\rho}}^\dagger(\vect{k},0) \right\rangle,
    \label{eq:def_hat_F}
\end{equation}
where $\Nions=2\Npairs$ is the total number of ions and $\dagger$ denotes the conjugate transpose and $\tilde{\vect{\rho}}(\vect{k},t)$ is the Fourier transform of the fluctuations of ionic number and charge distributions $\vect{\delta \rho}(\vect{r},t) = \vect{\rho}(\vect{r},t) - \left\langle \vect{\rho}(\vect{r},t) \right\rangle$, where
\begin{equation}
    \vect{\rho}(\vect{r},t) =
    \begin{bmatrix}
        \rho_N(\vect{r},t) \equiv \sum_i \, \delta^3(\vect{r}-\vect{r}_i(t)) \\
        \rho_Z(\vect{r},t) \equiv \sum_i Z_i \, \delta^3(\vect{r}-\vect{r}_i(t))
    \end{bmatrix}.
    \label{eq:dens_cz}
\end{equation}

\subsection{Brownian dynamics simulations}
\label{ssec:BD_sim}

In Brownian Dynamics (BD) simulations, the trajectory of ions is obtained by integrating numerically the overdamped Langevin equation Eq.~\ref{eq:BD} and the observable (densities and currents) are obtained from the positions of the ions and the forces acting on them. These results serve as the ``exact'' reference against which analytical predictions are tested. In practice, we use the Large-scale Atomic/Molecular Massively Parallel Simulator (LAMMPS) simulation package \cite{thompson_LAMMPS_2022}. The system consists of $\Npairs=7530$ ion pairs in a fixed cubic volume $V = L^3$ = (500~\AA)$^3$ ($C_s = 0.1$~mol/L) and periodic boundary conditions (PBC) in all directions are used. The ion positions are updated using the overdamped BAOAB integrator\cite{Leimkuhler_BAOAB_2013} with a time step of $\delta t =25 $~fs. Short-range interactions are computed using a real space cut-off of $15$~\AA, and long-range electrostatic interactions are computed via the P3M algorithm~\cite{pollock1996comments} with a relative root mean square error in per-atom forces below $10^{-5}$.

For transient currents presented in Sec.~\ref{ssec:res_currents}, the system is first equilibrated for $25~\mathrm{ns}$. Subsequently, a longitudinal external electric field of the form $\vect{E}^\mathrm{ext}_{\veckext}(\vect{r}) = E \, \sin(\kext \, z)\, \vecez$ (see Eq.~\ref{eq:Ert_complex}) is applied for $t \geq 0$. We consider several values of the wave vector $\kext$ and a fixed amplitude $E=2.68$~mV/\AA~$\approx E^{\rm th}=k_B T/e \ld$. We have checked that this value still falls within the linear response regime, while ensuring a sufficiently large signal to noise ratio. The corresponding components of the current, defined in Eq.~\ref{eq:current_k}, are recorded every 1~ps for $50$~ns and a running average over 50~ps windows is performed to filter out the high frequency fluctuations. Simulations are carried out independently for different wave vectors compatible with the PBC ($k\in \frac{2\pi}{L} \, \mathbb{N} $), ranging from approximately $\kappa/10$ to $10\kappa$. For each system (with diffusivity asymmetries $\gamma=0$ and $\gamma=0.5$), this full procedure is repeated for 10 independent initial conditions, and the transient currents are averaged over the 10 independent trajectories and results reported in Fig.~\ref{fig:currents} with 95\% confidence interval of a smoothing window.

For the stationary fluctuations presented in Sec.~\ref{ssec:res_Fkt}, the system is first equilibrated for $25~\mathrm{ns}$, either in the absence or in the presence of a homogeneous $(\veckext=\vect{0})$ electric field of amplitude $E=2.68$~mV/\AA. The partial Fourier components of the densities $\tilde{\vect{\rho}}_{\pm}(\vect{k},t) = \sum_{i\in\pm}\e^{-\im \vect{k} \cdot \vect{r}_i(t)} $ with $k\in \frac{2\pi}{L}  \mathbb{N}$, ranging from ($k_\mathrm{min} = \frac{2\pi}{L}\approx0.12\kappa$ to $k_\mathrm{max} = 256\,k_\mathrm{min}$) are sampled every 1~ps for a total of 1~$\mu$s. From these time series, the intermediate scattering matrix (see Eq.~\ref{eq:def_hat_F}) is computed for each sampled wave vector using Fast Fourier Transform algorithm over 5 block windows of 0.2~$\mu$s assumed to be statistically independent. This procedure is repeated for 5 independent initial conditions, and results are reported as averages with 95\% confidence intervals.

\subsection{Stochastic density functional theory}
\label{ssec:SDFT}

\subsubsection{Governing equations}
\label{ssec:SDFT:equ}

While BD simulations provide ``exact'' numerical results, they do not provide a simple interpretation of the various contributions to the observables and their scaling with the physical parameters defining the system. This goal can be achieved using analytical theories under certain approximations, whose reliability can be tested against a BD reference. Here, we use stochastic density functional theory (SDFT) for electrolytes in the mean-field approximation, which describes the evolution of the distributions of cations and anions
\begin{equation}
    \rho_\pm(\vect{r},t) = \sum_{i \in \left\lbrace \pm \right\rbrace} \, \delta^3(\vect{r}-\vect{r}_i(t) ) \, .
    \label{eq:dens_cat_ani}
\end{equation}
In the SDFT framework, the ions evolve according to Eq.~\ref{eq:BD} and the forces are limited to electrostatic interactions (no short-range WCA repulsion) treated at the mean-field level. Following the Dean-Kawasaki approach~\cite{kawasaki_stochastic_1994, dean_langevin_1996}, one obtains the coupled conservation equations
\begin{equation}
    \partial_t \rho_\pm(\vect{r},t)  + \vect{\nabla} \cdot \vect{j}_\pm(\vect{r},t)  = 0 \,
    \label{eq:conservation}
\end{equation}
where the fluxes are
\begin{equation}
    \begin{split}
         \vect{j}_\pm(\vect{r},t)  = & - D_\pm \vect{\nabla} \rho_\pm(\vect{r},t)  \\
         & \pm \beta D_\pm e \rho_\pm(\vect{r},t)  \left( \vect{E}^\mathrm{ext}(\vect{r},t) - \vect{\nabla}V^\mathrm{int}(\vect{r},t)  \right) \\
         & + \sqrt{ 2 D_\pm\rho_\pm(\vect{r},t) } \, \vect{\xi}_\pm(\vect{r},t) \ ,\\
    \end{split}
    \label{eq:sPNP_flux}
\end{equation}
where the migration of ions includes both the effect of the external field $\vect{E}(\vect{r},t)$ and that of the internal potential satisfying the Poisson equation
\begin{equation}
    \Delta V^\mathrm{int}(\vect{r},t) = - \frac{e}{\varepsilon_0\varepsilon_r} \left[ \rho_+(\vect{r},t) -  \rho_-(\vect{r},t) \right] \, .
    \label{eq:Poisson}
\end{equation}
Eq.~\ref{eq:sPNP_flux} is referred to as the stochastic Nernst-Planck flux and includes (last term) multiplicative white noises satisfying: 
\begin{align}
    \left\langle \xi_{a,\alpha}(\bfr,t) \right\rangle &= 0 \,, \\
    \left\langle \xi_{a,\alpha}(\bfr,t) \, \xi_{a',\alpha'}(\bfr',t') \right\rangle &=  \, \delta_{aa'} \, \delta_{\alpha\alpha'} \delta^3(\vect{r}-\vect{r}') \delta(t - t') \, ,
    \label{eq:noise_fields}
\end{align}
with $a,a'\in\left\{+,-\right\}$.

\subsubsection{Change of variables, linearization and Fourier space}
\label{ssec:SDFT:linearization}

The number and charge densities introduced in Eq.~\ref{eq:dens_cz} are straightforwardly related to the cation and anion densities via the following change of variable
\begin{equation}
    \vect{\rho}(\vect{r},t) =
    \begin{bmatrix}
        \rho_N \\
        \rho_Z \\
    \end{bmatrix}  
    = \hat{\vect{T}} 
    \begin{bmatrix}
        \rho_+ \\
        \rho_- \\
    \end{bmatrix}
    \; \textrm{with} \;
    \hat{\vect{T}} =
    \begin{bmatrix}
        & 1 & 1 \\
        & 1 & -1\\
    \end{bmatrix}
    \, .
    \label{eq:change_of_var}
\end{equation}
We then consider their deviations from their respective average, namely
$\delta\rho_N(\bfr,t)=\rho_N(\bfr,t)-2C_s$ and $\delta\rho_Z(\bfr,t)=\rho_Z(\bfr,t)$ and neglect terms beyond linear order in these deviations. Within this approximation, and using the change of variable for the diffusion coefficients Eq.~\ref{eq:def:D_ave} and Eq.~\ref{eq:def:gamma}, we can recast Eq.~\ref{eq:conservation} and \ref{eq:sPNP_flux} into the following evolution equations: 
\begin{align}
\partial_t\delta \rho_N &+ D \left[ - \Delta \delta\rho_N + \gamma  \beta e \vect{\nabla} \cdot (\vecEext\delta\rho_N) \right.  
\nonumber \\ \label{eq:evolrhoN} 
& \left. \hskip -0.5cm + \gamma(-\Delta + \kappa^2)\delta\rho_Z +   \beta e \vect{\nabla} \cdot (\vecEext\delta\rho_Z) \right] = s_N^{\rm ext} + s_N^{\rm ran} \\
\partial_t\delta \rho_Z &+ D \left[ - \gamma\Delta \delta\rho_N + \beta e \vect{\nabla} \cdot (\vecEext\delta\rho_N) \right. 
\nonumber \\ \label{eq:evolrhoZ} 
& \left. \hskip -0.5cm + (-\Delta + \kappa^2)\delta\rho_Z +  \gamma \beta e \vect{\nabla} \cdot (\vecEext\delta\rho_Z) \right] = s_Z^{\rm ext} + s_Z^{\rm ran}
\end{align}
where the r.h.s. includes a deterministic source term due to the external potential:
\begin{align}
    \vect{s}^{\rm ext}(\vect{r},t) = 
        \begin{bmatrix}
        s_N^{\rm ext} \\
        s_Z^{\rm ext}
    \end{bmatrix}
    = -2 \beta D e C_s \vect{\nabla} \cdot \vecEext 
    \begin{bmatrix}
        \gamma \\
        1
    \end{bmatrix} \, , 
    \label{eq:source_e_rt}
\end{align}
and a random one due to the stochastic fluxes in Eq.~\ref{eq:sPNP_flux}:
\begin{equation}
    \vect{s}^{\rm ran}(\vect{r},t) = 
        \begin{bmatrix}
        s_N^{\rm ran} \\
        s_Z^{\rm ran}
    \end{bmatrix}
    = \sqrt{2D C_s} \, \vect{\nabla} \cdot
    \begin{bmatrix}
        \sqrt{1+\gamma} \ \vect{\xi}_+ + \sqrt{1-\gamma}  \ \vect{\xi}_- \\
        \sqrt{1+\gamma}  \ \vect{\xi}_+ - \sqrt{1-\gamma} \ \vect{\xi}_-
    \end{bmatrix} \, .
    \label{eq:source_r_rt}
\end{equation}

Notably, $\vect{s}^{\rm ran}$ is a correlated noise with zero mean and covariance
\begin{equation}
    \left\langle \vect{s}^{\rm ran}(\vect{r},t) \, (\vect{s}^{\rm ran})^\dagger(\vect{r}',t') \right\rangle= 4 D C_s \, \vect{\Xi} \, \nabla^2 \delta^3\left(\vect{r}-\vect{r}'\right) \, \delta\left(t-t' \right) 
    \label{eq:source_r_rt_cov}
\end{equation}
where
\begin{equation}
    \vect{\Xi} =
    \begin{bmatrix}
        1  & \gamma \\
        \gamma  & 1 \\
    \end{bmatrix} \, .
    \label{eq:Xi}
\end{equation}
In particular, asymmetric diffusion coefficients ($\gamma\neq0$) induce terms that are often ignored in the literature. The solution of these coupled equations can conveniently be found and analyzed in reciprocal space. Introducing the dimensionless quantities $\bfK = \ld \bfk = \bfk/\kappa$ (with norm $K=k/\kappa$) and $\bmmcE(\bfk,t)$, the spatial Fourier transform of $ \ld \beta e \vecEext(\bfr,t) = \vecEext/E^{\rm th}$, Eqs.~\ref{eq:evolrhoN} and~\ref{eq:evolrhoZ} can be rewritten as
\begin{align}
\partial_t \tdrhoN &+ \frac{1}{\taud}\left[ K^2\tdrhoN - i \gamma \bfK  \cdot (\bmmcE \ast \tdrhoN) \right. 
\nonumber \\ \label{eq:evoltdrhoN} 
& \left. \hskip 0.5cm + \gamma(K^2 + 1)\tdrhoZ - i \bfK  \cdot (\bmmcE \ast \tdrhoZ) \right] = \tilde{s}_N^{\rm ext} + \tilde{s}_N^{\rm ran} \\
\partial_t \tdrhoZ &+ \frac{1}{\taud}\left[ \gamma K^2\tdrhoN - i \bfK  \cdot (\bmmcE \ast \tdrhoN) \right. 
\nonumber \\ \label{eq:evoltdrhoZ} 
& \left. \hskip 0.5cm + (K^2 + 1)\tdrhoZ - i \gamma \bfK  \cdot (\bmmcE \ast \tdrhoZ) \right] = \tilde{s}_Z^{\rm ext} + \tilde{s}_Z^{\rm ran}
\end{align}
where $\ast$ denotes the convolution product and the relaxation Debye time is defined as
\begin{equation}
    \taud^{-1} = D\kappa^2= \frac{D_++D_-}{2} \frac{2 C_s e^2}{\varepsilon_0\varepsilon_r k_B T} \ ,
    \label{eq:taud}
\end{equation}
which in the considered system (see Sec.~\ref{ssec:BD_model}) yields $\taud\approx0.62$~ns. Finally, the deterministic and random terms on the r.h.s. are characterized by 
\begin{equation}
    \tilde{\vect{s}}^{\rm ext}(\vect{k},t) = \im \frac{ 2 C_s }{\taud} \vect{K}\cdot \bmmcE
    \begin{bmatrix}
        \gamma \\
        1
    \end{bmatrix} \, 
    \label{eq:source_e_rt_k}
\end{equation}
and
\begin{equation}
    \left\langle \tilde{\vect{s}}^{\rm ran}(\vect{k},t) \,  (\tilde{\vect{s}}^{\rm ran})^\dagger(\vect{k}',t') \right\rangle=  \frac{(2\pi)^3 4 C_s K^2 }{\taud}  \, \vect{\Xi}  \,\delta^3\left(\vect{k}-\vect{k}'\right) \, \delta\left(t-t' \right) \, .
    \label{eq:source_r_kt_cov}
\end{equation}
Eqs.~\ref{eq:evoltdrhoN} to~\ref{eq:source_r_kt_cov} can be solved to predict the relevant Fourier component of transient currents in response to an inhomogeneous external field (see Eq.~\ref{eq:current_k}), as well as the intermediate scattering function characterizing the density fluctuations at equilibrium or under an applied homogeneous electric field (see Eq.~\ref{eq:def_hat_F}). Details of these derivations can be found in Appendix.~\ref{app:SDFT}.

\section{Results}
\label{sec:results}

We now combine BD simulations and SDFT calculations to analyze the transient currents in response to an inhomogeneous external field (section Sec.~\ref{ssec:res_currents}), the density fluctuations at equilibrium or under an applied homogeneous electric field (section Sec.~\ref{ssec:res_Fkt}).

\subsection{Transient currents in response to inhomogeneous electric fields}
\label{ssec:res_currents}

For the perturbation Eq.~\ref{eq:Ert_complex}, the Fourier transform (see Eq.~\ref{eq:space_FT}) is $\bmmcE(\bfk,t)= \mcE \, (2\pi)^3 \frac{\delta^3(\bfk - \bfk_E) - \delta^3(\bfk + \bfk_E)}{2 \im } H(t) \vecez$ or, under periodic boundary conditions, $\bmmcE(\bfk,t)= \mcE \, V \frac{\delta_{\bfk,\bfk_E} - \delta_{\bfk,-\bfk_E}}{2 \im } H(t) \vecez$, with $\mcE= \beta e E \ld = E/E^\mathrm{th}$ the reduced electric field, so that its convolution with $\tdrhoN$ and $\tdrhoZ$ is greatly simplified (it selects the modes $\bfk=\pm\veckext$). Eqs.~\ref{eq:evoltdrhoN} and~\ref{eq:evoltdrhoZ}, averaged over the realizations of the stochastic noise, form a system of coupled linear equations that can be solved for the fields $\tdrhoN(\veckext,t)$ and $\tdrhoZ(\veckext,t)$, which are then used to obtain the Fourier components of the number and charge currents at the wave vector $\kext$ corresponding to the applied external field (see Eq.~\ref{eq:current_k}) as:
\begin{equation}
        \tilde{\vect{j}}(\veckext,t) =  \left( 2 C_s \beta D e \vecEext(\veckext,t) + \im \frac{\veckext}{\kext^2} D\kappa^2  \, \tdrhoZ(\veckext,t) \right)
        \begin{bmatrix}
             \gamma \\
             1\\
        \end{bmatrix} \, .
        \label{eq_app_jkt}
\end{equation}
The final SDFT prediction (considering only electrostatic interactions at the mean-field level and under the linearization approximation) can be written as 
\begin{align}
    J_{N}^{\Kext}(t) = & \, \gamma J_{Z}^{\Kext}(t) \, = \,  \gamma J_{\mathrm{NE}}
    \left\lbrace S_{ZZ}(\Kext) \right. 
    \nonumber \\
    &\left. \hskip -1cm + A_s(\Kext) \, \mathrm{e}^{- t / \tau_s(\Kext)} 
    + A_f(\Kext) \, \mathrm{e}^{-t / \tau_f(\Kext)} \right\rbrace  \ ,
    \label{eq:res_current}
\end{align}
where
\begin{equation}
    J_\mathrm{NE} = 2C_s \beta D e \,\tilde{\bm{E}}(\bfk_E,t) \cdot \vecez = - \im \, \beta e D \Npairs \, E \, H(t)
    \label{eq:JNE}
\end{equation}
is the ideal Nernst-Einstein current, corresponding to the response of non-interacting ions to an uniform external field. Note that a factor of 1/2 arises because we consider a field $E\sin(k_E z)=E(e^{i \kext z}-e^{-i \kext z})/2i$, \textit{i.e.} two modes with wave vectors $\pm \kext$ and amplitude $E/2$. In addition, the currents defined in Eq.~\ref{eq:current_k} are not electric currents, so that $J_\mathrm{NE}$ also differs from the usual definition by a factor of $e$. The steady-state current reached after the transient regime (see below) is determined by the static charge-charge structure factor, which at this level of theory is given by its Debye-H\"uckel (DH) prediction,
\begin{equation}
    S^\mathrm{DH}_{ZZ}(K) = \frac{K^2}{K^2 + 1} = \frac{k^2}{k^2 + \kappa^2} \, ,
    \label{eq_Szz_DH}
\end{equation}
while the time-dependent relaxation consists of two decaying modes (with indices $s$ and $f$ for slow and fast, respectively), with amplitudes 
\begin{align}
    & A_s(K) =  \frac{ 2 \nu_0(K)-1}{\nu_0(K)(K^2+1)} \ ,
    \label{eq:A1}\\
    & A_f(K) =  \frac{2 \nu_0(K)+1}{\nu_0(K)(K^2+1)} \ ,
    \label{eq:A2}
\end{align}
and corresponding relaxation times
\begin{align}
    & \tau_s(K) = \frac{\taud}{\zeta_0(K) - \nu_0(K)} \label{eq:tau1}\\
    & \tau_f(K) = \frac{\taud}{\zeta_0(K) + \nu_0(K)} \ , \label{eq:tau2}
\end{align}
where we have introduced
\begin{align}
    \zeta_0(K) &= K^2 + \frac{1}{2} 
    \label{eq:zeta_zero} \\
    \nu_0(K) &= \sqrt{\frac{1}{4}+\gamma^2K^2(K^2+1)} \ .
    \label{eq:nu_zero}
\end{align}

\begin{figure}[ht!]
    \begin{center}
        \includegraphics[width=0.49\textwidth]{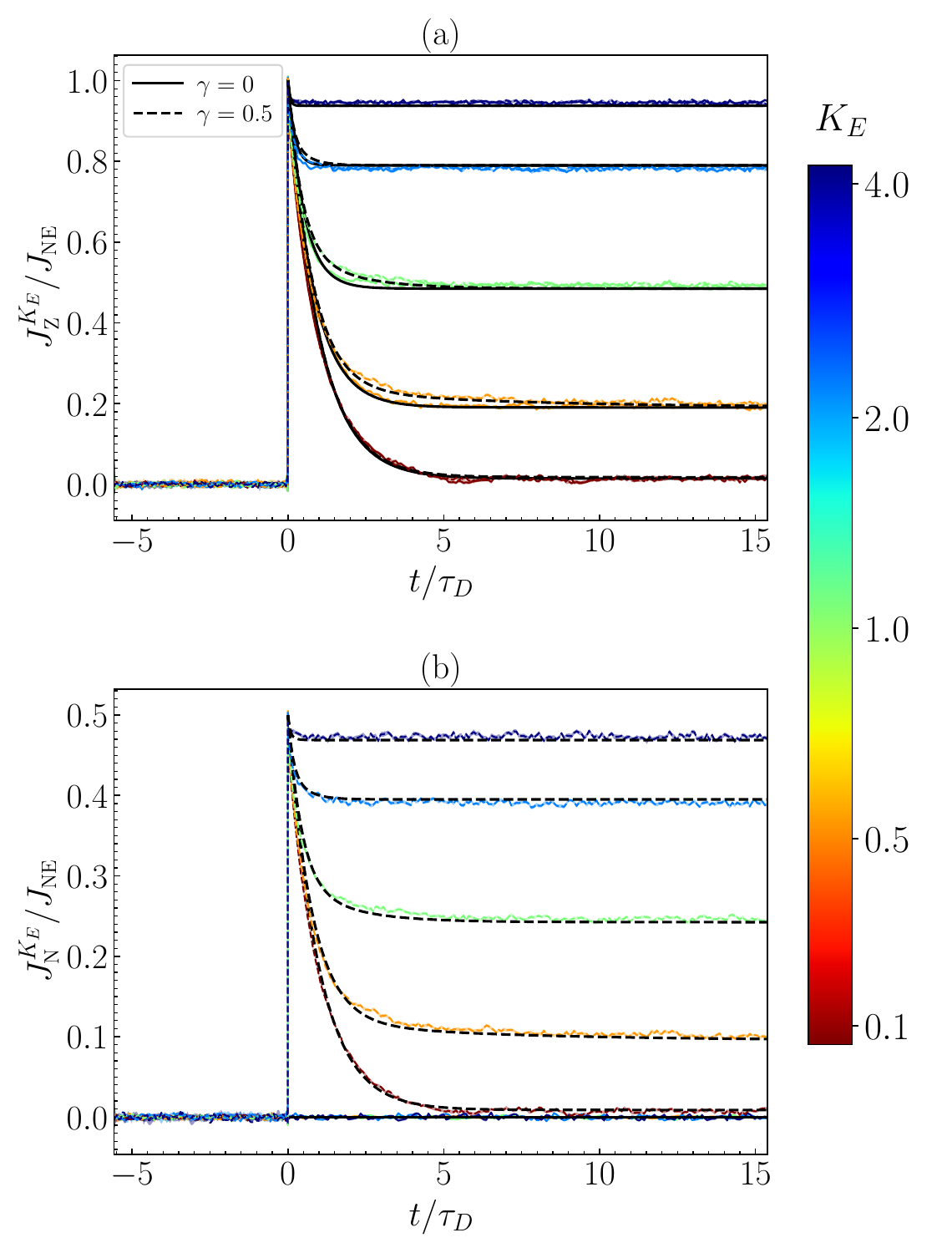} 
    \end{center}
    \caption{
    (a) Charge current, $J_{Z}^{\Kext}$, and (b) number current, $J_{N}^{\Kext}$, normalized by the ideal Nernst-Einstein current $J^{\rm NE}$ (see Eq.~\ref{eq:JNE}) and as a function of time normalized by the Debye relaxation time $\taud$ (see Eq.~\ref{eq:taud}), in response to the inhomogeneous electric field given by Eq.~\ref{eq:Ert_complex}. Colored lines indicate the BD simulation results for reduced wave vectors $\Kext=\kext/\kappa \in \left\{0.12, \, 0.5, \, 1, \, 2, \, 4\right\}$ (from dark red to dark blue), with shaded areas corresponding to 95\% confidence intervals. Analytical SDFT predictions are also indicated as black lines, for electrolytes with symmetric ($\gamma=0$, solid lines) and asymmetric ($\gamma=0.5$, dashed lines) diffusion coefficients.}
    \label{fig:currents}
\end{figure}

Figs.~\ref{fig:currents}a and ~\ref{fig:currents}b compare these SDFT predictions to BD simulations for symmetric ($\gamma=0$) and asymmetric ($\gamma = 0.5$) electrolytes over a range of reduced wave vectors $\Kext=\kext/\kappa \in \left\{0.12, \, 0.5, \, 1, \, 2, \, 4\right\}$. The excellent agreement between analytical predictions and numerical results indicates that neglecting short-range interactions and electrostatic correlations is sufficient to capture the dynamical response in this system. It also confirms that the applied external field is sufficiently small to remain in the linear response regime consistent with the linearization approximation required to obtain the analytical SDFT results.

Immediately after the electric field (Eq.~\ref{eq:Ert_complex}) is applied at $t=0^+$, the system responds with the ideal electrical current $J_{Z}^{\veckext}(0^+)=J_\mathrm{NE}$, as expected from the high-frequency limit of Brownian dynamics~\cite{hoang_frequency_field_2023}. On average, the fluid is homogeneously neutral with no net current before the external field is turned on, and only the latter contributes to the initial current.

At long times, the system reaches a steady state. As a consequence of linear response theory, the stationary electric current induced by an inhomogeneous field with wave vector $\kext$ is directly proportional to the static charge–charge structure factor (Eq.~\ref{eq_Szz_DH}):
\begin{equation}
    J_{Z}^{\kext}(t\rightarrow\infty)= J_\mathrm{NE} \, S_{ZZ}(\kext) \, .
    \label{eq:LRT_DH}
\end{equation}
Whereas for fields with wavelength much shorter than the Debye screening length the response at long times is that of an ideal electrolyte, in the opposite limit there is no steady-state current because of the local charge relaxation, which screens electrostatic effects:
\begin{empheq}[left=\empheqlbrace]{align}
    & J_{Z}^{\kext\gg \kappa}(t\rightarrow\infty) \sim  J_\mathrm{NE} \quad \quad  \text{(ideal)}\\
    & J_{Z}^{\kext\ll \kappa}(t\rightarrow\infty) \sim 0 \quad \quad  \text{(screened)}
\end{empheq}
This behavior reflects charge hyperuniformity in bulk electrolytes~\cite{blum_perfect_82,hoang_hyper_23}: only electric fields with spatial variations on scales smaller than the Debye length can drive a steady-state charge displacement on the same scale (and not beyond).

The dynamics at intermediate times ($t\sim\taud$) is a combination of two exponentially decaying modes. The faster mode has a characteristic time $\tau_f(\kext)$ given in Eq.~\ref{eq:tau2},  which in the ``hydrodynamic'' limit $\kext\to0$ (long-wavelength) becomes independent of $\kext$ -- a feature of relaxation modes. Furthermore, in the limit $\kext\ll\kappa$, $\tau_f$ coincides with the Debye relaxation time, $\taud$, corresponding to the relaxation of charge fluctuations at equilibrium. The slower mode has a characteristic time $\tau_s(\kext)$ given in Eq.~\ref{eq:tau1}, which for $\kext\ll\kappa$ behaves as $\tau_s \sim 1/(D_{NH}k_E^2)$ -- a feature of diffusive modes. The corresponding diffusion coefficient 
\begin{equation}
    D_\mathrm{NH} = (1-\gamma^2) \, D =\frac{2\, D_+\, D_-}{D_+ + D_-} \,
    \label{eq:D_NH}
\end{equation}
is the so-called Nernst-Hartley diffusion coefficient~\cite{Robinson1959}. Microscopically, if cations and anions have different diffusion coefficients ($\gamma\neq0$), a salt concentration would lead to transient charge separation, hence to internal electric field acting on the ions so as to restore local electroneutrality, within a time scale $\tau_f$. As a result, for $t\gg\tau_f$ the dominant transport mechanism is the diffusion of both species with diffusion coefficient $D_\mathrm{NH}$.

\begin{figure}[ht!]
    \begin{center}
    \includegraphics[width=0.49\textwidth]{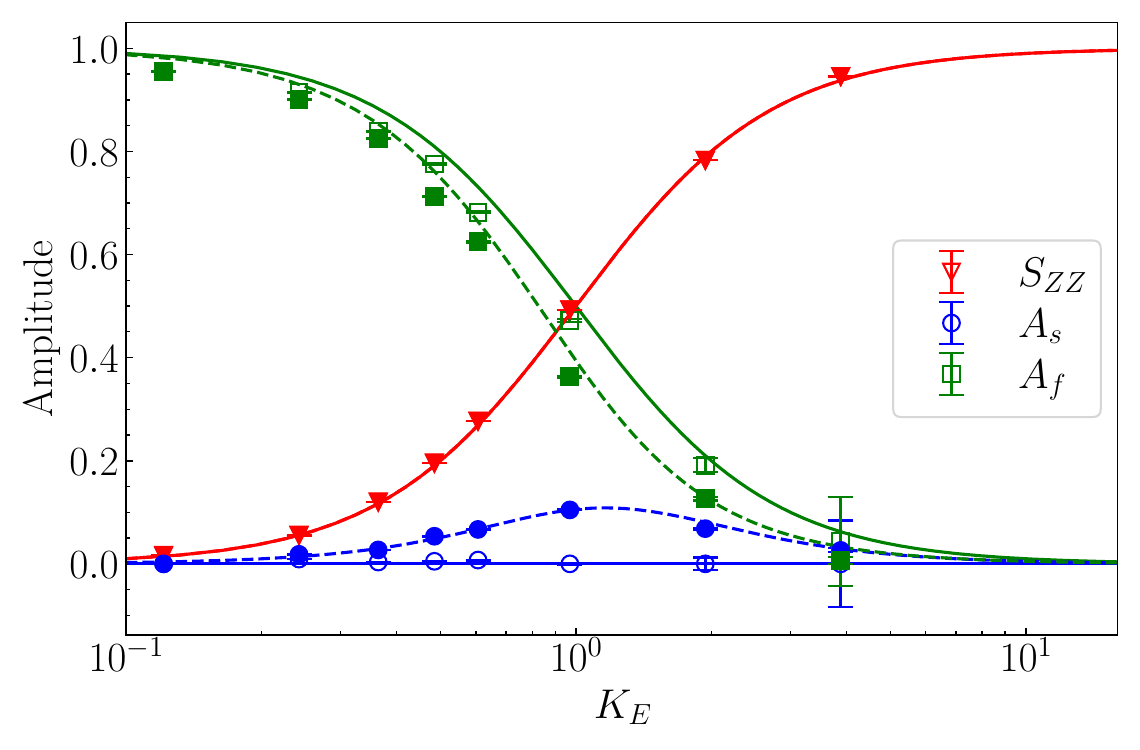} 
    \end{center}
    \caption{
    Relative weights of the steady-state ($S_{ZZ}$), slow ($A_s$) and fast ($A_f$) contributions to the charge current, as a function of the reduced wave vector $\kext/\kappa$. Open (resp. filled) symbols indicate the weights obtained by fitting the BD results for the current to Eq.~\ref{eq:res_current} for $\gamma=0$ (resp. $\gamma=0.5$). Solid (resp. dashed) lines indicate the predictions of SDFT for $\gamma=0$ (resp. $\gamma=0.5$). The errorbars correspond to one standard deviation of the fitted parameters, calculated from the covariance matrix of the regression fit.}
    \label{fig:modes}
\end{figure}

Although the asymmetry in the diffusion coefficients has a quantitative effect on the charge current $J_Z^{k}(t)$ (see Fig.~\ref{fig:currents}a), discussed below, it has a more dramatic impact on the number current $J_N^{k}(t)$: the latter vanishes in the symmetric case and becomes finite when $\gamma\neq0$  (see Fig.~\ref{fig:currents}b). The effect of the asymmetry is further visible in Fig.~\ref{fig:modes}, which reports the weights of the three contributions (steady-state, slow and fast) to the electric current (see Eq.~\ref{eq:res_current}) as a function of the wave vector. We first note the very good agreement between the weights obtained by fitting the BD results to Eq.~\ref{eq:res_current} (symbols) and that predicted by SDFT. This confirms the relevance of the latter for the system and conditions studied. 

For the symmetric electrolyte ($\gamma=0$), the slow mode is absent ($A_s\to0$ for all wave vectors). For external perturbations with a small wave length compared to the Debye length ($\kext\gg\kappa$), the current is dominated by the steady-state component, reflecting the individual response of ions, modulated by the static correlations. In the other limit (most easily achieved experimentally), the dominant mode is the fast relaxation of the charge leading to screening and a vanishing current. For the asymmetric electrolyte ($\gamma=0.5$), the steady-state component is unchanged with respect to the symmetric case, but the slow mode is now present. It only becomes significant with respect to the fast one for wave vectors $\kext\gtrsim \kappa$.

\subsection{Spatio-temporal correlations in non-equilibrium steady-state} 
\label{ssec:res_Fkt}

We now turn to the case of a constant and uniform perturbation ($\kext=0$), $\vecEext(\bfr,t)=E \, \vecez$, and consider the density fluctuations at steady-state (which includes the equilibrium for $E=0$) by computing the intermediate scattering matrix (see Eq.~\ref{eq:def_hat_F}). In that case the Fourier transform of the reduced field is $\bmmcE(\bfk,t)= \mcE\, (2\pi)^3\delta^3(\bfk) \vecez$ with $\mcE= \beta e E \ld = E/E^\mathrm{th}$.  The convolutions with the densities, followed by the dot product with the wave vector $\bfK=K\vecez$ then simplifies to $K\mcE\tilde{\rho}_{N,Z}$. As shown in Appendix~\ref{app:SDFT}, Eqs.~\ref{eq:evoltdrhoN} and~\ref{eq:evoltdrhoZ} can then be solved using linear algebra and the properties of the noise term $\vect{s}^\mathrm{ran}$ to compute the intermediate scattering matrix. In practice, this is done by considering its temporal Fourier transform (\textit{i.e.} in the frequency domain) and returning to the temporal domain via contour integration in the complex plane. This leads to a sum of two eigenmodes characterized by the complex frequencies $\omega_{s,f}(k) = D\kappa^{2}\,\Omega_{s,f}(K=k/\kappa)$, with
\begin{align}
    & \Omega_s(K) = i \left[ \zeta(K) - \nu(K) \right] \  \label{eq:Omega1}\\
    & \Omega_f(K) =  i \left[ \zeta(K) + \nu(K) \right]  \  \label{eq:Omega2}
\end{align}
where 
\begin{align}
     \zeta(K) = K^2+ \frac{1}{2} + \im \gamma \mcE K  
    \label{eq:zeta} 
\end{align}
and $\nu(K)$ is the principal square root:
\begin{align}
    \nu(K) =  \sqrt{\frac{1}{4} - \left[\mcE K - \im \gamma K^2 \right] \left[ \mcE K - \im \gamma (K^2 +1) \right]} \, .
    \label{eq:nu}
\end{align}

The dispersion relations Eqs.~\ref{eq:Omega1}, \ref{eq:Omega2}, \ref{eq:zeta} and~\ref{eq:nu} generalize Eqs.~\ref{eq:tau1}, \ref{eq:tau2}, \ref{eq:zeta_zero} and~\ref{eq:nu_zero}, which are immediately recovered in the form $D\kappa^2\Omega_{s,f}(K) = \im \, \tau_{s,f}^{-1}(K)$ upon setting $\mcE=0$. They carry key information on the intrinsic dynamics of the system, as will be discussed in more detail in the following. The final result for the intermediate scattering matrix can be written as 
\begin{align}
        \hat{\vect{F}}(K,t) &= \e^{- \zeta(K)t/\taud}
        \left[
            \hat{\vect{F}}(K,0) \, \cosh{\frac{\nu(K)t}{\taud}} 
            \right . \nonumber \\ 
            & \left. \hskip 3cm + \hat{\vect{G}}(K) \, \sinh{\frac{\nu(K)t}{\taud}}
        \right] \,,
        \label{eq:FULL_F}
\end{align}
where $\hat{\vect{F}}(K,0)$ is the static structure matrix, with elements
\begin{align}
    F_{NN}(K,0) & =  1 - \frac{\mcE^2}{(K^2+1+ \mcE^2) (2K^2+1)} \label{F_cc0}\\
    F_{ZZ}(K,0) & = \frac{2K^4 + 2 \mcE^2 K^2 + K^2}{(K^2+1+ \mcE^2) (2K^2+1)}   \label{F_zz0}\\
    F_{NZ}(K,0) & =  \frac{-\im \mcE K}{(K^2+1+ \mcE^2) (2K^2+1)}   \label{F_cz0}\\
    F_{ZN}(K,0) & =  \frac{+\im \mcE K}{(K^2+1+ \mcE^2) (2K^2+1)}   \label{F_zc0}
\end{align}
and finally $\hat{\vect{G}}(K)$ is a constant matrix defined in Appendix (Eq.~\ref{eq:formal_Gk}). At equilibrium ($\mcE=0$), Eq.~\ref{F_zz0} reduces to the static charge-charge structure factor Eq.~\ref{eq_Szz_DH}.

\subsubsection{Static correlations in non-equilibrium steady-state}
\label{sssec:Fk0}

\begin{figure}[ht!]
    \begin{center}
    \includegraphics[width=0.49\textwidth]{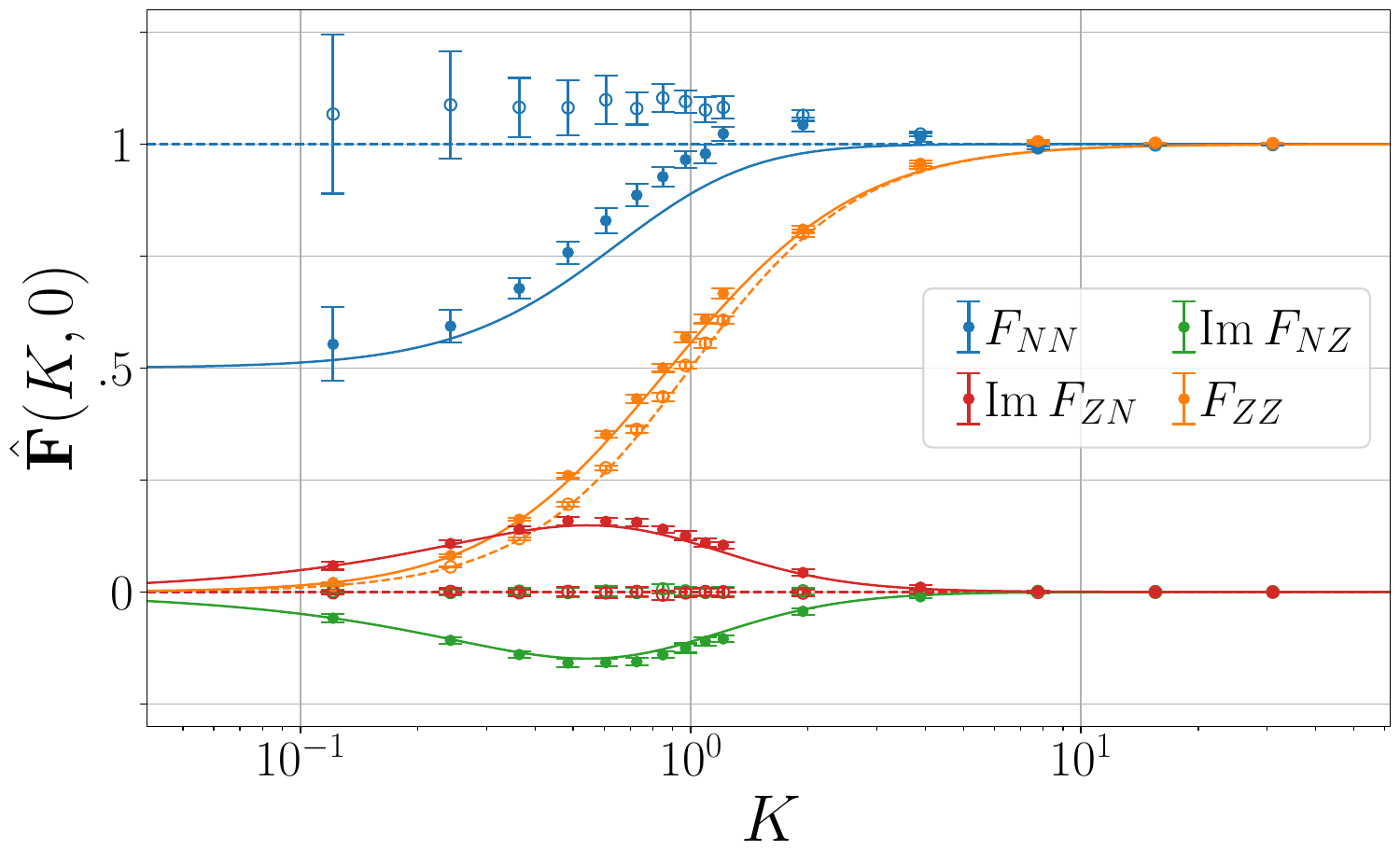} 
    \end{center}
    \caption{
    Elements of the static structure matrix $\hat{\vect{F}}(K,t=0)$ quantifying the static number-number ($F_{NN}$, blue), number-charge ($F_{NZ}$, green), charge-number ($F_{ZN}$, red) and charge-charge ($F_{ZZ}$, orange) correlations (see Eq.~\ref{eq:def_hat_F}), as a function of the reduced wave vector $K=k/\kappa$. Open and filled symbols are results from BD simulations at equilibrium ($\mcE=0$) and at NESS under a constant uniform electric field with reduced magnitude $\mcE=1$, respectively. The corresponding SDFT predictions Eqs.~\ref{F_cc0}-\ref{F_zc0} are shown as dashed (resp. solid) lines for $\mcE=0$ (resp. $\mcE=1$).
    }
    \label{fig:F0}
\end{figure}

Before diving into the dynamical analysis, we first examine the static structure matrix $\hat{\vect{F}}(\vect{K},0)$ Eqs.~\ref{F_cc0}-\ref{F_zc0}. Although the diffusion coefficients (hence their asymmetry $\gamma$) do not influence this quantity, the effect of an external electric field introduces important modifications, particularly relevant in the context of electrostatic screening and hyperuniformity in electrolytes. Eqs.~\ref{F_cc0}-\ref{F_zc0} match the results of D\'emery \textit{et al.}~\cite{demery_conductivity_2016} (see their Eq.~59), after a change of basis from anion/cation to number/charge variables: $\hat{\vect{F}}_{\{N,Z\}} = \hat{\vect{T}}\, \hat{\vect{F}}_{\{+,-\}}\, \hat{\vect{T}}^{-1}$ (see Eqs.~\ref{eq:def_hat_F} and \ref{eq:change_of_var}). Fig.~\ref{fig:F0} compares the prediction of Eqs.~\ref{F_cc0}–\ref{F_zc0} with the results of BD simulations, showing an overall excellent agreement. We first discuss the number-number correlations, before turning to charge-charge correlations and cross-correlations.

\paragraph*{Number-number correlations} In contrast to the classical result of Debye–Hückel theory in the absence of an electric field, where $F_{NN}^\mathrm{DH}(K,0) = 1$ at all wave vectors, number fluctuations in the direction of the field are partially suppressed in the small wave vector limit, where
\begin{equation}
        \lim_{K\rightarrow0} F_{NN}(K,0)  = \frac{1}{1+\mcE^2} \ ,
\end{equation}
indicating the emergence of field-induced long-range correlations~\cite{mahdisoltani_long_2021, du_2024_correlation}. Additionally, we observe a slight discrepancy between analytical predictions and simulations results, which we attribute to the finite ionic size and short-range interactions introducing additional correlations close to $k\sim\kappa$ and neglected in our mean-field approach~\cite{adar_2019_screening}. In the present system of implicit electrolytes at $C_s=0.1$~mol/L, the Debye screening length ($\ld\approx 1$~nm) is larger than ionic diameters ($\sigma_\mathrm{ion}\approx0.3$~nm), but the packing fraction $\phi_\mathrm{packing}\approx1.5\%$ is sufficient to induce effects not captured within the mean-field approximation~\cite{antypov_excluded_2005, adar_2019_screening, anousheh_ionic_structure_2020}.

\paragraph*{Charge-charge correlations} For $F_{ZZ}(K,0)$, the perfect screening limits $F_{ZZ}(K\ll1,0)\to0$ and $F_{ZZ}(K\gg1,0) \to 1$  remain valid even under an applied electric field (as assumed in the derivation), but the crossover is slightly shifted towards smaller wave vectors . Indeed, Eq.~\ref{F_zz0} can be rewritten using as
\begin{equation*}
    F_{ZZ}(K,0) = \frac{K^2}{K^2 + \ld^2/\lambda_{\mathrm{eff}}^{2}(K,\mcE)} \ ,
\end{equation*}
with an effective screening length
\begin{equation}
    \lambda_\mathrm{eff}(K,\mcE) = \frac{\ld}{\sqrt{1 - \dfrac{\mcE^2}{1 + 2K^2 + 2 \mcE^2}}} \ .
\end{equation}
The fact that $\lambda_{\rm{eff}} > \ld$ indicates that charge-charge correlations are longer-ranged than in the absence of the field. In the $K\to0$ limit, for small applied fields the screening length is increased by a factor $\sim~1+\mcE^2/2$. This finding also resonates with nonlinear water polarization or ionic cloud deformation effects responsible for the Wien effect~\cite{onsager_wien_1957, lesnicki_field_2020, lesnicki_molecular_2021, berthoumieux_2024_nonlinear_conductivity}, although hydrodynamic interactions are not taken into account in the present description. 

\paragraph*{Cross-correlations} The off-diagonal components $F_{NZ}(K,0)$ and $F_{ZN}(K,0)$, which quantify the cross-correlations between number and charge, are purely imaginary and satisfy Onsager’s reciprocal symmetry $F_{NZ}(K,0) = F_{ZN}^*(K,0)$. They vanish in both limits of small and large $K$, and display an extremum for a finite, field-dependent wave vector.

As expected, we observe no influence of asymmetry in ion mobility on the static structure matrix (not shown). In order to emphasize its effects on dynamical correlations, in the following we use $\hat{\vect{F}}(K,0)$ to normalize the time correlation functions at $t=0$ as 
\begin{equation}
    \frac{F_{ij}(K,t)}{ \sqrt{ F_{ii}(K,0) F_{jj}(K,0)} } \,
    \label{eq:norm_t0}
\end{equation}
with $i,j\in\{N,Z\}$ in Fig.~\ref{fig:Fkt_eq} and \ref{fig:Fkt_neq}. However, it remains important to account for these static correlations amplitudes / covariances when interpreting transport coefficients quantitatively.

\subsubsection{Dispersion relations}
\label{sssec:dispersion}

\paragraph*{Eigenmodes in the complex plane}

\begin{figure}[ht!]
    \begin{center}
    \includegraphics[width=0.49\textwidth]{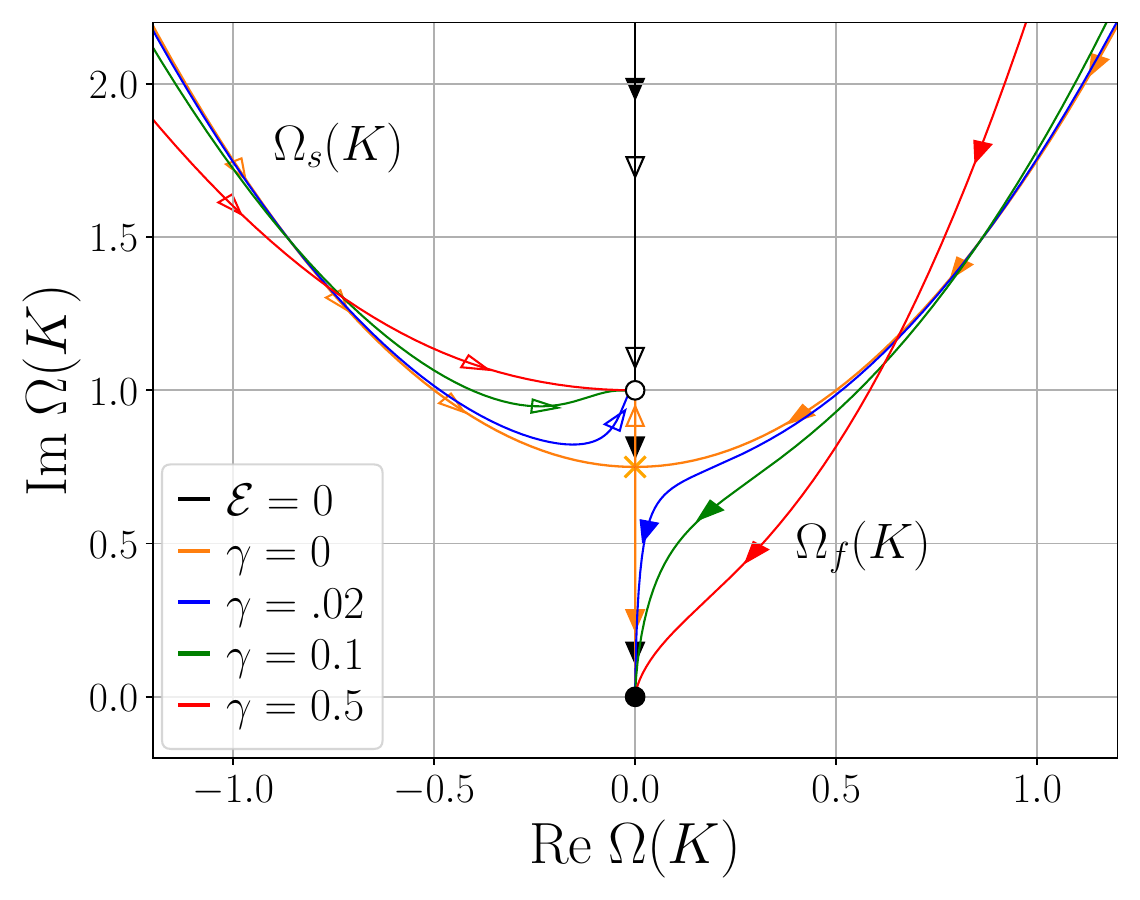} 
    \end{center}
    \caption{Parametric plot in the complex plane (as a function of the reduced wave number $K=k/\kappa$) of the frequencies $\Omega_{s,f}(K)$ (defined in Eqs.~\ref{eq:Omega1} and \ref{eq:Omega2}) characterizing the two modes of the coupled number and charge fluctuations. Arrows indicate decreasing wave vector $K\rightarrow0$, with filled and empty heads corresponding to $\Omega_s(K)$ (slow mode, closer to the origin of the complex plane) and $\Omega_f(K)$ (fast mode), respectively. The black filled and empty circles mark the ``hydrodynamic'' limits $\Omega_s(K\to0) = 0$ and $\Omega_f(K\to0) = i$ corresponding to salt diffusion and charge relaxation, respectively. The black lines correspond to equilibrium ($\mathcal{E} = 0$); the parametric plots for different values of asymmetry $\gamma$ overlap at equilibrium. The other curves correspond to NESS ($\mathcal{E} = 1$), with orange, blue, green and red for $\gamma = 0$, $0.02$, $0.1$, and $0.5$, respectively. 
    The orange cross at $(0,\frac{1}{2}+\frac{1}{4\mcE^2})$ indicates the bifurcation point at wave vector $K^*=1/2\mcE$ (see Eq.~\ref{eq:k_crit}) in the symmetric ($\gamma=0$) non-equilibrium case. 
    }
    \label{fig:dispersion}
\end{figure}

We now turn to dynamical properties. Before examining the intermediate scattering matrix, we consider the dispersion relations Eqs.~\ref{eq:Omega1} and~\ref{eq:Omega2} for the two eigenmodes. The two frequencies $\Omega_{s}(K)$ and $\Omega_{f}(K)$ are shown as a parametric plot in the complex plane, as a function of the reduced wave number $K$ in Fig.~\ref{fig:dispersion}, for the equilibrium ($\mcE=0$) and NESS with $\mcE=1$ of electrolytes symmetric ($\gamma=0$) and asymmetric ($\gamma\neq0$) diffusivities. We first examine the behavior of the eigenmodes in the hydrodynamic ($k\to0$, long-wavelength) limit, before discussing them in more details together with the intermediate structure matrix at equilibrium and under NESS in Sections~\ref{sssec:fkteq} and~\ref{sssec:fktness}, respectively.

\paragraph*{Hydrodynamic ($k\to0$) limit}

The two eigenmodes in the hydrodynamic regime ($K\to0$) are found near the black filled and empty circles in Fig.~\ref{fig:dispersion}, corresponding to  $\Omega_s(K\to0) = 0$ and $\Omega_f(K\to0) = \im$, respectively. Upon restoring physical units, we obtain the following approximate dispersion relations
\begin{empheq}[left=\empheqlbrace]{align}
    & \omega_s(k) \sim \mathrm{i} \, D^\mathrm{neq}_\mathrm{NH} \, k^2  \ ,
    \label{eq:mode_1} \\
    & \omega_f(k) \sim \mathrm{i} \, \taud^{-1} - v^\mathrm{neq} k \ ,
    \label{eq:mode_2}
\end{empheq}
which indicate, as in Section~\ref{ssec:res_currents}, a slow diffusive mode and a fast relaxation mode. However, the presence of an electric field leads to two important differences.

Firstly, the effective diffusivity of the salt in Eq.~\ref{eq:mode_1} is enhanced in the direction of the electric field compared to the equilibrium case (Eq.~\ref{eq:D_NH}) by a factor:
\begin{equation}
        D^\mathrm{neq}_\mathrm{NH} = D_\mathrm{NH}\times(1+\mcE^2)
        = D (1-\gamma^2) (1+\mcE^2)
        \label{eq:D_NH_ANI}
\end{equation}
This effective diffusion coefficient can be either larger or smaller than the average diffusion coefficient $D$, contrary to the equilibrium case. Secondly, although the relaxation mode corresponding to $\omega_f(k)$ again decays with a characteristic time $\taud$, it is now modulated by an oscillatory component with a frequency
\begin{equation}
        v^\mathrm{neq}k = \frac{2 \gamma K \mcE}{\taud} 
\end{equation}
that vanishes in the symmetric case ($\gamma=0$) as well as in the $k\to0$ limit.

\subsubsection{Dynamic correlations at equilibrium}
\label{sssec:fkteq}

\begin{figure*}[ht!]
    \centering
    \includegraphics[width=.90\textwidth]{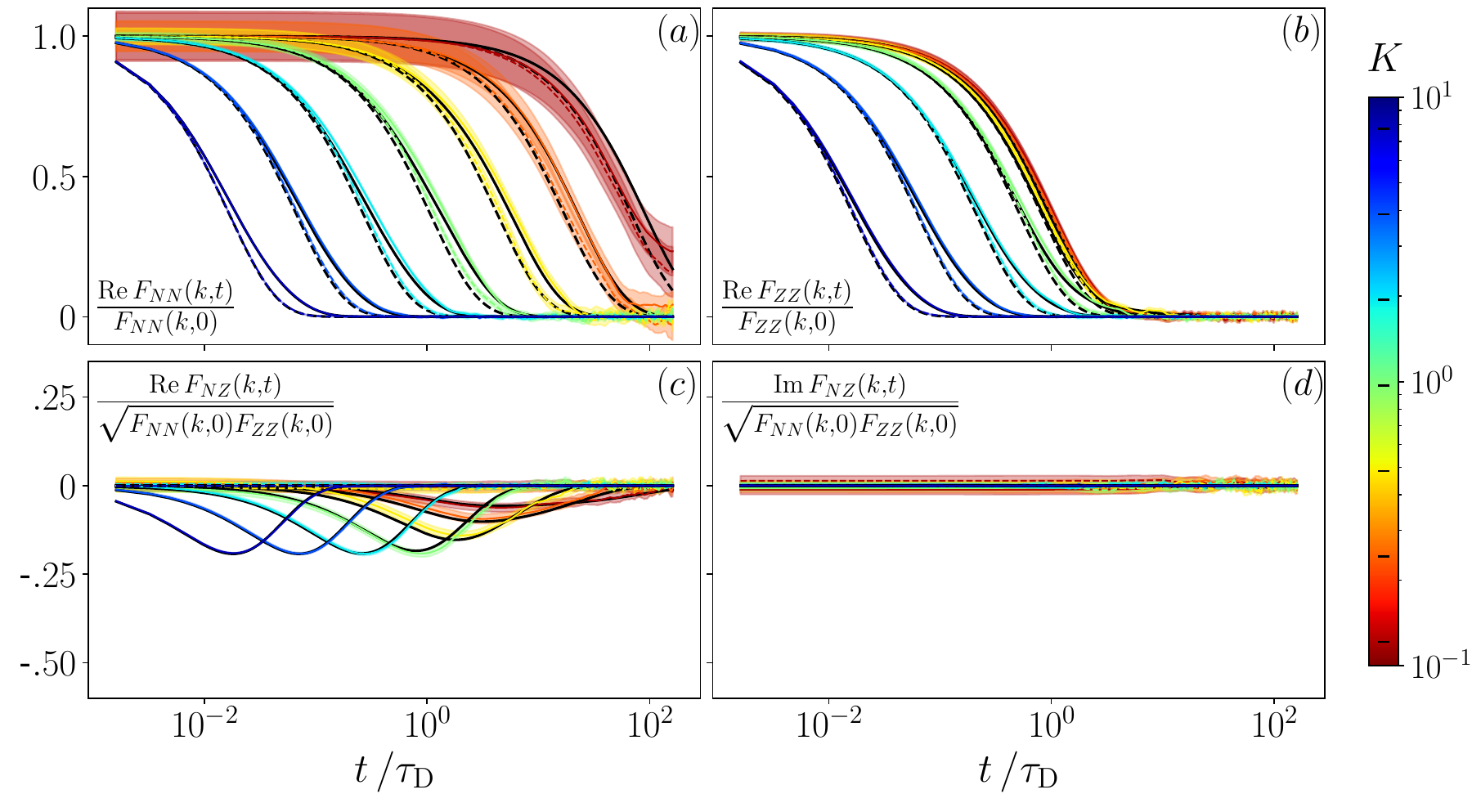}
    \caption{Elements of the intermediate scattering matrix Eq.~\ref{eq:def_hat_F} at equilibrium: $F_{NN}$ (a), $F_{ZZ}$ (b), $\operatorname{Re}F_{NZ}$ (c) and $\operatorname{Im}F_{NZ}$ (d), for reduced wave vectors in the range $K=k/\kappa\in[0.1,10]$ as indicated by the colorbar. Each component is normalized following Eq.~\ref{eq:norm_t0} and plotted as a function of the reduced time $t/\taud$, with $\taud=1/(D\kappa^{2})$.  Dashed and solid lines correspond to symmetric ($\gamma=0$) and asymmetric ($\gamma=0.5$) diffusion coefficients, respectively. Colored lines correspond to BD simulations (with shaded regions denoting 95\,\% confidence intervals obtained from independent replicas, see Sec.~\ref{ssec:BD_sim}), while black lines are the SDFT predictions: Eqs.~\ref{Fkt_cc_0}–\ref{Fkt_zc_0} for $\gamma=0$ and Eqs.~\ref{eq:Fkt_cc_A}–\ref{eq:Fkt_zc_A} for $\gamma=0.5$, respectively.
    }
    \label{fig:Fkt_eq}
\end{figure*}

\paragraph*{Symmetric electrolytes at equilibrium} 
In this simplest case ($\gamma=0$, $\mcE=0$), the dispersion relations Eqs.~\ref{eq:Omega1}-\ref{eq:nu} reduce to the well known results $\omega_s(k) = i D k^2$ and $\omega_f(k) = i D(k^2+\kappa^2)$, shown as black lines in Fig.~\ref{fig:dispersion}. The intermediate structure matrix given by Eq.~\ref{eq:FULL_F} simplifies to
\begin{align}
    F_{NN}(K,t) & =   \e^{-K^2\, t/\taud} \label{Fkt_cc_0}\\
    F_{ZZ}(K,t) & = \frac{K^2}{K^2+1} \, \e^{-(K^2+1) \, t/\taud}  \label{Fkt_zz_0}\\
    F_{NZ}(K,t) & = 0 \label{Fkt_cz_0}\\
    F_{ZN}(K,t) & = 0 \,. \label{Fkt_zc_0}
\end{align}
Fig.~\ref{fig:Fkt_eq} shows that these predictions (dashed black lines) describes the results well from BD simulations (dashed colored lines) over the whole time range (4 decades in $t/\taud$) for a wide range of reduced wave vectors $K=k/\kappa$. The number and charge fluctuations are decoupled and that the number-number and charge-charge fluctuations decay exponentially. Notably, while the timescale for the decay of diffusive relaxation of number fluctuations diverges in the $k\to0$ limit, it saturates to the Debye relaxation time $\taud$ for the charge fluctuations, which are suppressed beyond the Debye screening length.

\paragraph*{Asymmetric electrolytes at equilibrium} When $\gamma \neq 0$, as discussed in Sec.~\ref{ssec:res_currents}, the dynamics is modified due to the difference in diffusivity for the anion and the cation, which are coupled through electrostatic interactions. The dispersion relations are given by the relaxation times $\tau_{s,f}(k)$ defined in Eqs.~\ref{eq:tau1}–\ref{eq:tau2}, i.e., $\omega_{s,f}(k) = \im \, \tau_{s,f}^{-1}(k)$. The two eigenmodes do not correspond to either number or charge fluctuations, but to combinations of both. The matrix elements of the time-correlation matrix $\hat{\vect{F}}(k,t)$ now take the form:
\begin{align}
    F_{NN}(K,t) & =   \e^{-\zeta(K) \, t/\taud} \left[ \cosh{\frac{\nu(K)\,t}{\taud}} \right. \nonumber \\
    & \left. \hskip 2.5cm + \frac{1}{2\nu(K)} \sinh{\frac{\nu(K)\,t}{\taud}}\right] \label{eq:Fkt_cc_A} \\
    F_{ZZ}(K,t) & = \frac{K^2}{K^2+1} \, \e^{-\zeta(K) \, t/\taud} \left[ \cosh{\frac{\nu(K)\,t}{\taud}} \right. \nonumber \\
    & \left. \hskip 2.5cm- \frac{1}{2\nu(K)} \sinh{\frac{\nu(K)\,t}{\taud}} \right] \label{eq:Fkt_zz_A}\\
    F_{NZ}(K,t) & =  -\frac{\gamma K^2}{\nu(K)} \, \e^{-\zeta(K) \, t/\taud} \sinh{\frac{\nu(K)\,t}{\taud}} \label{eq:Fkt_cz_A}\\
    F_{ZN}(K,t) & = F_{NZ}(K,t) \label{eq:Fkt_zc_A} 
\end{align}
with (see Eqs.~\ref{eq:zeta} and \ref{eq:nu}):
\begin{align}
\zeta(K) &= K^2 + \frac{1}{2} \\
\nu(K) &= \sqrt{ \frac{1}{4} + \gamma^2 K^2 (K^2 + 1)} \; .
\end{align}
All correlation functions display two exponentially decaying modes, with the same relaxation times but different weights for the number, charge and cross-correlations. Mobility asymmetry, combined with electrostatic interactions, produces correlations that decay more slowly than in a symmetric electrolyte, as can be seen in Fig.~\ref{fig:Fkt_eq} (solid lines). The slowest decaying modes correspond to (see Eq.~\ref{eq:Omega1}) $\omega_s(k)= i D\kappa^2[\zeta(K)-\nu(K)]\sim i D_{\rm NH}k^2$ for $k\to0$, \textit{i.e.} a diffusive process with the Nernst-Hartley diffusion coefficient, given by Eq.~\ref{eq:D_NH}, which shows that $D_{\rm NH}<D$ when $\gamma\neq 0$. In the same $k\to0$ limit, the fast mode decays with the Debye relaxation time, as expected. As mentioned above, these modes do not correspond to the relaxation of number or charge fluctuations anymore, but rather to combinations of both. The real off‑diagonal elements $F_{NZ}(k,t)=F_{ZN}(k,t)$, which vanish at $\gamma=0$, reflect the Onsager symmetry at equilibrium, $\hat{\vect{F}}(k,t)=\hat{\vect{F}}^{\dagger}(k,t)$.

\subsubsection{Dynamic correlations in non-equilibrium steady-state}
\label{sssec:fktness}

\begin{figure*}[ht!]
    \centering
    \includegraphics[width=.90\textwidth]{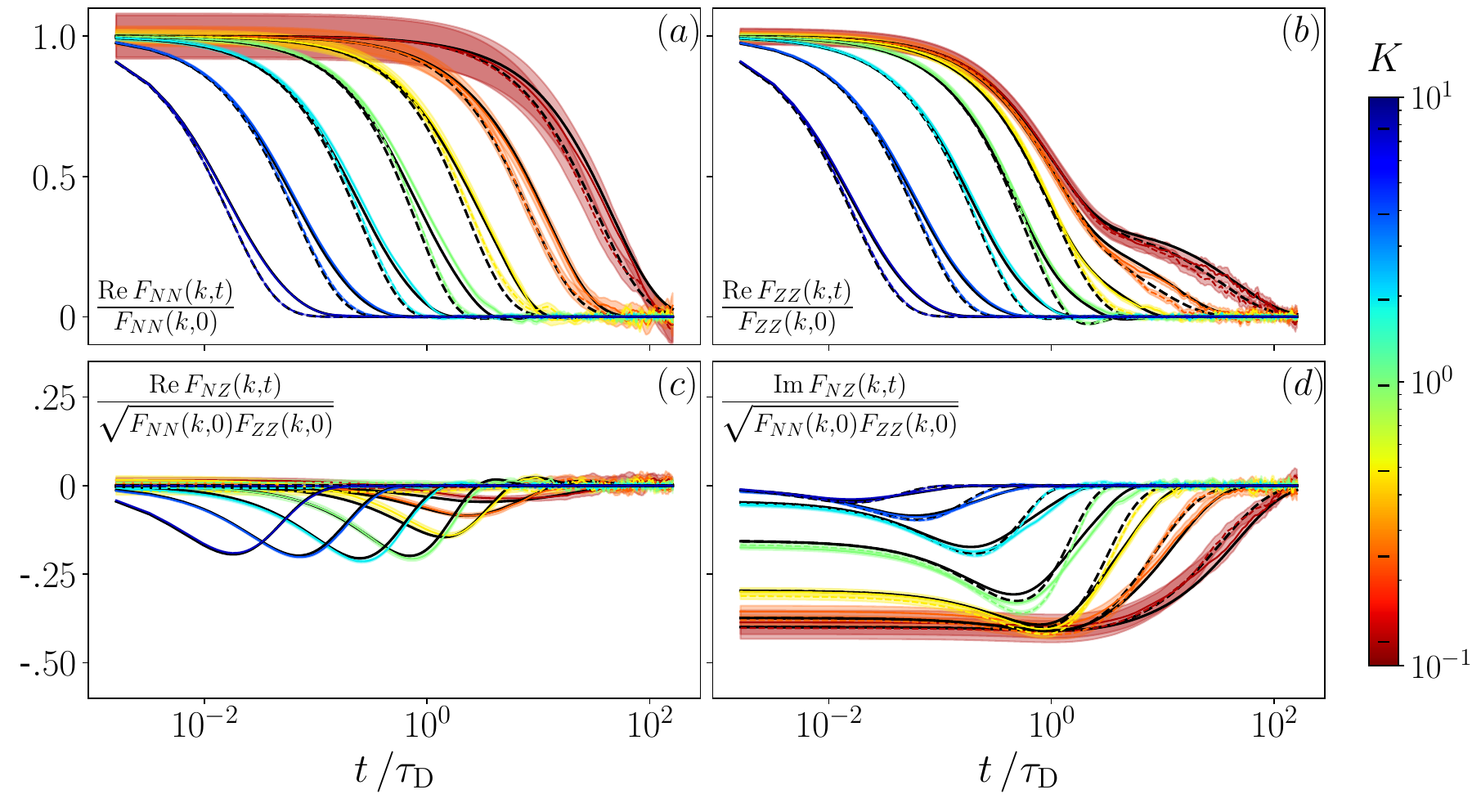}
    \caption{Elements of the intermediate scattering matrix Eq.~\ref{eq:def_hat_F} under NESS for a reduced electric field $\mcE=1$: $\operatorname{Re}F_{NN}$ (a), $\operatorname{Re}F_{ZZ}$ (b), $\operatorname{Re}F_{NZ}$ (c) and $\operatorname{Im}F_{NZ}$ (d), for reduced wave vectors in the range $K=k/\kappa\in[0.1,10]$ as indicated by the colorbar. Each component is normalized following Eq.~\ref{eq:norm_t0} and plotted as a function of the reduced time $t/\taud$, with $\taud=1/(D\kappa^{2})$.  Dashed and solid lines correspond to symmetric ($\gamma=0$) and asymmetric ($\gamma=0.5$) diffusion coefficients, respectively. Colored lines correspond to BD simulations (with shaded regions denoting 95\,\% confidence intervals obtained from independent replicas, see Sec.~\ref{ssec:BD_sim}), while black lines are the SDFT predictions Eq.~\ref{eq:FULL_F}, supplemented by Eqs.~\ref{G_cc0}–\ref{G_zc0} for $\gamma=0$ and by the general expression Eq.~\ref{eq:formal_Gk} for $\gamma\neq0$. 
    }
    \label{fig:Fkt_neq}
\end{figure*}

\paragraph*{Symmetric electrolytes in non-equilibrium steady-state} Under an external electric field, the time-correlation functions for a symmetric binary electrolyte ($E \neq 0$, $\gamma = 0$) are given by Eq.~\ref{eq:FULL_F} with the static structure factors $\hat{\vect{F}}(k,0)$ given by Eqs.~\ref{F_cc0}-\ref{F_zc0} with $\zeta(K) = K^2+1/2$ and $\nu(K) = \sqrt{1/4 - \mcE^2K^2}$, as well as the following elements of the $\hat{\vect{G}}(k)$ matrix:
\begin{align}
    G_{NN}(K) & =  \frac{K^2 + 1}{2 \nu(K)  \, [K^2+1+\mcE^2]} \label{G_cc0}\\
    G_{ZZ}(K) & = -\frac{K^2}{2 \nu(K)  \, [K^2+1+\mcE^2]}  \label{G_zz0}\\
    G_{NZ}(K) & =  \frac{-i \mcE K \left[ 2K^2(2K^2 + 2\mcE^2 + 3) + 1 \right]  }{2 \nu(K)  \, [K^2+1+\mcE^2] (2K^2+1)}   \label{G_cz0}\\
    G_{ZN}(K) & =  \frac{-i \mcE K \left[ 2K^2(2K^2 + 2\mcE^2 + 1) - 1 \right]  }{2 \nu(K)  \, [K^2+1+\mcE^2] (2K^2+1)}    \label{G_zc0}
\end{align}
Eqs.~\ref{G_cc0}-\ref{G_zz0} reproduce the results of Mahdisoltani \textit{et al.}~\cite{mahdisoltani_transient_2021} (specifically their Eqs. C2–C6). 

Fig.~\ref{fig:Fkt_neq} illustrates the same elements of the intermediate structure matrix as in Fig.~\ref{fig:Fkt_eq} but under a finite electric field $\mcE=1$. The remaining elements are provided in Fig.~\ref{fig:Fkt_neq2}. We first note again the very good agreement between the BD results and the analytical predictions (dashed black lines), which highlight the relevance of the latter. The two most striking differences with respect to the equilibrium case are the shoulder at long times for small $K$ in the charge-charge correlations ($\operatorname{Re}F_{ZZ}$ in panel b), and the appearance of an imaginary component in the number-charge correlations ($\operatorname{Im}F_{NZ}$ in panel d). The shoulder in $\operatorname{Re}F_{ZZ}$ corresponds to an exponential decay with a characteristic time that is only present in $\operatorname{Re}F_{NN}$ at equilibrium: the field-induced mode coupling results in a contribution of the slow mode to the charge-charge correlations. In fact both modes are present in all elements of the scattering matrix, with different weights, as can be observed by comparing the various panels. This slowest decay corresponds to a diffusive mode, with a diffusion coefficient enhanced by a factor $1+\mcE^2$ with respect to equilibrium (see Eq.~\ref{eq:D_NH_ANI}).

\begin{figure}[ht!]
    \centering
    \includegraphics[width=.45\textwidth]{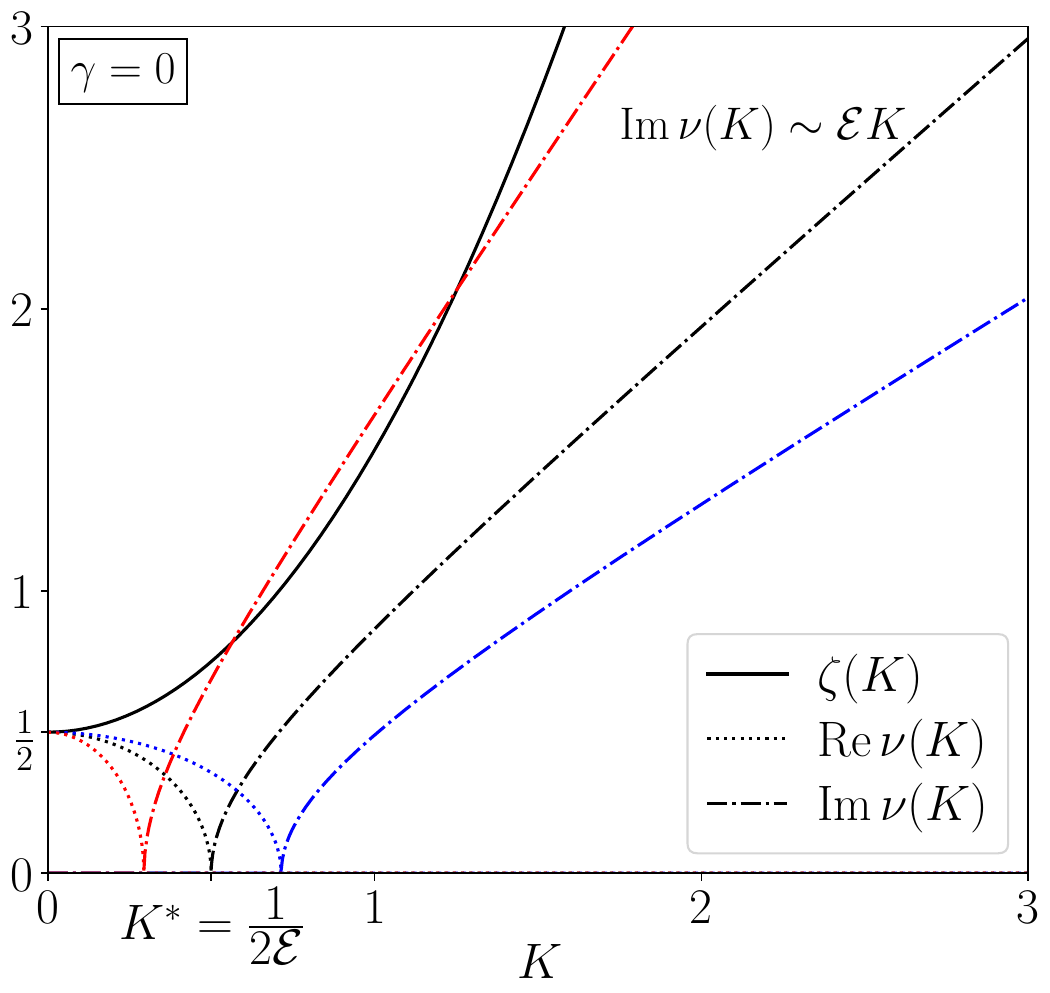}
    \caption{Functions $\zeta(K) = K^2+1/2$ and $\nu(K) = \sqrt{1/4 - \mcE^2K^2}$ as a function of the reduced wave vector $K=k/\kappa$ for a symmetric electrolyte ($\gamma=0$) in NESS under reduced electric field $\mcE=0.7, \, 1, \,1.7$ in blue, black and red, respectively. $\zeta(K)$ (full line) is always real and does not depend on $\mcE$, while $\nu(K)$ switches from real (dotted lines) to imaginary (dashed-dotted lines) at a critical wave vector $K^*=1/(2\mcE)$. For sufficiently large fields, there is a range of $K$ such that $\operatorname{Im}\nu(K)>\zeta(K)$. The imaginary branch of $\nu(K)$ is tangent to $\zeta(K)$ when $\mathcal{E}=\sqrt{1+\sqrt{2}}$, at $K=2^{-1/4}$.
    }
    \label{fig:zeta-nu}
\end{figure}

Notably, the dispersion relations $\Omega_{s,f}(K)$ exhibit a bifurcation at the critical reduced wave vector:
\begin{equation}
    K^*(\mcE) = \frac{1}{2 \mcE} \, .
    \label{eq:k_crit}
\end{equation}
This bifurcation marks a qualitative change in the relaxation dynamics, depending on the considered length scale. At small wave vectors ($K < K^*(\mcE)$), the poles $\Omega_{s,f}(K)$ are purely imaginary, corresponding to purely decaying (diffusive and relaxational) behavior. As $K$ increases, the poles coalesce at the critical wave vector to $\Omega_{s}(K^*)=\Omega_{f}(K^*)=i \left(\frac{1}{2}+\frac{1}{4\mcE^2}\right)$, before splitting for $K > K^*(E)$ into two branches with opposite real parts and identical imaginary parts (see orange lines in Fig.~\ref{fig:dispersion}).

The wave vector dependence of the two modes is further illustrated in Fig.~\ref{fig:zeta-nu}, which displays $\zeta(K)$ and $\nu(K)$ for several electric fields. While $\zeta(K)$ is always real and does not depend on $\mcE$ in the present case of symmetric mobilities ($\gamma=0$), $\nu(K)$ switches from real, $\nu(K<K^*(\mcE)) = \sqrt{1/4 - \mcE^2K^2}$, to purely imaginary, $\nu(K>K^*(\mcE)) = i\sqrt{\mcE^2K^2-1/4}$, indicating the coexistence in NESS of overdamped relaxation over large length scales and damped oscillatory features on short length scales. Such a coexistence cannot be observed at equilibrium ($\mcE=0$), since the critical wave vector above which oscillatory modes can be sustained diverges (see Eq.~\ref{eq:k_crit}). However, we note that in order for such oscillatory behavior to be observed, the characteristic time for the decay must be sufficiently long, typically $\operatorname{Im}\nu(K)>\zeta(K)$. Fig.~\ref{fig:zeta-nu} shows that a range of wave vectors satisfying this condition only exists for sufficiently large applied fields. We show in Appendix (see Fig.~\ref{fig:app_high}) that even though the analytical predictions are not as accurate in that case (especially for $K<1$), the simulation results for a larger field ($\mcE=3$) indeed display the predicted oscillatory behavior for a finite range of wave vectors.

Finally, in NESS, time-reversal symmetry is broken and the matrix $\vect{\hat{F}}(k,t)$ is not hermitian in general:  the Onsager reciprocal relations hold at $t=0$, $\hat{\vect{F}}(k,t=0) = \hat{\vect{F}}^\dagger(k,t=0)$, but this is not true for $t\neq0$, in particular $\operatorname{Im} F_{NZ}(k,t) \neq - \operatorname{Im} F_{ZN}(k,t)$. Therefore the real/imaginary components of the scattering matrix shown in Fig.~\ref{fig:Fkt_neq} are not sufficient to fully characterize it and the remaining terms are provided in Supplementary Fig.~\ref{fig:Fkt_neq2}. Even though this may not immediately appear from Eqs.~\ref{eq:FULL_F}, \ref{F_cc0}, \ref{F_zz0}, \ref{G_cc0} and~\ref{G_zz0}, $F_{NN}(k,t)$ and $F_{ZZ}(k,t)$ remain real for symmetric electrolyte even in NESS, even when $\nu(K)$ is purely imaginary, \textit{i.e.} for $K>K^*(\mcE)$. This is illustrated by their vanishing imaginary parts in Panels~\ref{fig:Fkt_neq2}a and~\ref{fig:Fkt_neq2}b (dashed lines).

\medskip

\paragraph*{Asymmetric electrolytes in non-equilibrium steady-state} 
With unequal diffusion coefficients ($\gamma\neq0$), the dispersion relations are marginally changed. As shown in Fig.~\ref{fig:dispersion} (colored lines), the bifurcation induced by the electric field is smoothed out: $\Omega_{s,f}(K)$ always displays a real part that continuously vanishes in the macroscopic limit, but the shape of the two branches depends on the value of the asymmetry $\gamma$. In this general case $\zeta(K)$ and $\nu(K)$ carry both a real and an imaginary part at finite wave vectors, and the matrix $\hat{\vect{G}}(K)$ is given by Eq.~\ref{eq:formal_Gk}. As in equilibrium, the asymmetry in NESS leads to additional mixing that slows down the decay of time-correlation functions (see Fig.~\ref{fig:Fkt_neq} solid lines) compared to symmetric electrolytes. Furthermore, the field-induced oscillations interfere destructively.  

Panels~\ref{fig:Fkt_neq2}a and~\ref{fig:Fkt_neq2}b indicate that, contrary to the symmetric case, in NESS $F_{NN}(k,t)$ and $F_{ZZ}(k,t)$ display a non-zero imaginary part that vanishes both at $t=0$ and $t\to\infty$ and displays minima/maxima at times that increase with decreasing $K$ (as the decay time of the real part) and a magnitude that varies non-monotonically with $K$. In addition, the comparison between panels~\ref{fig:Fkt_neq}c and~\ref{fig:Fkt_neq2}c show that the real parts of $F_{NZ}(k,t)$ and $F_{ZN}(k,t)$ are similar, while that between panels~\ref{fig:Fkt_neq}d and~\ref{fig:Fkt_neq2}d show that their imaginary parts are not opposite in general, except at $t\sim0$.

\section{Conclusion and perspectives}
\label{sec:conclusion}

In this study, we investigated the coupled dynamics of concentration and charge in asymmetric 1:1 electrolytes, focusing on the role of diffusion asymmetry and external electric fields on the relaxation of fluctuations across length- and time scales. Our findings reveal that asymmetry in ionic diffusion coefficients fundamentally alters the collective dynamics of electrolytes, introducing a non-trivial coupling between charge and number currents. This coupling manifests in both the transient response to inhomogeneous electric fields and the steady-state spatio-temporal fluctuations under uniform fields. The dynamics are governed by two relaxation modes -- a fast mode associated with charge relaxation and a slow mode linked to ambipolar diffusion -- whose characteristics depend sensitively on $\gamma=(D_+-D_-)/(D_++D_-)$ and the applied field strength.

Theoretically, we derived closed-form expressions for the intermediate scattering matrix, $\hat{\vect{F}}(k,t)$, capturing the dynamics of density fluctuations at equilibrium and in NESS. Our linearized stochastic density functional theory (SDFT) predictions, validated by Brownian dynamics simulations, demonstrate excellent agreement across a wide range of wave vectors, both at equilibrium and under a finite electric field. Notably, asymmetry in diffusion coefficients ($\gamma \neq 0$) induces cross-correlations between number and charge fluctuations, slows relaxation via ambipolar diffusion, and modifies the oscillatory signatures characteristic of non-equilibrium dynamics. Under an applied electric field, the system exhibits enhanced diffusion, field-dependent screening lengths, and a bifurcation in relaxation modes, highlighting the rich interplay between external driving and intrinsic asymmetry.

These results underscore the importance of diffusion asymmetry and external fields in tuning the transport properties of electrolytes. For example, it might be possible to exploit the wave vector dependence of relaxation modes by designing local electric field gradients via surface patterning to promote charge-concentration couplings. Such ideas are particularly relevant in the contexts of electrolyte transport at the nanoscale and iontronics, where confinement and interfacial effects would impact finite-scale fluctuations. The coupling between dynamical fluctuations in electrolytes investigated here and the electronic fluctuations in metals can also lead to interesting electro-hydrodynamic effects at electrode/electrolyte interfaces~\cite{herrero_fluids_2026}. 

While our mean-field SDFT framework provides quantitative insights for dilute electrolytes, future work could extend this approach to include the effects of an explicit solvent and hydrodynamic interactions, in order to describe more concentrated electrolytes. A natural extension is therefore to incorporate these effects using frequency/wave vector dependent dielectric constant of water, explicit solvent descriptions~\cite{illien2024stochastic, varghese_dynamic_2025}, long-range hydrodynamic interactions~\cite{jardat_brownian_2000, donev_fluctuating_hydro_2019, bonneau_2024_frequency, bonneau_2025_stationary}, or short-range correlations~\cite{bernard_binding_msa_1996, sammuller_neural_functional_2023, bui_learning_classical_2025} into the present asymmetric framework and to benchmark the resulting theory against all‑atom molecular dynamics simulations. Such extensions would bridge the gap between continuum models and atomistic realism, offering new avenues for designing and optimizing electrolyte-based systems.


\section*{Acknowledgements}

This work is dedicated to the memory of Rudi Podgornik, whose papers on fluctuations in charged and soft matter inspired many of our recent works. The authors thank Pierre Illien, Sophie Marbach and David Andelman for many discussions on SDFT for electrolytes. This project received funding from the European Research Council under the European Union’s Horizon 2020 research and innovation program (grant agreement no. 863473).

\section*{Author declarations}

\subsection*{Conflict of interest}
There is no conflict of interest to declare.

\subsection*{Author contributions}

\textbf{Th\^e Hoang Ngoc Minh} Conceptualization (equal); Formal analysis (equal); Investigation (lead); Methodology (equal); Writing/Original Draft Preparation (lead); Writing – review \& editing (equal). \textbf{Sleeba Varghese} Formal analysis (equal); Investigation (supporting); Methodology (supporting); Writing – review \& editing (equal). \textbf{Benjamin Rotenberg:} Conceptualization (equal); Formal analysis (equal); Funding Acquisition (lead); Investigation (supporting); Methodology (equal); Supervision (lead); Writing/Original Draft Preparation (lead); Writing – review \& editing (lead).

\section*{Data availability}
The data supporting the findings of this study will be made openly available on Zenodo upon final acceptance of the manuscript.

\appendix

\section{Derivation of the intermediate scattering matrix in non-equilibrium steady-state}
\label{app:SDFT}

In this Appendix, we derive the intermediate scattering matrix, $\hat{\vect{F}}(\vect{k},t)$, defined in Eq.~\ref{eq:def_hat_F} and characterizing the spatio-temporal density fluctuations, in the specific case of a steady homogeneous electric field (see Eq.~\ref{eq:E_bias}). 
This derivation is conveniently done in the frequency domain, and we introduce the temporal Fourier transform
\begin{equation}
  \hat{\vect{S}}(\vect{k},\omega)
    =\int_{-\infty}^{\infty}\hat{\vect{F}}(\vect{k},t)\,
      e^{-i\omega t}\,{\rm d}t .
  \label{eq:def_hat_S}
\end{equation}
By construction, $\hat{\vect{S}}(\vect{k},\omega)$ is also the power–spectral density of the density fluctuations:
\begin{equation}
  \bigl\langle\widetilde{\vect{\rho}}(\vect{k},\omega)\,
               \widetilde{\vect{\rho}}^{\dagger}(\vect{k}',\omega')\bigr\rangle
  = 2C_{s}\,\hat{\vect{S}}(\vect{k},\omega)\,
    (2\pi)^{4}\,\delta^{3}(\vect{k}-\vect{k}')\,
    \delta(\omega-\omega').
  \label{eq:def_hat_S_2}
\end{equation}

The linearized stochastic PNP equations Eqs.~\ref{eq:evolrhoN}-\ref{eq:evolrhoZ} simplify in that case since $\nabla\cdot\vect{E}^\mathrm{ext}_{\vect{0}}=0$ -- in particular $\vect{s}^{\rm ext}(\vect{r},t)=0$, see Eq.~\ref{eq:source_e_rt}. 
Eqs.~\ref{eq:evoltdrhoN}-\ref{eq:evoltdrhoZ} can then be written as:
\begin{equation} 
\label{eq:evolmatform}
\frac{\partial \widetilde{\vect{\rho}}(\vect{k},t)}{\partial t}
  +\widetilde{\vect{\mathcal{L}}}(\vect{k})\,
   \widetilde{\vect{\rho}}(\vect{k},t)
  = \widetilde{\vect{s}}_{r}(\vect{k},t)    
\end{equation}
with the time-independent matrix
\begin{equation} 
\label{eq:Lofk}
\widetilde{\vect{\mathcal{L}}}(\vect{k})= \frac{1}{\taud}
\begin{bmatrix}
  K^{2}-i\gamma\,\vect{\mathcal{E}}\!\cdot\!\vect{K} &
  \gamma(K^{2}+1)-i\,\vect{\mathcal{E}}\!\cdot\!\vect{K} \\[4pt]
  \gamma K^{2}-i\,\vect{\mathcal{E}}\!\cdot\!\vect{K} &
  K^{2}+1-i\gamma\,\vect{\mathcal{E}}\!\cdot\!\vect{K}
\end{bmatrix} 
\; .
\end{equation}
Taking the temporal Fourier transform, we obtain
\begin{equation}
-i\omega\,\widetilde{\vect{\rho}}(\vect{k},\omega)
   +\widetilde{\vect{\mathcal{L}}}(\vect{k})\,
    \widetilde{\vect{\rho}}(\vect{k},\omega)
  =\widetilde{\vect{s}}_{r}(\vect{k},\omega),
\end{equation}
whose solution is formally written as 
\begin{equation}
    \widetilde{\vect{\rho}}(\vect{k},\omega)
    =\vect{\hat{R}}(\vect{k},\omega)\,
    \widetilde{\vect{s}}_{r}(\vect{k},\omega)
    \label{eq:solution_formal}
\end{equation}
with the resolvent
\begin{equation}
\vect{\hat{R}}(\vect{k},\omega)
    \equiv\bigl[\widetilde{\vect{\mathcal{L}}}(\vect{k})-i\omega\mathbb{I}\bigr]^{\!-1}
    \; .
    \label{eq:resolvent}
\end{equation}
To compute the power-spectral density (Eq.~\ref{eq:def_hat_S_2}), we consider the noise covariance relation (Eq.~\ref{eq:source_r_kt_cov}), whose frequency counterpart reads
\begin{equation}
    \bigl\langle\widetilde{\vect{s}}_{r}(\vect{k},\omega)\,
          \widetilde{\vect{s}}_{r}^{\dagger}(\vect{k}',\omega')\bigr\rangle
  =(2\pi)^{4}\,4D C_{s}k^{2}\,\vect{\hat{\Xi}}\,
    \delta^{3}(\vect{k}-\vect{k}')\,
    \delta(\omega-\omega')
    \; ,
    \label{eq:source_r_kw_cov}
\end{equation}
with $\vect{\hat{\Xi}}$ provided in Eq.~\ref{eq:Xi}.
Substituting the relation Eq.~\ref{eq:source_r_kw_cov} and the solution Eq.~\ref{eq:solution_formal} into Eq.~\ref{eq:def_hat_S_2} yields
\begin{equation}
  \hat{\vect{S}}(\vect{k},\omega)
    =2Dk^{2}\,
      \vect{\hat{R}}(\vect{k},\omega)\cdot
      \vect{\hat{\Xi}}\cdot
      \vect{\hat{R}}^{\dagger}(\vect{k},\omega)
      \; .
  \label{eq:res:S_matrix}
\end{equation}
Finally, $\hat{\vect{S}}(\vect{k},\omega)$ and $\hat{\vect{F}}(\vect{k},t)$, can be expressed by considering the poles, $\Omega_{s,f}(K)$, of the resolvent Eq.~\ref{eq:resolvent}. These poles, found by solving $\det\bigl[\widetilde{\vect{\mathcal{L}}}(\vect{k})-i\omega\mathbb{I}\bigr]=0$, are given in Eqs.~\ref{eq:Omega1}-\ref{eq:nu} of the main text. Eq.~\ref{eq:source_r_kw_cov} then leads to
\begin{equation}
  \hat{\vect{S}}(k,\omega)
    = 2 \taud K^2 \, \frac{\vect{H}(\Omega)}
           {(\Omega-\Omega_{s})(\Omega-\Omega_{f})
            (\Omega-\overline{\Omega}_{s})(\Omega-\overline{\Omega}_{f})},
  \label{eq:S_ALL}
\end{equation}
where the matrix $\vect{H}(\Omega)$ is
\begin{align}
  H_{NN}&=\Omega^{2}+(1-\gamma^{2})\bigl[\mcE^2K^2+(K^{2}+1)^{2}\bigr],\\
  H_{ZZ}&=\Omega^{2}+(1-\gamma^{2})\bigl[\mcE^2K^2+K^{4}\bigr],\\
  \nonumber
  H_{NZ}&=\gamma\Omega^{2}
          -(1-\gamma^{2})\Bigl[\mcE K(2\Omega+i) \\
             &+\gamma\bigl(K^{4}+\mcE^2K^2+K^{2}\bigr)\Bigr],\\
    \nonumber
  H_{ZN}&=\gamma\Omega^{2}
          -(1-\gamma^{2})\Bigl[\mcE K(2\Omega-i) \\
             &+ \gamma\bigl(K^{4}+\mcE^2K^2+K^{2}\bigr)\Bigr].
\end{align}
To obtain the intermediate scattering matrix, we perform the inverse Fourier transform by considering $z\mapsto\hat{\vect{S}}(k,z)\,e^{izt}$ and using the residue theorem over half-circles with radius $R$, $\mathcal{C}_{R}^{+}$, in the upper half complex plane:
\begin{align}
  \hat{\vect{F}}(\vect{k},t)
   &=\frac{1}{2\pi}\int_{-\infty}^{\infty}
       \hat{\vect{S}}(\vect{k},\omega)\,e^{i\omega t}\,\diff \omega \\
   &=\lim_{R\to\infty}\frac{1}{2\pi}
       \oint_{\mathcal{C}_{R}^{+}}
       \hat{\vect{S}}(\vect{k},z)\,e^{izt}\,\diff z \\
   &=i\sum_{n=1,2}
       \operatorname{Res}\!\Bigl[
       \hat{\vect{S}}(\vect{k},z)\,e^{izt},
       \,z=\omega_{n}(\vect{k})\Bigr] \, .
\end{align}
The sum over the residues yields the final result Eq.~\ref{eq:FULL_F}, with the two damping rates $\zeta(K)$ and $\nu(K)$ defined in Eqs.~\ref{eq:zeta} and~\ref{eq:nu}, and the two matrices
\begin{equation}
  \hat{\vect{F}}(K,0)=
    \frac{K^{2}\bigl[\zeta_0(K) \, \hat{\vect{H}}_+
          +\frac{\zeta_0(K)^{2}+i\nu'(K) \nu''(K)}{\nu(K)}\,\hat{\vect{H}}_-\bigr]}
         {4\,\bigl[\zeta_0(K)^{2}-\nu'(K)^{2}\bigr]\bigl[\zeta_0(K)^{2}+\nu''(K)^{2}\bigr]},
  \label{eq:formal_Fk}
\end{equation}
\begin{equation}
  \hat{\vect{G}}(K)=
    \frac{K^{2}\bigl[\zeta_0(K) \, \hat{\vect{H}}_-
          +\frac{\zeta_0(K)^{2}+i\nu'(K) \nu''(K)}{\nu(K)}\,\hat{\vect{H}}_+\bigr]}
         {4\,\bigl[\zeta_0(K)^{2}-\nu'(K)^{2}\bigr]\bigl[\zeta_0(K)^{2}+\nu''(K)^{2}\bigr]},
  \label{eq:formal_Gk}
\end{equation}
where $\nu(K)=\nu'(K) + \im \nu''(K)$ is given in Eq.~\ref{eq:nu}, $\hat{\vect{H}}_+ = \hat{\vect{H}}(\Omega_s) + \hat{\vect{H}}(\Omega_f)$ and $\hat{\vect{H}}_- = \hat{\vect{H}}(\Omega_s) - \hat{\vect{H}}(\Omega_f)$ with the matrix $\hat{\vect{H}}$ described above, $\Omega_{s,f}$ given in Eqs.~\ref{eq:Omega1}-\ref{eq:Omega2}, and $\zeta_0(K)$ is given in Eq.~\ref{eq:zeta_zero}. Notably, Eq.~\ref{eq:formal_Fk} simplifies to Eqs.~\ref{F_cc0}-\ref{F_zc0}, while no further simplification was found for $\hat{\vect{G}}(K)$ (Eq.~\ref{eq:formal_Gk}) in the general case.

\section{Large applied electric field}
\label{app:high_field}

\begin{figure*}[ht!]
    \centering
    \includegraphics[width=.80\textwidth]{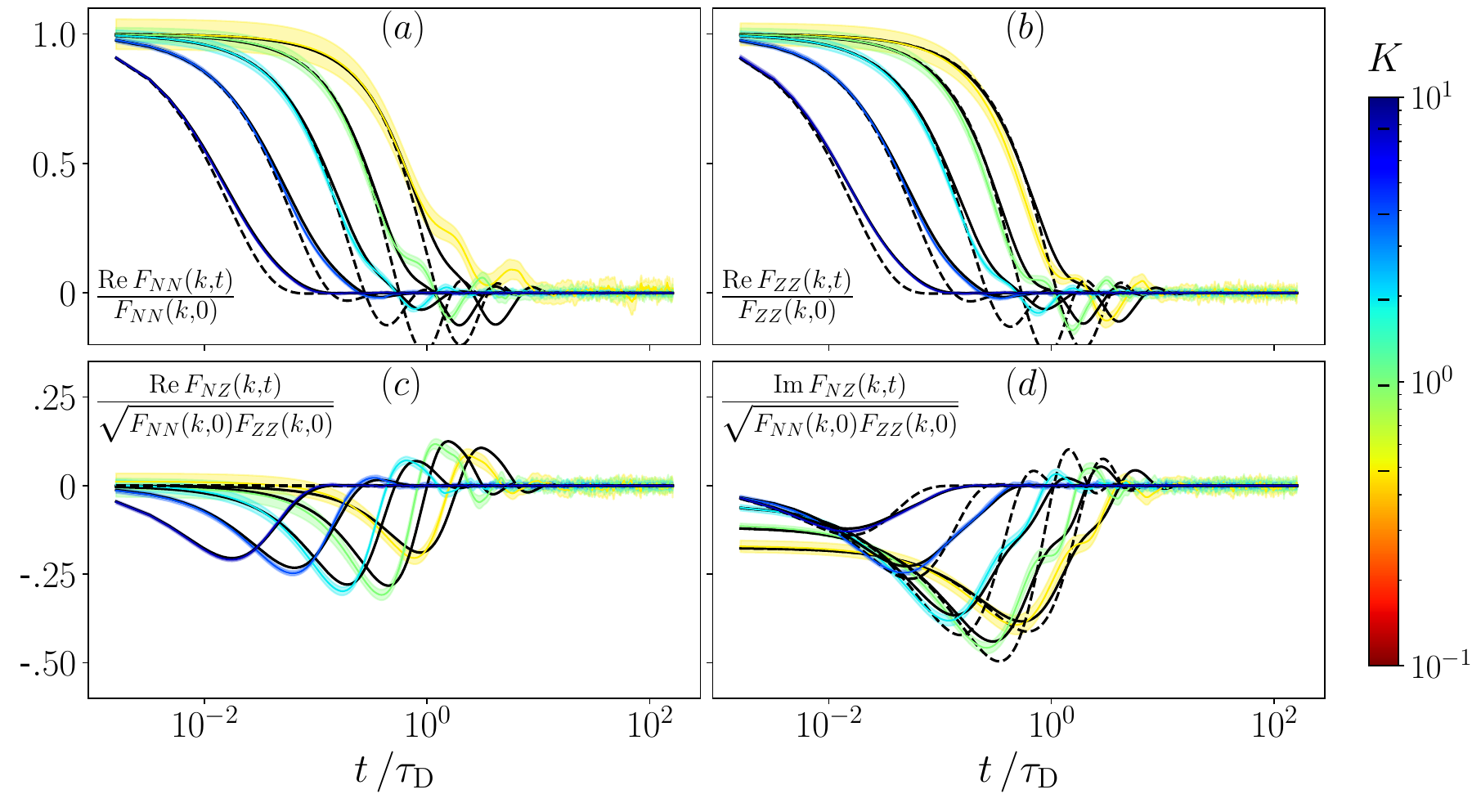}
    \caption{
    Elements of the intermediate scattering matrix Eq.~\ref{eq:def_hat_F} under NESS for a reduced electric field $\mcE=3$: $\operatorname{Re}F_{NN}$ (a), $\operatorname{Re}F_{ZZ}$ (b), $\operatorname{Re}F_{NZ}$ (c) and $\operatorname{Im}F_{NZ}$ (d), for reduced wave vectors in the range $K=k/\kappa\in[0.1,10]$ as indicated by the colorbar. Each component is normalized following Eq.~\ref{eq:norm_t0} and plotted as a function of the reduced time $t/\taud$, with $\taud=1/(D\kappa^{2})$.  Dashed and solid lines correspond to symmetric ($\gamma=0$) and asymmetric ($\gamma=0.5$) diffusion coefficients, respectively. Colored lines correspond to BD simulations in the symmetric case (with shaded regions denoting 95\,\% confidence intervals obtained from independent replicas, see Sec.~\ref{ssec:BD_sim}), while black lines are the SDFT predictions Eq.~\ref{eq:FULL_F}, supplemented by Eqs.~\ref{G_cc0}–\ref{G_zc0} for $\gamma=0$ and by the general expression Eq.~\ref{eq:formal_Gk} for $\gamma\neq0$. While the predictions are not as good as for the small $\mcE$ regime (especially for $K<1$), qualitative changes due to asymmetry are well captured.}
    \label{fig:app_high}
\end{figure*}

Although the SDFT predictions are obtained after linearizing the ionic densities around their average value, they can be compared to the BD results beyond the limit of small applied electric field ($\mcE\ll1$). Fig.~\ref{fig:app_high} compares the simulation results for a reduced electric field $\mcE=3$ with the SDFT prediction Eq.~\ref{eq:FULL_F}. While the agreement is not as quantitative as in the small field regime (especially for $K<1$), SDFT still captures the main features of BD simulations. In addition, the simulations confirm the existence of the oscillatory regime for intermediate wave numbers discussed in Section~\ref{sssec:fktness}.

\section{Complementary components of the intermediate scattering matrix in non-equilibrium steady-state}
\label{app:complementary}

\begin{figure*}[ht!]
    \centering
    \includegraphics[width=0.8\textwidth]{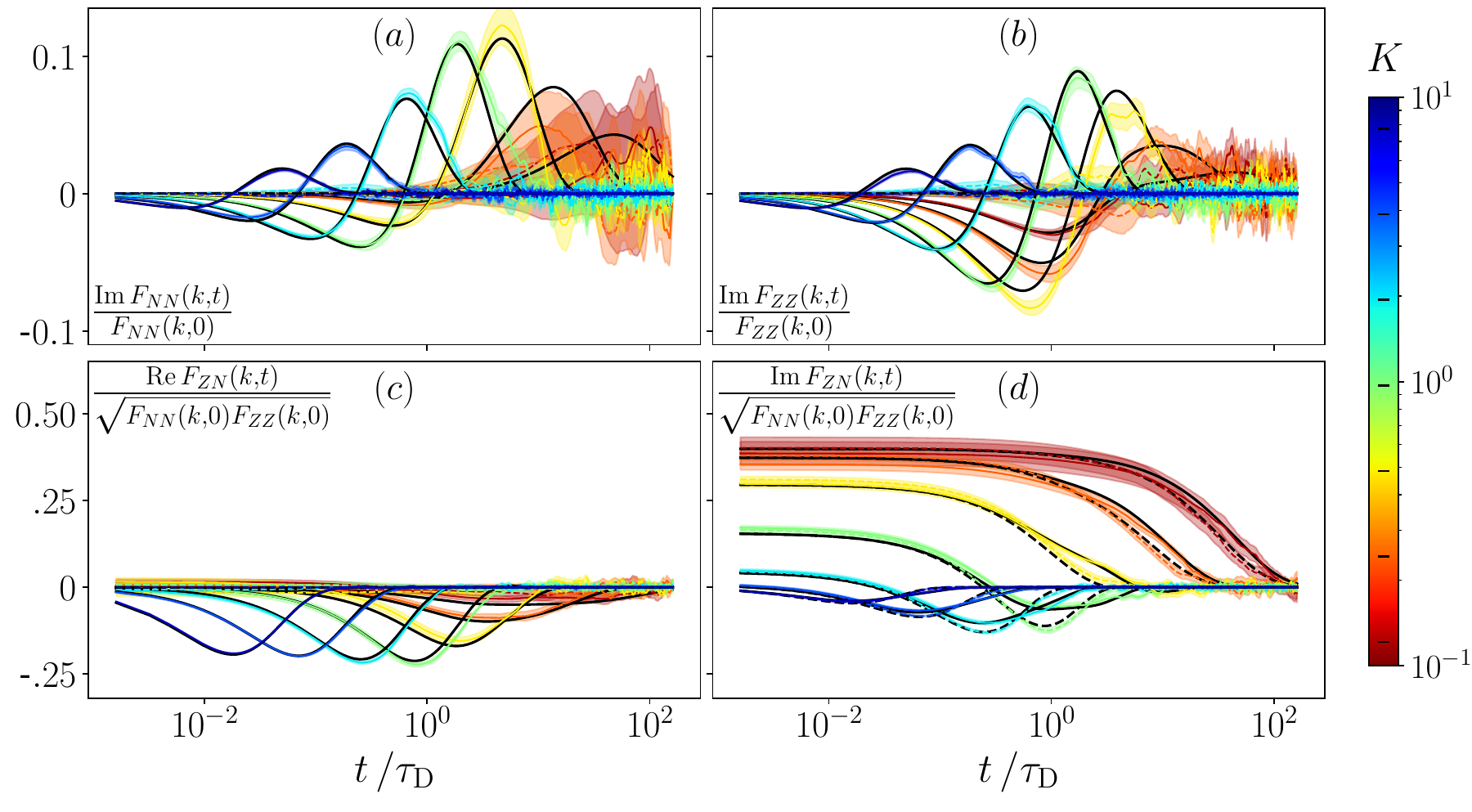}
    \caption{Elements of the intermediate scattering matrix Eq.~\ref{eq:def_hat_F} under NESS for a reduced electric field $\mcE=1$ (complementary to Fig.~\ref{fig:Fkt_neq}): $\operatorname{Im}F_{NN}$ (a), $\operatorname{Im}F_{ZZ}$ (b), $\operatorname{Re}F_{ZN}$ (c) and $\operatorname{Im}F_{ZN}$ (d), for reduced wave vectors in the range $K=k/\kappa\in[0.1,10]$ as indicated by the colorbar. Each component is normalized following Eq.~\ref{eq:norm_t0} and plotted as a function of the reduced time $t/\taud$, with $\taud=1/(D\kappa^{2})$.  Dashed and solid lines correspond to symmetric ($\gamma=0$) and asymmetric ($\gamma=0.5$) diffusion coefficients, respectively. Colored lines correspond to BD simulations (with shaded regions denoting 95\,\% confidence intervals obtained from independent replicas, see Sec.~\ref{ssec:BD_sim}), while black lines are the SDFT predictions Eq.~\ref{eq:FULL_F}, supplemented by Eqs.~\ref{G_cc0}–\ref{G_zc0} for $\gamma=0$ and by the general expression Eq.~\ref{eq:formal_Gk} for $\gamma\neq0$. 
    }
    \label{fig:Fkt_neq2}
\end{figure*}

In NESS, time-reversal symmetry is broken and the matrix $\vect{\hat{F}}(k,t)$ is not hermitian. Therefore the real/imaginary components of the scattering matrix shown in Fig.~\ref{fig:Fkt_neq} are not sufficient to fully characterize it. The other components are reported in Fig.~\ref{fig:Fkt_neq2}.

\bibliographystyle{aipnum4-1}
\bibliography{biblio}

\begin{thebibliography}{90}%
\makeatletter
\providecommand \@ifxundefined [1]{%
 \@ifx{#1\undefined}
}%
\providecommand \@ifnum [1]{%
 \ifnum #1\expandafter \@firstoftwo
 \else \expandafter \@secondoftwo
 \fi
}%
\providecommand \@ifx [1]{%
 \ifx #1\expandafter \@firstoftwo
 \else \expandafter \@secondoftwo
 \fi
}%
\providecommand \natexlab [1]{#1}%
\providecommand \enquote  [1]{``#1''}%
\providecommand \bibnamefont  [1]{#1}%
\providecommand \bibfnamefont [1]{#1}%
\providecommand \citenamefont [1]{#1}%
\providecommand \href@noop [0]{\@secondoftwo}%
\providecommand \href [0]{\begingroup \@sanitize@url \@href}%
\providecommand \@href[1]{\@@startlink{#1}\@@href}%
\providecommand \@@href[1]{\endgroup#1\@@endlink}%
\providecommand \@sanitize@url [0]{\catcode `\\12\catcode `\$12\catcode `\&12\catcode `\#12\catcode `\^12\catcode `\_12\catcode `\%12\relax}%
\providecommand \@@startlink[1]{}%
\providecommand \@@endlink[0]{}%
\providecommand \url  [0]{\begingroup\@sanitize@url \@url }%
\providecommand \@url [1]{\endgroup\@href {#1}{\urlprefix }}%
\providecommand \urlprefix  [0]{URL }%
\providecommand \Eprint [0]{\href }%
\providecommand \doibase [0]{http://dx.doi.org/}%
\providecommand \selectlanguage [0]{\@gobble}%
\providecommand \bibinfo  [0]{\@secondoftwo}%
\providecommand \bibfield  [0]{\@secondoftwo}%
\providecommand \translation [1]{[#1]}%
\providecommand \BibitemOpen [0]{}%
\providecommand \bibitemStop [0]{}%
\providecommand \bibitemNoStop [0]{.\EOS\space}%
\providecommand \EOS [0]{\spacefactor3000\relax}%
\providecommand \BibitemShut  [1]{\csname bibitem#1\endcsname}%
\let\auto@bib@innerbib\@empty
\bibitem [{\citenamefont {Hille}(2001)}]{Hille2001-es}%
  \BibitemOpen
  \bibfield  {author} {\bibinfo {author} {\bibfnamefont {B.}~\bibnamefont {Hille}},\ }\href@noop {} {\emph {\bibinfo {title} {Ionic channels of excitable membranes}}},\ \bibinfo {edition} {3rd}\ ed.\ (\bibinfo  {publisher} {Oxford University Press},\ \bibinfo {address} {New York, NY},\ \bibinfo {year} {2001})\BibitemShut {NoStop}%
\bibitem [{\citenamefont {Liu}\ \emph {et~al.}(2022)\citenamefont {Liu}, \citenamefont {Tournassat}, \citenamefont {Grangeon}, \citenamefont {Kalinichev}, \citenamefont {Takahashi},\ and\ \citenamefont {Marques~Fernandes}}]{liu_molecular-level_2022}%
  \BibitemOpen
  \bibfield  {author} {\bibinfo {author} {\bibfnamefont {X.}~\bibnamefont {Liu}}, \bibinfo {author} {\bibfnamefont {C.}~\bibnamefont {Tournassat}}, \bibinfo {author} {\bibfnamefont {S.}~\bibnamefont {Grangeon}}, \bibinfo {author} {\bibfnamefont {A.~G.}\ \bibnamefont {Kalinichev}}, \bibinfo {author} {\bibfnamefont {Y.}~\bibnamefont {Takahashi}}, \ and\ \bibinfo {author} {\bibfnamefont {M.}~\bibnamefont {Marques~Fernandes}},\ }\href {\doibase 10.1038/s43017-022-00301-z} {\bibfield  {journal} {\bibinfo  {journal} {Nat. Rev. Earth Environ.}\ }\textbf {\bibinfo {volume} {3}},\ \bibinfo {pages} {461} (\bibinfo {year} {2022})}\BibitemShut {NoStop}%
\bibitem [{\citenamefont {Appelo}\ and\ \citenamefont {Postma}(2005)}]{Appelo2005-qm}%
  \BibitemOpen
  \bibfield  {author} {\bibinfo {author} {\bibfnamefont {C.~A.~J.}\ \bibnamefont {Appelo}}\ and\ \bibinfo {author} {\bibfnamefont {D.}~\bibnamefont {Postma}},\ }\href@noop {} {\emph {\bibinfo {title} {Geochemistry, Groundwater and Pollution}}},\ \bibinfo {edition} {2nd}\ ed.,\ edited by\ \bibinfo {editor} {\bibfnamefont {C.~A.~J.}\ \bibnamefont {Appelo}}\ and\ \bibinfo {editor} {\bibfnamefont {D.}~\bibnamefont {Postma}}\ (\bibinfo  {publisher} {CRC Press},\ \bibinfo {address} {London, England},\ \bibinfo {year} {2005})\BibitemShut {NoStop}%
\bibitem [{\citenamefont {Simon}\ and\ \citenamefont {Gogotsi}(2020)}]{simon_perspectives_2020}%
  \BibitemOpen
  \bibfield  {author} {\bibinfo {author} {\bibfnamefont {P.}~\bibnamefont {Simon}}\ and\ \bibinfo {author} {\bibfnamefont {Y.}~\bibnamefont {Gogotsi}},\ }\href {\doibase 10.1038/s41563-020-0747-z} {\bibfield  {journal} {\bibinfo  {journal} {Nat. Mater.}\ }\textbf {\bibinfo {volume} {19}},\ \bibinfo {pages} {1151} (\bibinfo {year} {2020})}\BibitemShut {NoStop}%
\bibitem [{\citenamefont {Siria}, \citenamefont {Bocquet},\ and\ \citenamefont {Bocquet}(2017)}]{siria_new_2017}%
  \BibitemOpen
  \bibfield  {author} {\bibinfo {author} {\bibfnamefont {A.}~\bibnamefont {Siria}}, \bibinfo {author} {\bibfnamefont {M.-L.}\ \bibnamefont {Bocquet}}, \ and\ \bibinfo {author} {\bibfnamefont {L.}~\bibnamefont {Bocquet}},\ }\href {\doibase 10.1038/s41570-017-0091} {\bibfield  {journal} {\bibinfo  {journal} {Nat Rev Chem}\ }\textbf {\bibinfo {volume} {1}},\ \bibinfo {pages} {0091} (\bibinfo {year} {2017})}\BibitemShut {NoStop}%
\bibitem [{\citenamefont {Laszlo}\ \emph {et~al.}(8 01)\citenamefont {Laszlo}, \citenamefont {Derrington}, \citenamefont {Ross}, \citenamefont {Brinkerhoff}, \citenamefont {Adey}, \citenamefont {Nova}, \citenamefont {Craig}, \citenamefont {Langford}, \citenamefont {Samson}, \citenamefont {Daza}, \citenamefont {Doering}, \citenamefont {Shendure},\ and\ \citenamefont {Gundlach}}]{laszlo_decoding_2014}%
  \BibitemOpen
  \bibfield  {author} {\bibinfo {author} {\bibfnamefont {A.~H.}\ \bibnamefont {Laszlo}}, \bibinfo {author} {\bibfnamefont {I.~M.}\ \bibnamefont {Derrington}}, \bibinfo {author} {\bibfnamefont {B.~C.}\ \bibnamefont {Ross}}, \bibinfo {author} {\bibfnamefont {H.}~\bibnamefont {Brinkerhoff}}, \bibinfo {author} {\bibfnamefont {A.}~\bibnamefont {Adey}}, \bibinfo {author} {\bibfnamefont {I.~C.}\ \bibnamefont {Nova}}, \bibinfo {author} {\bibfnamefont {J.~M.}\ \bibnamefont {Craig}}, \bibinfo {author} {\bibfnamefont {K.~W.}\ \bibnamefont {Langford}}, \bibinfo {author} {\bibfnamefont {J.~M.}\ \bibnamefont {Samson}}, \bibinfo {author} {\bibfnamefont {R.}~\bibnamefont {Daza}}, \bibinfo {author} {\bibfnamefont {K.}~\bibnamefont {Doering}}, \bibinfo {author} {\bibfnamefont {J.}~\bibnamefont {Shendure}}, \ and\ \bibinfo {author} {\bibfnamefont {J.~H.}\ \bibnamefont {Gundlach}},\ }\href {\doibase 10.1038/nbt.2950} {\bibfield  {journal} {\bibinfo  {journal} {Nat. Biotechnol.}\ }\textbf {\bibinfo {volume} {32}},\ \bibinfo
  {pages} {829} (\bibinfo {year} {2014-08-01})}\BibitemShut {NoStop}%
\bibitem [{\citenamefont {Kavokine}, \citenamefont {Netz},\ and\ \citenamefont {Bocquet}(2021)}]{kavokine_fluids_at_the_nanoscale_2021}%
  \BibitemOpen
  \bibfield  {author} {\bibinfo {author} {\bibfnamefont {N.}~\bibnamefont {Kavokine}}, \bibinfo {author} {\bibfnamefont {R.~R.}\ \bibnamefont {Netz}}, \ and\ \bibinfo {author} {\bibfnamefont {L.}~\bibnamefont {Bocquet}},\ }\href {\doibase https://doi.org/10.1146/annurev-fluid-071320-095958} {\bibfield  {journal} {\bibinfo  {journal} {Annu. Rev. Fluid Mech.}\ }\textbf {\bibinfo {volume} {53}},\ \bibinfo {pages} {377} (\bibinfo {year} {2021})}\BibitemShut {NoStop}%
\bibitem [{\citenamefont {Blum}\ \emph {et~al.}(1982)\citenamefont {Blum}, \citenamefont {Gruber}, \citenamefont {Lebowitz},\ and\ \citenamefont {Martin}}]{blum_perfect_82}%
  \BibitemOpen
  \bibfield  {author} {\bibinfo {author} {\bibfnamefont {L.}~\bibnamefont {Blum}}, \bibinfo {author} {\bibfnamefont {C.}~\bibnamefont {Gruber}}, \bibinfo {author} {\bibfnamefont {J.~L.}\ \bibnamefont {Lebowitz}}, \ and\ \bibinfo {author} {\bibfnamefont {P.}~\bibnamefont {Martin}},\ }\href {\doibase 10.1103/PhysRevLett.48.1769} {\bibfield  {journal} {\bibinfo  {journal} {Phys. Rev. Lett.}\ }\textbf {\bibinfo {volume} {48}},\ \bibinfo {pages} {1769} (\bibinfo {year} {1982})}\BibitemShut {NoStop}%
\bibitem [{\citenamefont {Hoang Ngoc~Minh}, \citenamefont {Rotenberg},\ and\ \citenamefont {Marbach}(2023)}]{hoang_hyper_23}%
  \BibitemOpen
  \bibfield  {author} {\bibinfo {author} {\bibfnamefont {T.}~\bibnamefont {Hoang Ngoc~Minh}}, \bibinfo {author} {\bibfnamefont {B.}~\bibnamefont {Rotenberg}}, \ and\ \bibinfo {author} {\bibfnamefont {S.}~\bibnamefont {Marbach}},\ }\href {\doibase 10.1039/D3FD00031A} {\bibfield  {journal} {\bibinfo  {journal} {Faraday Discuss.}\ }\textbf {\bibinfo {volume} {246}},\ \bibinfo {pages} {225} (\bibinfo {year} {2023})}\BibitemShut {NoStop}%
\bibitem [{\citenamefont {Lee}\ \emph {et~al.}(2018)\citenamefont {Lee}, \citenamefont {Hansen}, \citenamefont {Bernard},\ and\ \citenamefont {and}}]{lee_2018_casimir}%
  \BibitemOpen
  \bibfield  {author} {\bibinfo {author} {\bibfnamefont {A.~A.}\ \bibnamefont {Lee}}, \bibinfo {author} {\bibfnamefont {J.-P.}\ \bibnamefont {Hansen}}, \bibinfo {author} {\bibfnamefont {O.}~\bibnamefont {Bernard}}, \ and\ \bibinfo {author} {\bibfnamefont {B.~R.}\ \bibnamefont {and}},\ }\href {\doibase 10.1080/00268976.2018.1478137} {\bibfield  {journal} {\bibinfo  {journal} {Mol. Phys.}\ }\textbf {\bibinfo {volume} {116}},\ \bibinfo {pages} {3147} (\bibinfo {year} {2018})}\BibitemShut {NoStop}%
\bibitem [{\citenamefont {Siria}\ \emph {et~al.}(2 01)\citenamefont {Siria}, \citenamefont {Poncharal}, \citenamefont {Biance}, \citenamefont {Fulcrand}, \citenamefont {Blase}, \citenamefont {Purcell},\ and\ \citenamefont {Bocquet}}]{siria_giant_2013}%
  \BibitemOpen
  \bibfield  {author} {\bibinfo {author} {\bibfnamefont {A.}~\bibnamefont {Siria}}, \bibinfo {author} {\bibfnamefont {P.}~\bibnamefont {Poncharal}}, \bibinfo {author} {\bibfnamefont {A.-L.}\ \bibnamefont {Biance}}, \bibinfo {author} {\bibfnamefont {R.}~\bibnamefont {Fulcrand}}, \bibinfo {author} {\bibfnamefont {X.}~\bibnamefont {Blase}}, \bibinfo {author} {\bibfnamefont {S.~T.}\ \bibnamefont {Purcell}}, \ and\ \bibinfo {author} {\bibfnamefont {L.}~\bibnamefont {Bocquet}},\ }\href {\doibase 10.1038/nature11876} {\bibfield  {journal} {\bibinfo  {journal} {Nature}\ }\textbf {\bibinfo {volume} {494}},\ \bibinfo {pages} {455} (\bibinfo {year} {2013-02-01})}\BibitemShut {NoStop}%
\bibitem [{\citenamefont {Balos}\ \emph {et~al.}(2020)\citenamefont {Balos}, \citenamefont {Imoto}, \citenamefont {Netz}, \citenamefont {Bonn}, \citenamefont {Bonthuis}, \citenamefont {Nagata},\ and\ \citenamefont {Hunger}}]{Balos2020-gf}%
  \BibitemOpen
  \bibfield  {author} {\bibinfo {author} {\bibfnamefont {V.}~\bibnamefont {Balos}}, \bibinfo {author} {\bibfnamefont {S.}~\bibnamefont {Imoto}}, \bibinfo {author} {\bibfnamefont {R.~R.}\ \bibnamefont {Netz}}, \bibinfo {author} {\bibfnamefont {M.}~\bibnamefont {Bonn}}, \bibinfo {author} {\bibfnamefont {D.~J.}\ \bibnamefont {Bonthuis}}, \bibinfo {author} {\bibfnamefont {Y.}~\bibnamefont {Nagata}}, \ and\ \bibinfo {author} {\bibfnamefont {J.}~\bibnamefont {Hunger}},\ }\href@noop {} {\bibfield  {journal} {\bibinfo  {journal} {Nat. Commun.}\ }\textbf {\bibinfo {volume} {11}},\ \bibinfo {pages} {1611} (\bibinfo {year} {2020})}\BibitemShut {NoStop}%
\bibitem [{\citenamefont {Mondal}\ and\ \citenamefont {Bagchi}(2021)}]{mondal_2021_anomalous_dielectric}%
  \BibitemOpen
  \bibfield  {author} {\bibinfo {author} {\bibfnamefont {S.}~\bibnamefont {Mondal}}\ and\ \bibinfo {author} {\bibfnamefont {B.}~\bibnamefont {Bagchi}},\ }\href {\doibase 10.1063/5.0032879} {\bibfield  {journal} {\bibinfo  {journal} {J. Chem. Phys.}\ }\textbf {\bibinfo {volume} {154}},\ \bibinfo {pages} {044501} (\bibinfo {year} {2021})}\BibitemShut {NoStop}%
\bibitem [{\citenamefont {Smeets}\ \emph {et~al.}(2008)\citenamefont {Smeets}, \citenamefont {Keyser}, \citenamefont {Dekker},\ and\ \citenamefont {Dekker}}]{smeets_2008_noise_in_ssn}%
  \BibitemOpen
  \bibfield  {author} {\bibinfo {author} {\bibfnamefont {R.~M.~M.}\ \bibnamefont {Smeets}}, \bibinfo {author} {\bibfnamefont {U.~F.}\ \bibnamefont {Keyser}}, \bibinfo {author} {\bibfnamefont {N.~H.}\ \bibnamefont {Dekker}}, \ and\ \bibinfo {author} {\bibfnamefont {C.}~\bibnamefont {Dekker}},\ }\href@noop {} {\bibfield  {journal} {\bibinfo  {journal} {Proc. Natl. Acad. Sci. U. S. A.}\ }\textbf {\bibinfo {volume} {105}},\ \bibinfo {pages} {417} (\bibinfo {year} {2008})}\BibitemShut {NoStop}%
\bibitem [{\citenamefont {Secchi}\ \emph {et~al.}(2016)\citenamefont {Secchi}, \citenamefont {Nigu\`es}, \citenamefont {Jubin}, \citenamefont {Siria},\ and\ \citenamefont {Bocquet}}]{secchi_scaling_2016}%
  \BibitemOpen
  \bibfield  {author} {\bibinfo {author} {\bibfnamefont {E.}~\bibnamefont {Secchi}}, \bibinfo {author} {\bibfnamefont {A.}~\bibnamefont {Nigu\`es}}, \bibinfo {author} {\bibfnamefont {L.}~\bibnamefont {Jubin}}, \bibinfo {author} {\bibfnamefont {A.}~\bibnamefont {Siria}}, \ and\ \bibinfo {author} {\bibfnamefont {L.}~\bibnamefont {Bocquet}},\ }\href {\doibase 10.1103/PhysRevLett.116.154501} {\bibfield  {journal} {\bibinfo  {journal} {Phys. Rev. Lett.}\ }\textbf {\bibinfo {volume} {116}},\ \bibinfo {pages} {154501} (\bibinfo {year} {2016})}\BibitemShut {NoStop}%
\bibitem [{\citenamefont {Knowles}, \citenamefont {Keyser},\ and\ \citenamefont {Thorneywork}(2019)}]{knowles_2020_noise}%
  \BibitemOpen
  \bibfield  {author} {\bibinfo {author} {\bibfnamefont {S.~F.}\ \bibnamefont {Knowles}}, \bibinfo {author} {\bibfnamefont {U.~F.}\ \bibnamefont {Keyser}}, \ and\ \bibinfo {author} {\bibfnamefont {A.~L.}\ \bibnamefont {Thorneywork}},\ }\href {\doibase 10.1088/1361-6528/ab5be3} {\bibfield  {journal} {\bibinfo  {journal} {Nanotechnology}\ }\textbf {\bibinfo {volume} {31}},\ \bibinfo {pages} {10LT01} (\bibinfo {year} {2019})}\BibitemShut {NoStop}%
\bibitem [{\citenamefont {Gravelle}, \citenamefont {Netz},\ and\ \citenamefont {Bocquet}(2019)}]{gravelle_adsorption_2019}%
  \BibitemOpen
  \bibfield  {author} {\bibinfo {author} {\bibfnamefont {S.}~\bibnamefont {Gravelle}}, \bibinfo {author} {\bibfnamefont {R.~R.}\ \bibnamefont {Netz}}, \ and\ \bibinfo {author} {\bibfnamefont {L.}~\bibnamefont {Bocquet}},\ }\href {\doibase 10.1021/acs.nanolett.9b02858} {\bibfield  {journal} {\bibinfo  {journal} {Nano Lett.}\ }\textbf {\bibinfo {volume} {19}},\ \bibinfo {pages} {7265} (\bibinfo {year} {2019})}\BibitemShut {NoStop}%
\bibitem [{\citenamefont {Marbach}(2021)}]{marbach_intrinsic_2021}%
  \BibitemOpen
  \bibfield  {author} {\bibinfo {author} {\bibfnamefont {S.}~\bibnamefont {Marbach}},\ }\href {\doibase 10.1063/5.0047380} {\bibfield  {journal} {\bibinfo  {journal} {J. Chem. Phys.}\ }\textbf {\bibinfo {volume} {154}},\ \bibinfo {pages} {171101} (\bibinfo {year} {2021})}\BibitemShut {NoStop}%
\bibitem [{\citenamefont {Robin}\ \emph {et~al.}(2023)\citenamefont {Robin}, \citenamefont {Lizée}, \citenamefont {Yang}, \citenamefont {Emmerich}, \citenamefont {Siria},\ and\ \citenamefont {Bocquet}}]{robin_disentangling_1_f_2023}%
  \BibitemOpen
  \bibfield  {author} {\bibinfo {author} {\bibfnamefont {P.}~\bibnamefont {Robin}}, \bibinfo {author} {\bibfnamefont {M.}~\bibnamefont {Lizée}}, \bibinfo {author} {\bibfnamefont {Q.}~\bibnamefont {Yang}}, \bibinfo {author} {\bibfnamefont {T.}~\bibnamefont {Emmerich}}, \bibinfo {author} {\bibfnamefont {A.}~\bibnamefont {Siria}}, \ and\ \bibinfo {author} {\bibfnamefont {L.}~\bibnamefont {Bocquet}},\ }\href {\doibase 10.1039/D3FD00035D} {\bibfield  {journal} {\bibinfo  {journal} {Faraday Discuss.}\ }\textbf {\bibinfo {volume} {246}},\ \bibinfo {pages} {556} (\bibinfo {year} {2023})}\BibitemShut {NoStop}%
\bibitem [{\citenamefont {Hansen}\ and\ \citenamefont {McDonald}(2013)}]{hansen_mcdonald_theory_of_simple_liquids_book}%
  \BibitemOpen
  \bibfield  {author} {\bibinfo {author} {\bibfnamefont {J.~P.}\ \bibnamefont {Hansen}}\ and\ \bibinfo {author} {\bibfnamefont {I.~R.}\ \bibnamefont {McDonald}},\ }\href@noop {} {\emph {\bibinfo {title} {Theory of {Simple} {Liquids}}}},\ \bibinfo {edition} {4th}\ ed.\ (\bibinfo  {publisher} {Elsevier},\ \bibinfo {address} {Amsterdam},\ \bibinfo {year} {2013})\BibitemShut {NoStop}%
\bibitem [{\citenamefont {Yamaguchi}, \citenamefont {Matsuoka},\ and\ \citenamefont {Koda}(2007{\natexlab{a}})}]{yamaguchi_theoretical_frequency_dependent_conductivity_2007}%
  \BibitemOpen
  \bibfield  {author} {\bibinfo {author} {\bibfnamefont {T.}~\bibnamefont {Yamaguchi}}, \bibinfo {author} {\bibfnamefont {T.}~\bibnamefont {Matsuoka}}, \ and\ \bibinfo {author} {\bibfnamefont {S.}~\bibnamefont {Koda}},\ }\href {\doibase 10.1063/1.2806289} {\bibfield  {journal} {\bibinfo  {journal} {J. Chem. Phys.}\ }\textbf {\bibinfo {volume} {127}},\ \bibinfo {pages} {234501} (\bibinfo {year} {2007}{\natexlab{a}})}\BibitemShut {NoStop}%
\bibitem [{\citenamefont {Hoang Ngoc~Minh}\ \emph {et~al.}(2023)\citenamefont {Hoang Ngoc~Minh}, \citenamefont {Kim}, \citenamefont {Pireddu}, \citenamefont {Chubak}, \citenamefont {Nair},\ and\ \citenamefont {Rotenberg}}]{hoang_electrical_noise_2023}%
  \BibitemOpen
  \bibfield  {author} {\bibinfo {author} {\bibfnamefont {T.}~\bibnamefont {Hoang Ngoc~Minh}}, \bibinfo {author} {\bibfnamefont {J.}~\bibnamefont {Kim}}, \bibinfo {author} {\bibfnamefont {G.}~\bibnamefont {Pireddu}}, \bibinfo {author} {\bibfnamefont {I.}~\bibnamefont {Chubak}}, \bibinfo {author} {\bibfnamefont {S.}~\bibnamefont {Nair}}, \ and\ \bibinfo {author} {\bibfnamefont {B.}~\bibnamefont {Rotenberg}},\ }\href {\doibase 10.1039/D3FD00026E} {\bibfield  {journal} {\bibinfo  {journal} {Faraday Discuss.}\ }\textbf {\bibinfo {volume} {246}},\ \bibinfo {pages} {198} (\bibinfo {year} {2023})}\BibitemShut {NoStop}%
\bibitem [{\citenamefont {Hoang Ngoc~Minh}, \citenamefont {Stoltz},\ and\ \citenamefont {Rotenberg}(2023)}]{hoang_frequency_field_2023}%
  \BibitemOpen
  \bibfield  {author} {\bibinfo {author} {\bibfnamefont {T.}~\bibnamefont {Hoang Ngoc~Minh}}, \bibinfo {author} {\bibfnamefont {G.}~\bibnamefont {Stoltz}}, \ and\ \bibinfo {author} {\bibfnamefont {B.}~\bibnamefont {Rotenberg}},\ }\href {\doibase 10.1063/5.0139258} {\bibfield  {journal} {\bibinfo  {journal} {J. Chem. Phys.}\ }\textbf {\bibinfo {volume} {158}},\ \bibinfo {pages} {104103} (\bibinfo {year} {2023})}\BibitemShut {NoStop}%
\bibitem [{\citenamefont {Yamaguchi}, \citenamefont {Matsuoka},\ and\ \citenamefont {Koda}(2007{\natexlab{b}})}]{yamaguchi_theoretical_2007}%
  \BibitemOpen
  \bibfield  {author} {\bibinfo {author} {\bibfnamefont {T.}~\bibnamefont {Yamaguchi}}, \bibinfo {author} {\bibfnamefont {T.}~\bibnamefont {Matsuoka}}, \ and\ \bibinfo {author} {\bibfnamefont {S.}~\bibnamefont {Koda}},\ }\href {\doibase doi:10.1063/1.2722261} {\bibfield  {journal} {\bibinfo  {journal} {J. Chem. Phys.}\ }\textbf {\bibinfo {volume} {126}},\ \bibinfo {pages} {144505} (\bibinfo {year} {2007}{\natexlab{b}})}\BibitemShut {NoStop}%
\bibitem [{\citenamefont {Sedlmeier}\ \emph {et~al.}(2014)\citenamefont {Sedlmeier}, \citenamefont {Shadkhoo}, \citenamefont {Bruinsma},\ and\ \citenamefont {Netz}}]{sedlmeier_2014_charge_mass}%
  \BibitemOpen
  \bibfield  {author} {\bibinfo {author} {\bibfnamefont {F.}~\bibnamefont {Sedlmeier}}, \bibinfo {author} {\bibfnamefont {S.}~\bibnamefont {Shadkhoo}}, \bibinfo {author} {\bibfnamefont {R.}~\bibnamefont {Bruinsma}}, \ and\ \bibinfo {author} {\bibfnamefont {R.~R.}\ \bibnamefont {Netz}},\ }\href {\doibase 10.1063/1.4863444} {\bibfield  {journal} {\bibinfo  {journal} {J. Chem. Phys.}\ }\textbf {\bibinfo {volume} {140}},\ \bibinfo {pages} {054512} (\bibinfo {year} {2014})}\BibitemShut {NoStop}%
\bibitem [{\citenamefont {Debye}\ and\ \citenamefont {H{\"u}ckel}(1923)}]{DebyeHuckel1923}%
  \BibitemOpen
  \bibfield  {author} {\bibinfo {author} {\bibfnamefont {P.}~\bibnamefont {Debye}}\ and\ \bibinfo {author} {\bibfnamefont {E.}~\bibnamefont {H{\"u}ckel}},\ }\href@noop {} {\bibfield  {journal} {\bibinfo  {journal} {Physikalische Zeitschrift}\ }\textbf {\bibinfo {volume} {24}},\ \bibinfo {pages} {185} (\bibinfo {year} {1923})},\ \bibinfo {note} {english translation: \emph{The theory of electrolytes. I. Freezing point depression and related phenomena}.}\BibitemShut {Stop}%
\bibitem [{\citenamefont {Onsager}(1927)}]{onsager_report_1927}%
  \BibitemOpen
  \bibfield  {author} {\bibinfo {author} {\bibfnamefont {L.}~\bibnamefont {Onsager}},\ }\href {\doibase 10.1039/TF9272300341} {\bibfield  {journal} {\bibinfo  {journal} {Trans. Faraday Soc.}\ }\textbf {\bibinfo {volume} {23}},\ \bibinfo {pages} {341} (\bibinfo {year} {1927})}\BibitemShut {NoStop}%
\bibitem [{\citenamefont {Onsager}(1933)}]{onsager_theories_1933}%
  \BibitemOpen
  \bibfield  {author} {\bibinfo {author} {\bibfnamefont {L.}~\bibnamefont {Onsager}},\ }\href {\doibase 10.1021/cr60044a006} {\bibfield  {journal} {\bibinfo  {journal} {Chem. Rev.}\ }\textbf {\bibinfo {volume} {13}},\ \bibinfo {pages} {73} (\bibinfo {year} {1933})}\BibitemShut {NoStop}%
\bibitem [{\citenamefont {Onsager}\ and\ \citenamefont {Kim}(1957{\natexlab{a}})}]{onsager_relaxatio_1957}%
  \BibitemOpen
  \bibfield  {author} {\bibinfo {author} {\bibfnamefont {L.}~\bibnamefont {Onsager}}\ and\ \bibinfo {author} {\bibfnamefont {S.~K.}\ \bibnamefont {Kim}},\ }\href {\doibase 10.1021/j150548a016} {\bibfield  {journal} {\bibinfo  {journal} {J. Phys. Chem.}\ }\textbf {\bibinfo {volume} {61}},\ \bibinfo {pages} {215} (\bibinfo {year} {1957}{\natexlab{a}})}\BibitemShut {NoStop}%
\bibitem [{\citenamefont {Bernard}\ \emph {et~al.}(1992{\natexlab{a}})\citenamefont {Bernard}, \citenamefont {Kunz}, \citenamefont {Turq},\ and\ \citenamefont {Blum}}]{bernard_conductance_1992}%
  \BibitemOpen
  \bibfield  {author} {\bibinfo {author} {\bibfnamefont {O.}~\bibnamefont {Bernard}}, \bibinfo {author} {\bibfnamefont {W.}~\bibnamefont {Kunz}}, \bibinfo {author} {\bibfnamefont {P.}~\bibnamefont {Turq}}, \ and\ \bibinfo {author} {\bibfnamefont {L.}~\bibnamefont {Blum}},\ }\href {\doibase 10.1021/j100188a049} {\bibfield  {journal} {\bibinfo  {journal} {J. Phys. Chem.}\ }\textbf {\bibinfo {volume} {96}},\ \bibinfo {pages} {3833} (\bibinfo {year} {1992}{\natexlab{a}})}\BibitemShut {NoStop}%
\bibitem [{\citenamefont {Bernard}\ \emph {et~al.}(1992{\natexlab{b}})\citenamefont {Bernard}, \citenamefont {Kunz}, \citenamefont {Turq},\ and\ \citenamefont {Blum}}]{bernard_self_diffusion_1992}%
  \BibitemOpen
  \bibfield  {author} {\bibinfo {author} {\bibfnamefont {O.}~\bibnamefont {Bernard}}, \bibinfo {author} {\bibfnamefont {W.}~\bibnamefont {Kunz}}, \bibinfo {author} {\bibfnamefont {P.}~\bibnamefont {Turq}}, \ and\ \bibinfo {author} {\bibfnamefont {L.}~\bibnamefont {Blum}},\ }\href {\doibase 10.1021/j100180a074} {\bibfield  {journal} {\bibinfo  {journal} {J. Phys. Chem.}\ }\textbf {\bibinfo {volume} {96}},\ \bibinfo {pages} {398} (\bibinfo {year} {1992}{\natexlab{b}})}\BibitemShut {NoStop}%
\bibitem [{\citenamefont {Bernard}\ and\ \citenamefont {Blum}(1996)}]{bernard_binding_msa_1996}%
  \BibitemOpen
  \bibfield  {author} {\bibinfo {author} {\bibfnamefont {O.}~\bibnamefont {Bernard}}\ and\ \bibinfo {author} {\bibfnamefont {L.}~\bibnamefont {Blum}},\ }\href {\doibase 10.1063/1.471168} {\bibfield  {journal} {\bibinfo  {journal} {J. Chem. Phys.}\ }\textbf {\bibinfo {volume} {104}},\ \bibinfo {pages} {4746} (\bibinfo {year} {1996})}\BibitemShut {NoStop}%
\bibitem [{\citenamefont {Dufrêche}, \citenamefont {Bernard},\ and\ \citenamefont {Turq}(2005)}]{dufreche_2005_transport}%
  \BibitemOpen
  \bibfield  {author} {\bibinfo {author} {\bibfnamefont {J.-F.}\ \bibnamefont {Dufrêche}}, \bibinfo {author} {\bibfnamefont {O.}~\bibnamefont {Bernard}}, \ and\ \bibinfo {author} {\bibfnamefont {P.}~\bibnamefont {Turq}},\ }\href {\doibase https://doi.org/10.1016/j.molliq.2004.07.036} {\bibfield  {journal} {\bibinfo  {journal} {J. Mol. Liq.}\ }\textbf {\bibinfo {volume} {118}},\ \bibinfo {pages} {189} (\bibinfo {year} {2005})},\ \bibinfo {note} {contributions to the 28th International Conference on Solution Chemistry}\BibitemShut {NoStop}%
\bibitem [{\citenamefont {Dufr\^eche}\ \emph {et~al.}(2002)\citenamefont {Dufr\^eche}, \citenamefont {Bernard}, \citenamefont {Turq}, \citenamefont {Mukherjee},\ and\ \citenamefont {Bagchi}}]{dufreche_2002_ionic_self}%
  \BibitemOpen
  \bibfield  {author} {\bibinfo {author} {\bibfnamefont {J.-F.}\ \bibnamefont {Dufr\^eche}}, \bibinfo {author} {\bibfnamefont {O.}~\bibnamefont {Bernard}}, \bibinfo {author} {\bibfnamefont {P.}~\bibnamefont {Turq}}, \bibinfo {author} {\bibfnamefont {A.}~\bibnamefont {Mukherjee}}, \ and\ \bibinfo {author} {\bibfnamefont {B.}~\bibnamefont {Bagchi}},\ }\href {\doibase 10.1103/PhysRevLett.88.095902} {\bibfield  {journal} {\bibinfo  {journal} {Phys. Rev. Lett.}\ }\textbf {\bibinfo {volume} {88}},\ \bibinfo {pages} {095902} (\bibinfo {year} {2002})}\BibitemShut {NoStop}%
\bibitem [{\citenamefont {Dufrêche}\ \emph {et~al.}(2008)\citenamefont {Dufrêche}, \citenamefont {Jardat}, \citenamefont {Turq},\ and\ \citenamefont {Bagchi}}]{dufreche_2008_electrostatic_relaxation}%
  \BibitemOpen
  \bibfield  {author} {\bibinfo {author} {\bibfnamefont {J.-F.}\ \bibnamefont {Dufrêche}}, \bibinfo {author} {\bibfnamefont {M.}~\bibnamefont {Jardat}}, \bibinfo {author} {\bibfnamefont {P.}~\bibnamefont {Turq}}, \ and\ \bibinfo {author} {\bibfnamefont {B.}~\bibnamefont {Bagchi}},\ }\href {\doibase 10.1021/jp801796g} {\bibfield  {journal} {\bibinfo  {journal} {J. Phys. Chem. B}\ }\textbf {\bibinfo {volume} {112}},\ \bibinfo {pages} {10264} (\bibinfo {year} {2008})}\BibitemShut {NoStop}%
\bibitem [{\citenamefont {Chandra}, \citenamefont {Wei},\ and\ \citenamefont {Patey}(1993)}]{chandra_frequency_1993}%
  \BibitemOpen
  \bibfield  {author} {\bibinfo {author} {\bibfnamefont {A.}~\bibnamefont {Chandra}}, \bibinfo {author} {\bibfnamefont {D.}~\bibnamefont {Wei}}, \ and\ \bibinfo {author} {\bibfnamefont {G.~N.}\ \bibnamefont {Patey}},\ }\href {\doibase 10.1063/1.465274} {\bibfield  {journal} {\bibinfo  {journal} {J. Chem. Phys.}\ }\textbf {\bibinfo {volume} {99}},\ \bibinfo {pages} {2083} (\bibinfo {year} {1993})}\BibitemShut {NoStop}%
\bibitem [{\citenamefont {Chandra}\ and\ \citenamefont {Bagchi}(2000)}]{chandra_frequency_2000}%
  \BibitemOpen
  \bibfield  {author} {\bibinfo {author} {\bibfnamefont {A.}~\bibnamefont {Chandra}}\ and\ \bibinfo {author} {\bibfnamefont {B.}~\bibnamefont {Bagchi}},\ }\href {\doibase 10.1063/1.480751} {\bibfield  {journal} {\bibinfo  {journal} {J. Chem. Phys.}\ }\textbf {\bibinfo {volume} {112}},\ \bibinfo {pages} {1876} (\bibinfo {year} {2000})}\BibitemShut {NoStop}%
\bibitem [{\citenamefont {Banerjee}\ and\ \citenamefont {Bagchi}(2019)}]{banerjee_ions_2019}%
  \BibitemOpen
  \bibfield  {author} {\bibinfo {author} {\bibfnamefont {P.}~\bibnamefont {Banerjee}}\ and\ \bibinfo {author} {\bibfnamefont {B.}~\bibnamefont {Bagchi}},\ }\href {\doibase 10.1063/1.5090765} {\bibfield  {journal} {\bibinfo  {journal} {J. Chem. Phys.}\ }\textbf {\bibinfo {volume} {150}},\ \bibinfo {pages} {190901} (\bibinfo {year} {2019})}\BibitemShut {NoStop}%
\bibitem [{\citenamefont {Jardat}, \citenamefont {Hribar-Lee},\ and\ \citenamefont {Vlachy}(2012)}]{jardat_self-diffusion_2012}%
  \BibitemOpen
  \bibfield  {author} {\bibinfo {author} {\bibfnamefont {M.}~\bibnamefont {Jardat}}, \bibinfo {author} {\bibfnamefont {B.}~\bibnamefont {Hribar-Lee}}, \ and\ \bibinfo {author} {\bibfnamefont {V.}~\bibnamefont {Vlachy}},\ }\href {\doibase 10.1039/C1SM05985H} {\bibfield  {journal} {\bibinfo  {journal} {Soft Matter}\ }\textbf {\bibinfo {volume} {8}},\ \bibinfo {pages} {954} (\bibinfo {year} {2012})}\BibitemShut {NoStop}%
\bibitem [{\citenamefont {Yamaguchi}, \citenamefont {Matsuoka},\ and\ \citenamefont {Koda}(2011)}]{yamaguchi_dynamic_2011}%
  \BibitemOpen
  \bibfield  {author} {\bibinfo {author} {\bibfnamefont {T.}~\bibnamefont {Yamaguchi}}, \bibinfo {author} {\bibfnamefont {T.}~\bibnamefont {Matsuoka}}, \ and\ \bibinfo {author} {\bibfnamefont {S.}~\bibnamefont {Koda}},\ }\href {\doibase 10.1063/1.3657401} {\bibfield  {journal} {\bibinfo  {journal} {J. Chem. Phys.}\ }\textbf {\bibinfo {volume} {135}},\ \bibinfo {pages} {164511} (\bibinfo {year} {2011})}\BibitemShut {NoStop}%
\bibitem [{\citenamefont {te~Vrugt}, \citenamefont {Löwen},\ and\ \citenamefont {Wittkowski}(0020)}]{te_vrugt_classical_2020}%
  \BibitemOpen
  \bibfield  {author} {\bibinfo {author} {\bibfnamefont {M.}~\bibnamefont {te~Vrugt}}, \bibinfo {author} {\bibfnamefont {H.}~\bibnamefont {Löwen}}, \ and\ \bibinfo {author} {\bibfnamefont {R.}~\bibnamefont {Wittkowski}},\ }\href {\doibase 10.1080/00018732.2020.1854965} {\bibfield  {journal} {\bibinfo  {journal} {Adv. Phys.}\ }\textbf {\bibinfo {volume} {69}},\ \bibinfo {pages} {121} (\bibinfo {year} {20020})}\BibitemShut {NoStop}%
\bibitem [{\citenamefont {Bazant}, \citenamefont {Thornton},\ and\ \citenamefont {Ajdari}(2004)}]{bazant_diffuse-charge_2004}%
  \BibitemOpen
  \bibfield  {author} {\bibinfo {author} {\bibfnamefont {M.~Z.}\ \bibnamefont {Bazant}}, \bibinfo {author} {\bibfnamefont {K.}~\bibnamefont {Thornton}}, \ and\ \bibinfo {author} {\bibfnamefont {A.}~\bibnamefont {Ajdari}},\ }\href {\doibase 10.1103/PhysRevE.70.021506} {\bibfield  {journal} {\bibinfo  {journal} {Phys. Rev. E}\ }\textbf {\bibinfo {volume} {70}},\ \bibinfo {pages} {021506} (\bibinfo {year} {2004})}\BibitemShut {NoStop}%
\bibitem [{\citenamefont {Donev}\ \emph {et~al.}(2019)\citenamefont {Donev}, \citenamefont {Garcia}, \citenamefont {Péraud}, \citenamefont {Nonaka},\ and\ \citenamefont {Bell}}]{donev_fluctuating_hydro_2019}%
  \BibitemOpen
  \bibfield  {author} {\bibinfo {author} {\bibfnamefont {A.}~\bibnamefont {Donev}}, \bibinfo {author} {\bibfnamefont {A.~L.}\ \bibnamefont {Garcia}}, \bibinfo {author} {\bibfnamefont {J.-P.}\ \bibnamefont {Péraud}}, \bibinfo {author} {\bibfnamefont {A.~J.}\ \bibnamefont {Nonaka}}, \ and\ \bibinfo {author} {\bibfnamefont {J.~B.}\ \bibnamefont {Bell}},\ }\href {\doibase 10.1016/j.coelec.2018.09.004} {\bibfield  {journal} {\bibinfo  {journal} {Curr. Opin. Electrochem.}\ }\textbf {\bibinfo {volume} {13}},\ \bibinfo {pages} {1 } (\bibinfo {year} {2019})}\BibitemShut {NoStop}%
\bibitem [{\citenamefont {Péraud}\ \emph {et~al.}(0 10)\citenamefont {Péraud}, \citenamefont {Nonaka}, \citenamefont {Bell}, \citenamefont {Donev},\ and\ \citenamefont {Garcia}}]{peraud_fluctuating_2017}%
  \BibitemOpen
  \bibfield  {author} {\bibinfo {author} {\bibfnamefont {J.-P.}\ \bibnamefont {Péraud}}, \bibinfo {author} {\bibfnamefont {A.~J.}\ \bibnamefont {Nonaka}}, \bibinfo {author} {\bibfnamefont {J.~B.}\ \bibnamefont {Bell}}, \bibinfo {author} {\bibfnamefont {A.}~\bibnamefont {Donev}}, \ and\ \bibinfo {author} {\bibfnamefont {A.~L.}\ \bibnamefont {Garcia}},\ }\href {\doibase 10.1073/pnas.1714464114} {\bibfield  {journal} {\bibinfo  {journal} {Proc. Natl. Acad. Sci. U.S.A.}\ }\textbf {\bibinfo {volume} {114}},\ \bibinfo {pages} {10829} (\bibinfo {year} {2017-10-10})}\BibitemShut {NoStop}%
\bibitem [{\citenamefont {Mahdisoltani}\ and\ \citenamefont {Golestanian}(2021{\natexlab{a}})}]{mahdisoltani_transient_2021}%
  \BibitemOpen
  \bibfield  {author} {\bibinfo {author} {\bibfnamefont {S.}~\bibnamefont {Mahdisoltani}}\ and\ \bibinfo {author} {\bibfnamefont {R.}~\bibnamefont {Golestanian}},\ }\href {\doibase 10.1088/1367-2630/ac0f1a} {\bibfield  {journal} {\bibinfo  {journal} {New J. Phys.}\ }\textbf {\bibinfo {volume} {23}},\ \bibinfo {pages} {073034} (\bibinfo {year} {2021}{\natexlab{a}})}\BibitemShut {NoStop}%
\bibitem [{\citenamefont {Illien}, \citenamefont {Carof},\ and\ \citenamefont {Rotenberg}(2024{\natexlab{a}})}]{illien_2024_stochastic}%
  \BibitemOpen
  \bibfield  {author} {\bibinfo {author} {\bibfnamefont {P.}~\bibnamefont {Illien}}, \bibinfo {author} {\bibfnamefont {A.}~\bibnamefont {Carof}}, \ and\ \bibinfo {author} {\bibfnamefont {B.}~\bibnamefont {Rotenberg}},\ }\href {\doibase 10.1103/PhysRevLett.133.268002} {\bibfield  {journal} {\bibinfo  {journal} {Phys. Rev. Lett.}\ }\textbf {\bibinfo {volume} {133}},\ \bibinfo {pages} {268002} (\bibinfo {year} {2024}{\natexlab{a}})}\BibitemShut {NoStop}%
\bibitem [{\citenamefont {D{\'{e}}mery}\ and\ \citenamefont {Dean}(2016)}]{demery_conductivity_2016}%
  \BibitemOpen
  \bibfield  {author} {\bibinfo {author} {\bibfnamefont {V.}~\bibnamefont {D{\'{e}}mery}}\ and\ \bibinfo {author} {\bibfnamefont {D.~S.}\ \bibnamefont {Dean}},\ }\href {\doibase 10.1088/1742-5468/2016/02/023106} {\bibfield  {journal} {\bibinfo  {journal} {J. Stat. Mech.}\ }\textbf {\bibinfo {volume} {2016}},\ \bibinfo {pages} {023106} (\bibinfo {year} {2016})}\BibitemShut {NoStop}%
\bibitem [{\citenamefont {Bonneau}, \citenamefont {Démery},\ and\ \citenamefont {Raphaël}(2023)}]{bonneau_2023_temporal}%
  \BibitemOpen
  \bibfield  {author} {\bibinfo {author} {\bibfnamefont {H.}~\bibnamefont {Bonneau}}, \bibinfo {author} {\bibfnamefont {V.}~\bibnamefont {Démery}}, \ and\ \bibinfo {author} {\bibfnamefont {E.}~\bibnamefont {Raphaël}},\ }\href {\doibase 10.1088/1742-5468/acdced} {\bibfield  {journal} {\bibinfo  {journal} {J. Stat. Mech.:Theory Exp.}\ }\textbf {\bibinfo {volume} {2023}},\ \bibinfo {pages} {073205} (\bibinfo {year} {2023})}\BibitemShut {NoStop}%
\bibitem [{\citenamefont {Bonneau}\ \emph {et~al.}(2024)\citenamefont {Bonneau}, \citenamefont {Avni}, \citenamefont {Andelman},\ and\ \citenamefont {Orland}}]{bonneau_2024_frequency}%
  \BibitemOpen
  \bibfield  {author} {\bibinfo {author} {\bibfnamefont {H.}~\bibnamefont {Bonneau}}, \bibinfo {author} {\bibfnamefont {Y.}~\bibnamefont {Avni}}, \bibinfo {author} {\bibfnamefont {D.}~\bibnamefont {Andelman}}, \ and\ \bibinfo {author} {\bibfnamefont {H.}~\bibnamefont {Orland}},\ }\href {\doibase 10.1063/5.0236073} {\bibfield  {journal} {\bibinfo  {journal} {J. Chem. Phys.}\ }\textbf {\bibinfo {volume} {161}},\ \bibinfo {pages} {244501} (\bibinfo {year} {2024})}\BibitemShut {NoStop}%
\bibitem [{\citenamefont {Bonneau}, \citenamefont {D{\'e}mery},\ and\ \citenamefont {Rapha{\"e}l}(2025)}]{bonneau_2025_stationary}%
  \BibitemOpen
  \bibfield  {author} {\bibinfo {author} {\bibfnamefont {H.}~\bibnamefont {Bonneau}}, \bibinfo {author} {\bibfnamefont {V.}~\bibnamefont {D{\'e}mery}}, \ and\ \bibinfo {author} {\bibfnamefont {E.}~\bibnamefont {Rapha{\"e}l}},\ }\href@noop {} {\bibfield  {journal} {\bibinfo  {journal} {J. Stat. Mech.}\ }\textbf {\bibinfo {volume} {2025}},\ \bibinfo {pages} {033201} (\bibinfo {year} {2025})}\BibitemShut {NoStop}%
\bibitem [{\citenamefont {Poncet}\ \emph {et~al.}(2017)\citenamefont {Poncet}, \citenamefont {B\'enichou}, \citenamefont {D\'emery},\ and\ \citenamefont {Oshanin}}]{poncet_2017_universal}%
  \BibitemOpen
  \bibfield  {author} {\bibinfo {author} {\bibfnamefont {A.}~\bibnamefont {Poncet}}, \bibinfo {author} {\bibfnamefont {O.}~\bibnamefont {B\'enichou}}, \bibinfo {author} {\bibfnamefont {V.}~\bibnamefont {D\'emery}}, \ and\ \bibinfo {author} {\bibfnamefont {G.}~\bibnamefont {Oshanin}},\ }\href {\doibase 10.1103/PhysRevLett.118.118002} {\bibfield  {journal} {\bibinfo  {journal} {Phys. Rev. Lett.}\ }\textbf {\bibinfo {volume} {118}},\ \bibinfo {pages} {118002} (\bibinfo {year} {2017})}\BibitemShut {NoStop}%
\bibitem [{\citenamefont {Berthoumieux}, \citenamefont {Démery},\ and\ \citenamefont {Maggs}(2024)}]{berthoumieux_2024_nonlinear_conductivity}%
  \BibitemOpen
  \bibfield  {author} {\bibinfo {author} {\bibfnamefont {H.}~\bibnamefont {Berthoumieux}}, \bibinfo {author} {\bibfnamefont {V.}~\bibnamefont {Démery}}, \ and\ \bibinfo {author} {\bibfnamefont {A.~C.}\ \bibnamefont {Maggs}},\ }\href {\doibase 10.1063/5.0226773} {\bibfield  {journal} {\bibinfo  {journal} {J. Chem. Phys.}\ }\textbf {\bibinfo {volume} {161}},\ \bibinfo {pages} {184504} (\bibinfo {year} {2024})}\BibitemShut {NoStop}%
\bibitem [{\citenamefont {Bernard}\ \emph {et~al.}(2023)\citenamefont {Bernard}, \citenamefont {Jardat}, \citenamefont {Rotenberg},\ and\ \citenamefont {Illien}}]{bernard_2023_analytical}%
  \BibitemOpen
  \bibfield  {author} {\bibinfo {author} {\bibfnamefont {O.}~\bibnamefont {Bernard}}, \bibinfo {author} {\bibfnamefont {M.}~\bibnamefont {Jardat}}, \bibinfo {author} {\bibfnamefont {B.}~\bibnamefont {Rotenberg}}, \ and\ \bibinfo {author} {\bibfnamefont {P.}~\bibnamefont {Illien}},\ }\href {\doibase 10.1063/5.0165533} {\bibfield  {journal} {\bibinfo  {journal} {J. Chem. Phys.}\ }\textbf {\bibinfo {volume} {159}},\ \bibinfo {pages} {164105} (\bibinfo {year} {2023})}\BibitemShut {NoStop}%
\bibitem [{\citenamefont {Zorkot}, \citenamefont {Golestanian},\ and\ \citenamefont {Bonthuis}(2016)}]{zorkot_current_2016}%
  \BibitemOpen
  \bibfield  {author} {\bibinfo {author} {\bibfnamefont {M.}~\bibnamefont {Zorkot}}, \bibinfo {author} {\bibfnamefont {R.}~\bibnamefont {Golestanian}}, \ and\ \bibinfo {author} {\bibfnamefont {D.~J.}\ \bibnamefont {Bonthuis}},\ }\href {\doibase 10.1140/epjst/e2016-60152-y} {\bibfield  {journal} {\bibinfo  {journal} {Eur. Phys. J. Spec. Top.}\ }\textbf {\bibinfo {volume} {225}},\ \bibinfo {pages} {1583} (\bibinfo {year} {2016})}\BibitemShut {NoStop}%
\bibitem [{\citenamefont {Zorkot}\ and\ \citenamefont {Golestanian}(8 03)}]{zorkot_current_sPNP_2018}%
  \BibitemOpen
  \bibfield  {author} {\bibinfo {author} {\bibfnamefont {M.}~\bibnamefont {Zorkot}}\ and\ \bibinfo {author} {\bibfnamefont {R.}~\bibnamefont {Golestanian}},\ }\href {\doibase 10.1088/1361-648X/aab016} {\bibfield  {journal} {\bibinfo  {journal} {J. Phys.: Condens. Matter}\ }\textbf {\bibinfo {volume} {30}},\ \bibinfo {pages} {134001} (\bibinfo {year} {2018-03})}\BibitemShut {NoStop}%
\bibitem [{\citenamefont {Kawasaki}(1994)}]{kawasaki_stochastic_1994}%
  \BibitemOpen
  \bibfield  {author} {\bibinfo {author} {\bibfnamefont {K.}~\bibnamefont {Kawasaki}},\ }\href {\doibase https://doi.org/10.1016/0378-4371(94)90533-9} {\bibfield  {journal} {\bibinfo  {journal} {Physica A}\ }\textbf {\bibinfo {volume} {208}},\ \bibinfo {pages} {35} (\bibinfo {year} {1994})}\BibitemShut {NoStop}%
\bibitem [{\citenamefont {Dean}(1996)}]{dean_langevin_1996}%
  \BibitemOpen
  \bibfield  {author} {\bibinfo {author} {\bibfnamefont {D.~S.}\ \bibnamefont {Dean}},\ }\href {\doibase 10.1088/0305-4470/29/24/001} {\bibfield  {journal} {\bibinfo  {journal} {J. Phys. A:Math. Gen.}\ }\textbf {\bibinfo {volume} {29}},\ \bibinfo {pages} {L613} (\bibinfo {year} {1996})}\BibitemShut {NoStop}%
\bibitem [{\citenamefont {Illien}(2025)}]{illien2025dean}%
  \BibitemOpen
  \bibfield  {author} {\bibinfo {author} {\bibfnamefont {P.}~\bibnamefont {Illien}},\ }\href@noop {} {\bibfield  {journal} {\bibinfo  {journal} {Rep. Prog. Phys.}\ }\textbf {\bibinfo {volume} {88}},\ \bibinfo {pages} {086601} (\bibinfo {year} {2025})}\BibitemShut {NoStop}%
\bibitem [{\citenamefont {Avni}\ \emph {et~al.}(2022)\citenamefont {Avni}, \citenamefont {Adar}, \citenamefont {Andelman},\ and\ \citenamefont {Orland}}]{avni_conductivity_2022}%
  \BibitemOpen
  \bibfield  {author} {\bibinfo {author} {\bibfnamefont {Y.}~\bibnamefont {Avni}}, \bibinfo {author} {\bibfnamefont {R.~M.}\ \bibnamefont {Adar}}, \bibinfo {author} {\bibfnamefont {D.}~\bibnamefont {Andelman}}, \ and\ \bibinfo {author} {\bibfnamefont {H.}~\bibnamefont {Orland}},\ }\href {\doibase 10.1103/PhysRevLett.128.098002} {\bibfield  {journal} {\bibinfo  {journal} {Phys. Rev. Lett.}\ }\textbf {\bibinfo {volume} {128}},\ \bibinfo {pages} {098002} (\bibinfo {year} {2022})}\BibitemShut {NoStop}%
\bibitem [{\citenamefont {Giaquinta}, \citenamefont {Parrinello},\ and\ \citenamefont {Tosi}(1978)}]{giaqinta_1978_generalized_hydro}%
  \BibitemOpen
  \bibfield  {author} {\bibinfo {author} {\bibfnamefont {P.}~\bibnamefont {Giaquinta}}, \bibinfo {author} {\bibfnamefont {M.}~\bibnamefont {Parrinello}}, \ and\ \bibinfo {author} {\bibfnamefont {M.}~\bibnamefont {Tosi}},\ }\href {\doibase https://doi.org/10.1016/0378-4371(78)90027-4} {\bibfield  {journal} {\bibinfo  {journal} {Physica A}\ }\textbf {\bibinfo {volume} {92}},\ \bibinfo {pages} {185} (\bibinfo {year} {1978})}\BibitemShut {NoStop}%
\bibitem [{\citenamefont {Caillol}, \citenamefont {Levesque},\ and\ \citenamefont {Weis}(1986)}]{caillol_theoretical_1986}%
  \BibitemOpen
  \bibfield  {author} {\bibinfo {author} {\bibfnamefont {J.~M.}\ \bibnamefont {Caillol}}, \bibinfo {author} {\bibfnamefont {D.}~\bibnamefont {Levesque}}, \ and\ \bibinfo {author} {\bibfnamefont {J.~J.}\ \bibnamefont {Weis}},\ }\href {\doibase 10.1063/1.451446} {\bibfield  {journal} {\bibinfo  {journal} {J. Chem. Phys.}\ }\textbf {\bibinfo {volume} {85}},\ \bibinfo {pages} {6645} (\bibinfo {year} {1986})}\BibitemShut {NoStop}%
\bibitem [{\citenamefont {Caillol}(1987)}]{caillol_dielectric_1987}%
  \BibitemOpen
  \bibfield  {author} {\bibinfo {author} {\bibfnamefont {J.~M.}\ \bibnamefont {Caillol}},\ }\href {\doibase 10.1209/0295-5075/4/2/006} {\bibfield  {journal} {\bibinfo  {journal} {Europhys. Lett.}\ }\textbf {\bibinfo {volume} {4}},\ \bibinfo {pages} {159} (\bibinfo {year} {1987})}\BibitemShut {NoStop}%
\bibitem [{\citenamefont {Felderhof}(1980)}]{felderhof_fluctuation_1980}%
  \BibitemOpen
  \bibfield  {author} {\bibinfo {author} {\bibfnamefont {B.~U.}\ \bibnamefont {Felderhof}},\ }\href {\doibase 10.1016/0378-4371(80)90114-4} {\bibfield  {journal} {\bibinfo  {journal} {Physica A}\ }\textbf {\bibinfo {volume} {101}},\ \bibinfo {pages} {275} (\bibinfo {year} {1980})}\BibitemShut {NoStop}%
\bibitem [{\citenamefont {Pollock}\ and\ \citenamefont {Alder}(1981)}]{pollock_frequency-dependent_1981}%
  \BibitemOpen
  \bibfield  {author} {\bibinfo {author} {\bibfnamefont {E.~L.}\ \bibnamefont {Pollock}}\ and\ \bibinfo {author} {\bibfnamefont {B.~J.}\ \bibnamefont {Alder}},\ }\href {\doibase 10.1103/PhysRevLett.46.950} {\bibfield  {journal} {\bibinfo  {journal} {Phys. Rev. Lett.}\ }\textbf {\bibinfo {volume} {46}},\ \bibinfo {pages} {950} (\bibinfo {year} {1981})}\BibitemShut {NoStop}%
\bibitem [{\citenamefont {Yamaguchi}, \citenamefont {Akatsuka},\ and\ \citenamefont {Koda}(2011)}]{yamaguchi_brownian_2011}%
  \BibitemOpen
  \bibfield  {author} {\bibinfo {author} {\bibfnamefont {T.}~\bibnamefont {Yamaguchi}}, \bibinfo {author} {\bibfnamefont {T.}~\bibnamefont {Akatsuka}}, \ and\ \bibinfo {author} {\bibfnamefont {S.}~\bibnamefont {Koda}},\ }\href {\doibase 10.1063/1.3604532} {\bibfield  {journal} {\bibinfo  {journal} {J. Chem. Phys.}\ }\textbf {\bibinfo {volume} {134}},\ \bibinfo {pages} {244506} (\bibinfo {year} {2011})}\BibitemShut {NoStop}%
\bibitem [{\citenamefont {Cheng}\ and\ \citenamefont {Frenkel}(2020)}]{cheng_computing_2020}%
  \BibitemOpen
  \bibfield  {author} {\bibinfo {author} {\bibfnamefont {B.}~\bibnamefont {Cheng}}\ and\ \bibinfo {author} {\bibfnamefont {D.}~\bibnamefont {Frenkel}},\ }\href {\doibase 10.1103/PhysRevLett.125.130602} {\bibfield  {journal} {\bibinfo  {journal} {Phys. Rev. Lett.}\ }\textbf {\bibinfo {volume} {125}},\ \bibinfo {pages} {130602} (\bibinfo {year} {2020})}\BibitemShut {NoStop}%
\bibitem [{\citenamefont {Robinson}\ and\ \citenamefont {Stokes}(2002)}]{Robinson1959}%
  \BibitemOpen
  \bibfield  {author} {\bibinfo {author} {\bibfnamefont {R.~A.}\ \bibnamefont {Robinson}}\ and\ \bibinfo {author} {\bibfnamefont {R.~H.}\ \bibnamefont {Stokes}},\ }\href@noop {} {\emph {\bibinfo {title} {{Electrolyte solutions: Second Revised Edition}}}}\ (\bibinfo  {publisher} {Dover Publications},\ \bibinfo {year} {2002})\ pp.\ \bibinfo {pages} {286--292}\BibitemShut {NoStop}%
\bibitem [{\citenamefont {Lelidis}\ and\ \citenamefont {Barbero}(2005)}]{lelidis_effect_2005}%
  \BibitemOpen
  \bibfield  {author} {\bibinfo {author} {\bibfnamefont {I.}~\bibnamefont {Lelidis}}\ and\ \bibinfo {author} {\bibfnamefont {G.}~\bibnamefont {Barbero}},\ }\href {\doibase 10.1016/j.physleta.2005.06.038} {\bibfield  {journal} {\bibinfo  {journal} {Phys. Lett. A}\ }\textbf {\bibinfo {volume} {343}},\ \bibinfo {pages} {440} (\bibinfo {year} {2005})}\BibitemShut {NoStop}%
\bibitem [{\citenamefont {Barbero}, \citenamefont {Batalioto},\ and\ \citenamefont {Figueiredo~Neto}(2008)}]{Barbero2008}%
  \BibitemOpen
  \bibfield  {author} {\bibinfo {author} {\bibfnamefont {G.}~\bibnamefont {Barbero}}, \bibinfo {author} {\bibfnamefont {F.}~\bibnamefont {Batalioto}}, \ and\ \bibinfo {author} {\bibfnamefont {A.~M.}\ \bibnamefont {Figueiredo~Neto}},\ }\href {\doibase 10.1063/1.2908044} {\bibfield  {journal} {\bibinfo  {journal} {Appl. Phys. Lett.}\ }\textbf {\bibinfo {volume} {92}},\ \bibinfo {pages} {172908} (\bibinfo {year} {2008})}\BibitemShut {NoStop}%
\bibitem [{\citenamefont {Antonova}\ \emph {et~al.}(2020)\citenamefont {Antonova}, \citenamefont {Barbero}, \citenamefont {Evangelista},\ and\ \citenamefont {Tilli}}]{Antonova2020}%
  \BibitemOpen
  \bibfield  {author} {\bibinfo {author} {\bibfnamefont {A.}~\bibnamefont {Antonova}}, \bibinfo {author} {\bibfnamefont {G.}~\bibnamefont {Barbero}}, \bibinfo {author} {\bibfnamefont {L.~R.}\ \bibnamefont {Evangelista}}, \ and\ \bibinfo {author} {\bibfnamefont {P.}~\bibnamefont {Tilli}},\ }\href {\doibase 10.1088/1742-5468/ab7a23} {\bibfield  {journal} {\bibinfo  {journal} {J. Stat. Mech.: Theory Exp.}\ }\textbf {\bibinfo {volume} {2020}},\ \bibinfo {pages} {043202} (\bibinfo {year} {2020})}\BibitemShut {NoStop}%
\bibitem [{\citenamefont {Palaia}\ \emph {et~al.}(2025{\natexlab{a}})\citenamefont {Palaia}, \citenamefont {Asta}, \citenamefont {Dutta}, \citenamefont {Warren}, \citenamefont {Rotenberg},\ and\ \citenamefont {Trizac}}]{palaia_charging_2025}%
  \BibitemOpen
  \bibfield  {author} {\bibinfo {author} {\bibfnamefont {I.}~\bibnamefont {Palaia}}, \bibinfo {author} {\bibfnamefont {A.~J.}\ \bibnamefont {Asta}}, \bibinfo {author} {\bibfnamefont {M.}~\bibnamefont {Dutta}}, \bibinfo {author} {\bibfnamefont {P.~B.}\ \bibnamefont {Warren}}, \bibinfo {author} {\bibfnamefont {B.}~\bibnamefont {Rotenberg}}, \ and\ \bibinfo {author} {\bibfnamefont {E.}~\bibnamefont {Trizac}},\ }\href {\doibase 10.1103/72b9-c8cq} {\bibfield  {journal} {\bibinfo  {journal} {Phys. Rev. Lett.}\ }\textbf {\bibinfo {volume} {135}},\ \bibinfo {pages} {148002} (\bibinfo {year} {2025}{\natexlab{a}})}\BibitemShut {NoStop}%
\bibitem [{\citenamefont {Palaia}\ \emph {et~al.}(2025{\natexlab{b}})\citenamefont {Palaia}, \citenamefont {Asta}, \citenamefont {Dutta}, \citenamefont {Warren}, \citenamefont {Rotenberg},\ and\ \citenamefont {Trizac}}]{palaia_poisson-nernst-planck_2025}%
  \BibitemOpen
  \bibfield  {author} {\bibinfo {author} {\bibfnamefont {I.}~\bibnamefont {Palaia}}, \bibinfo {author} {\bibfnamefont {A.~J.}\ \bibnamefont {Asta}}, \bibinfo {author} {\bibfnamefont {M.}~\bibnamefont {Dutta}}, \bibinfo {author} {\bibfnamefont {P.~B.}\ \bibnamefont {Warren}}, \bibinfo {author} {\bibfnamefont {B.}~\bibnamefont {Rotenberg}}, \ and\ \bibinfo {author} {\bibfnamefont {E.}~\bibnamefont {Trizac}},\ }\href {\doibase 10.1103/p4dg-snqf} {\bibfield  {journal} {\bibinfo  {journal} {Physical Review E}\ }\textbf {\bibinfo {volume} {112}},\ \bibinfo {pages} {035417} (\bibinfo {year} {2025}{\natexlab{b}})}\BibitemShut {NoStop}%
\bibitem [{\citenamefont {Mahdisoltani}\ and\ \citenamefont {Golestanian}(2021{\natexlab{b}})}]{mahdisoltani_long_2021}%
  \BibitemOpen
  \bibfield  {author} {\bibinfo {author} {\bibfnamefont {S.}~\bibnamefont {Mahdisoltani}}\ and\ \bibinfo {author} {\bibfnamefont {R.}~\bibnamefont {Golestanian}},\ }\href {\doibase 10.1103/PhysRevLett.126.158002} {\bibfield  {journal} {\bibinfo  {journal} {Phys. Rev. Lett.}\ }\textbf {\bibinfo {volume} {126}},\ \bibinfo {pages} {158002} (\bibinfo {year} {2021}{\natexlab{b}})}\BibitemShut {NoStop}%
\bibitem [{\citenamefont {Du}\ \emph {et~al.}(2024)\citenamefont {Du}, \citenamefont {Dean}, \citenamefont {Miao},\ and\ \citenamefont {Podgornik}}]{du_2024_correlation}%
  \BibitemOpen
  \bibfield  {author} {\bibinfo {author} {\bibfnamefont {G.}~\bibnamefont {Du}}, \bibinfo {author} {\bibfnamefont {D.~S.}\ \bibnamefont {Dean}}, \bibinfo {author} {\bibfnamefont {B.}~\bibnamefont {Miao}}, \ and\ \bibinfo {author} {\bibfnamefont {R.}~\bibnamefont {Podgornik}},\ }\href {\doibase 10.1103/PhysRevLett.133.238002} {\bibfield  {journal} {\bibinfo  {journal} {Phys. Rev. Lett.}\ }\textbf {\bibinfo {volume} {133}},\ \bibinfo {pages} {238002} (\bibinfo {year} {2024})}\BibitemShut {NoStop}%
\bibitem [{\citenamefont {Du}\ \emph {et~al.}(2025)\citenamefont {Du}, \citenamefont {Dean}, \citenamefont {Miao},\ and\ \citenamefont {Podgornik}}]{du_repulsive_2025}%
  \BibitemOpen
  \bibfield  {author} {\bibinfo {author} {\bibfnamefont {G.}~\bibnamefont {Du}}, \bibinfo {author} {\bibfnamefont {D.~S.}\ \bibnamefont {Dean}}, \bibinfo {author} {\bibfnamefont {B.}~\bibnamefont {Miao}}, \ and\ \bibinfo {author} {\bibfnamefont {R.}~\bibnamefont {Podgornik}},\ }\href {\doibase 10.1103/PhysRevE.111.044108} {\bibfield  {journal} {\bibinfo  {journal} {Phys. Rev. E}\ }\textbf {\bibinfo {volume} {111}},\ \bibinfo {pages} {044108} (\bibinfo {year} {2025})}\BibitemShut {NoStop}%
\bibitem [{\citenamefont {Thompson}\ \emph {et~al.}(2022)\citenamefont {Thompson}, \citenamefont {Aktulga}, \citenamefont {Berger}, \citenamefont {Bolintineanu}, \citenamefont {Brown}, \citenamefont {Crozier}, \citenamefont {{in 't Veld}}, \citenamefont {Kohlmeyer}, \citenamefont {Moore}, \citenamefont {Nguyen}, \citenamefont {Shan}, \citenamefont {Stevens}, \citenamefont {Tranchida}, \citenamefont {Trott},\ and\ \citenamefont {Plimpton}}]{thompson_LAMMPS_2022}%
  \BibitemOpen
  \bibfield  {author} {\bibinfo {author} {\bibfnamefont {A.~P.}\ \bibnamefont {Thompson}}, \bibinfo {author} {\bibfnamefont {H.~M.}\ \bibnamefont {Aktulga}}, \bibinfo {author} {\bibfnamefont {R.}~\bibnamefont {Berger}}, \bibinfo {author} {\bibfnamefont {D.~S.}\ \bibnamefont {Bolintineanu}}, \bibinfo {author} {\bibfnamefont {W.~M.}\ \bibnamefont {Brown}}, \bibinfo {author} {\bibfnamefont {P.~S.}\ \bibnamefont {Crozier}}, \bibinfo {author} {\bibfnamefont {P.~J.}\ \bibnamefont {{in 't Veld}}}, \bibinfo {author} {\bibfnamefont {A.}~\bibnamefont {Kohlmeyer}}, \bibinfo {author} {\bibfnamefont {S.~G.}\ \bibnamefont {Moore}}, \bibinfo {author} {\bibfnamefont {T.~D.}\ \bibnamefont {Nguyen}}, \bibinfo {author} {\bibfnamefont {R.}~\bibnamefont {Shan}}, \bibinfo {author} {\bibfnamefont {M.~J.}\ \bibnamefont {Stevens}}, \bibinfo {author} {\bibfnamefont {J.}~\bibnamefont {Tranchida}}, \bibinfo {author} {\bibfnamefont {C.}~\bibnamefont {Trott}}, \ and\ \bibinfo {author} {\bibfnamefont {S.~J.}\ \bibnamefont {Plimpton}},\
  }\href {\doibase https://doi.org/10.1016/j.cpc.2021.108171} {\bibfield  {journal} {\bibinfo  {journal} {Comput. Phys. Commun.}\ }\textbf {\bibinfo {volume} {271}},\ \bibinfo {pages} {108171} (\bibinfo {year} {2022})}\BibitemShut {NoStop}%
\bibitem [{\citenamefont {Leimkuhler}\ and\ \citenamefont {Matthews}(2013)}]{Leimkuhler_BAOAB_2013}%
  \BibitemOpen
  \bibfield  {author} {\bibinfo {author} {\bibfnamefont {B.}~\bibnamefont {Leimkuhler}}\ and\ \bibinfo {author} {\bibfnamefont {C.}~\bibnamefont {Matthews}},\ }\href {\doibase 10.1093/amrx/abs010} {\bibfield  {journal} {\bibinfo  {journal} {Appl. Math. Res. eXpress}\ }\textbf {\bibinfo {volume} {2013}},\ \bibinfo {pages} {34} (\bibinfo {year} {2013})}\BibitemShut {NoStop}%
\bibitem [{\citenamefont {Pollock}\ and\ \citenamefont {Glosli}(1996)}]{pollock1996comments}%
  \BibitemOpen
  \bibfield  {author} {\bibinfo {author} {\bibfnamefont {E.}~\bibnamefont {Pollock}}\ and\ \bibinfo {author} {\bibfnamefont {J.}~\bibnamefont {Glosli}},\ }\href@noop {} {\bibfield  {journal} {\bibinfo  {journal} {Comput. Phys. Commun.}\ }\textbf {\bibinfo {volume} {95}},\ \bibinfo {pages} {93} (\bibinfo {year} {1996})}\BibitemShut {NoStop}%
\bibitem [{\citenamefont {Adar}\ \emph {et~al.}(2019)\citenamefont {Adar}, \citenamefont {Safran}, \citenamefont {Diamant},\ and\ \citenamefont {Andelman}}]{adar_2019_screening}%
  \BibitemOpen
  \bibfield  {author} {\bibinfo {author} {\bibfnamefont {R.~M.}\ \bibnamefont {Adar}}, \bibinfo {author} {\bibfnamefont {S.~A.}\ \bibnamefont {Safran}}, \bibinfo {author} {\bibfnamefont {H.}~\bibnamefont {Diamant}}, \ and\ \bibinfo {author} {\bibfnamefont {D.}~\bibnamefont {Andelman}},\ }\href {\doibase 10.1103/PhysRevE.100.042615} {\bibfield  {journal} {\bibinfo  {journal} {Phys. Rev. E}\ }\textbf {\bibinfo {volume} {100}},\ \bibinfo {pages} {042615} (\bibinfo {year} {2019})}\BibitemShut {NoStop}%
\bibitem [{\citenamefont {Antypov}, \citenamefont {Barbosa},\ and\ \citenamefont {Holm}(2005)}]{antypov_excluded_2005}%
  \BibitemOpen
  \bibfield  {author} {\bibinfo {author} {\bibfnamefont {D.}~\bibnamefont {Antypov}}, \bibinfo {author} {\bibfnamefont {M.~C.}\ \bibnamefont {Barbosa}}, \ and\ \bibinfo {author} {\bibfnamefont {C.}~\bibnamefont {Holm}},\ }\href {\doibase 10.1103/PhysRevE.71.061106} {\bibfield  {journal} {\bibinfo  {journal} {Phys. Rev. E}\ }\textbf {\bibinfo {volume} {71}},\ \bibinfo {pages} {061106} (\bibinfo {year} {2005})}\BibitemShut {NoStop}%
\bibitem [{\citenamefont {Anousheh}, \citenamefont {Solis},\ and\ \citenamefont {Jadhao}(2020)}]{anousheh_ionic_structure_2020}%
  \BibitemOpen
  \bibfield  {author} {\bibinfo {author} {\bibfnamefont {N.}~\bibnamefont {Anousheh}}, \bibinfo {author} {\bibfnamefont {F.~J.}\ \bibnamefont {Solis}}, \ and\ \bibinfo {author} {\bibfnamefont {V.}~\bibnamefont {Jadhao}},\ }\href {\doibase 10.1063/5.0028003} {\bibfield  {journal} {\bibinfo  {journal} {AIP Adv.}\ }\textbf {\bibinfo {volume} {10}},\ \bibinfo {pages} {125312} (\bibinfo {year} {2020})}\BibitemShut {NoStop}%
\bibitem [{\citenamefont {Onsager}\ and\ \citenamefont {Kim}(1957{\natexlab{b}})}]{onsager_wien_1957}%
  \BibitemOpen
  \bibfield  {author} {\bibinfo {author} {\bibfnamefont {L.}~\bibnamefont {Onsager}}\ and\ \bibinfo {author} {\bibfnamefont {S.~K.}\ \bibnamefont {Kim}},\ }\href {\doibase 10.1021/j150548a015} {\bibfield  {journal} {\bibinfo  {journal} {J. Phys. Chem.}\ }\textbf {\bibinfo {volume} {61}},\ \bibinfo {pages} {198} (\bibinfo {year} {1957}{\natexlab{b}})}\BibitemShut {NoStop}%
\bibitem [{\citenamefont {Lesnicki}\ \emph {et~al.}(2020)\citenamefont {Lesnicki}, \citenamefont {Gao}, \citenamefont {Rotenberg},\ and\ \citenamefont {Limmer}}]{lesnicki_field_2020}%
  \BibitemOpen
  \bibfield  {author} {\bibinfo {author} {\bibfnamefont {D.}~\bibnamefont {Lesnicki}}, \bibinfo {author} {\bibfnamefont {C.~Y.}\ \bibnamefont {Gao}}, \bibinfo {author} {\bibfnamefont {B.}~\bibnamefont {Rotenberg}}, \ and\ \bibinfo {author} {\bibfnamefont {D.~T.}\ \bibnamefont {Limmer}},\ }\href {\doibase 10.1103/PhysRevLett.124.206001} {\bibfield  {journal} {\bibinfo  {journal} {Phys. Rev. Lett.}\ }\textbf {\bibinfo {volume} {124}},\ \bibinfo {pages} {206001} (\bibinfo {year} {2020})}\BibitemShut {NoStop}%
\bibitem [{\citenamefont {Lesnicki}\ \emph {et~al.}(2021)\citenamefont {Lesnicki}, \citenamefont {Gao}, \citenamefont {Limmer},\ and\ \citenamefont {Rotenberg}}]{lesnicki_molecular_2021}%
  \BibitemOpen
  \bibfield  {author} {\bibinfo {author} {\bibfnamefont {D.}~\bibnamefont {Lesnicki}}, \bibinfo {author} {\bibfnamefont {C.~Y.}\ \bibnamefont {Gao}}, \bibinfo {author} {\bibfnamefont {D.~T.}\ \bibnamefont {Limmer}}, \ and\ \bibinfo {author} {\bibfnamefont {B.}~\bibnamefont {Rotenberg}},\ }\href {\doibase 10.1063/5.0052860} {\bibfield  {journal} {\bibinfo  {journal} {J. Chem. Phys.}\ }\textbf {\bibinfo {volume} {155}},\ \bibinfo {pages} {014507} (\bibinfo {year} {2021})}\BibitemShut {NoStop}%
\bibitem [{\citenamefont {Herrero}, \citenamefont {Bocquet},\ and\ \citenamefont {Coasne}(2026)}]{herrero_fluids_2026}%
  \BibitemOpen
  \bibfield  {author} {\bibinfo {author} {\bibfnamefont {C.}~\bibnamefont {Herrero}}, \bibinfo {author} {\bibfnamefont {L.}~\bibnamefont {Bocquet}}, \ and\ \bibinfo {author} {\bibfnamefont {B.}~\bibnamefont {Coasne}},\ }\href {https://arxiv.org/abs/2601.02539} {\enquote {\bibinfo {title} {Fluids at an electrostatically active surface: Optimum in interfacial friction and electrohydrodynamic drag},}\ } (\bibinfo {year} {2026}),\ \Eprint {http://arxiv.org/abs/2601.02539} {arXiv:2601.02539 [cond-mat.soft]} \BibitemShut {NoStop}%
\bibitem [{\citenamefont {Illien}, \citenamefont {Carof},\ and\ \citenamefont {Rotenberg}(2024{\natexlab{b}})}]{illien2024stochastic}%
  \BibitemOpen
  \bibfield  {author} {\bibinfo {author} {\bibfnamefont {P.}~\bibnamefont {Illien}}, \bibinfo {author} {\bibfnamefont {A.}~\bibnamefont {Carof}}, \ and\ \bibinfo {author} {\bibfnamefont {B.}~\bibnamefont {Rotenberg}},\ }\href@noop {} {\bibfield  {journal} {\bibinfo  {journal} {Phys. Rev. Lett.}\ }\textbf {\bibinfo {volume} {133}},\ \bibinfo {pages} {268002} (\bibinfo {year} {2024}{\natexlab{b}})}\BibitemShut {NoStop}%
\bibitem [{\citenamefont {Varghese}, \citenamefont {Illien},\ and\ \citenamefont {Rotenberg}(2025)}]{varghese_dynamic_2025}%
  \BibitemOpen
  \bibfield  {author} {\bibinfo {author} {\bibfnamefont {S.}~\bibnamefont {Varghese}}, \bibinfo {author} {\bibfnamefont {P.}~\bibnamefont {Illien}}, \ and\ \bibinfo {author} {\bibfnamefont {B.}~\bibnamefont {Rotenberg}},\ }\href {\doibase 10.1063/5.0292306} {\bibfield  {journal} {\bibinfo  {journal} {J. Chem. Phys.}\ }\textbf {\bibinfo {volume} {163}},\ \bibinfo {pages} {124107} (\bibinfo {year} {2025})}\BibitemShut {NoStop}%
\bibitem [{\citenamefont {Jardat}\ \emph {et~al.}(2000)\citenamefont {Jardat}, \citenamefont {Durand-Vidal}, \citenamefont {Turq},\ and\ \citenamefont {Kneller}}]{jardat_brownian_2000}%
  \BibitemOpen
  \bibfield  {author} {\bibinfo {author} {\bibfnamefont {M.}~\bibnamefont {Jardat}}, \bibinfo {author} {\bibfnamefont {S.}~\bibnamefont {Durand-Vidal}}, \bibinfo {author} {\bibfnamefont {P.}~\bibnamefont {Turq}}, \ and\ \bibinfo {author} {\bibfnamefont {G.~R.}\ \bibnamefont {Kneller}},\ }\href {\doibase 10.1016/S0167-7322(99)00163-4} {\bibfield  {journal} {\bibinfo  {journal} {J. Mol. Liq.}\ }\textbf {\bibinfo {volume} {85}},\ \bibinfo {pages} {45} (\bibinfo {year} {2000})}\BibitemShut {NoStop}%
\bibitem [{\citenamefont {Sammüller}\ \emph {et~al.}(2023)\citenamefont {Sammüller}, \citenamefont {Hermann}, \citenamefont {de~las Heras},\ and\ \citenamefont {Schmidt}}]{sammuller_neural_functional_2023}%
  \BibitemOpen
  \bibfield  {author} {\bibinfo {author} {\bibfnamefont {F.}~\bibnamefont {Sammüller}}, \bibinfo {author} {\bibfnamefont {S.}~\bibnamefont {Hermann}}, \bibinfo {author} {\bibfnamefont {D.}~\bibnamefont {de~las Heras}}, \ and\ \bibinfo {author} {\bibfnamefont {M.}~\bibnamefont {Schmidt}},\ }\href {\doibase 10.1073/pnas.2312484120} {\bibfield  {journal} {\bibinfo  {journal} {Proc. Natl. Acad. Sci. U.S.A.}\ }\textbf {\bibinfo {volume} {120}},\ \bibinfo {pages} {e2312484120} (\bibinfo {year} {2023})}\BibitemShut {NoStop}%
\bibitem [{\citenamefont {Bui}\ and\ \citenamefont {Cox}(2025)}]{bui_learning_classical_2025}%
  \BibitemOpen
  \bibfield  {author} {\bibinfo {author} {\bibfnamefont {A.~T.}\ \bibnamefont {Bui}}\ and\ \bibinfo {author} {\bibfnamefont {S.~J.}\ \bibnamefont {Cox}},\ }\href {\doibase 10.1103/PhysRevLett.134.148001} {\bibfield  {journal} {\bibinfo  {journal} {Phys. Rev. Lett.}\ }\textbf {\bibinfo {volume} {134}},\ \bibinfo {pages} {148001} (\bibinfo {year} {2025})}\BibitemShut {NoStop}%
\end{thebibliography}%

\end{document}